\documentclass[pra,aps,psfig,epsf,twocolumn,amsmath,amssymb,groupedaddress]{revtex4}

\newcommand{\rv}{{\bf r}}

\newcommand{\kv}{{\bf k}}

\newcommand{\eps}{{\varepsilon}}

\newcommand{\qv}{{\bf q}}

\newcommand{\jv}{{\bf j}}

\newcommand{\Nv}{\vec{N}}

\newcommand{\nh}{{\hat n}}

\newcommand{\mh}{{\hat m}}
\newcommand{\lh}{{\hat \ell}}

\newcommand{\oh}{{\frac{1}{2}}}

\newcommand{\cH}{{\mathcal H}}
\newcommand{\cL}{{\mathcal L}}
\def\rf#1{(\ref{#1})}
\def\rfs#1{Eq.~\rf{#1}}

\newcommand{\as}{a_s}

\newcommand{\curH}{{\mathcal H}}
\newcommand{\phdag}{{\phantom{\dagger}}}

\newcommand{\kf}{k_{\rm F}}

\newcommand{\tphi}{\tilde{\phi}}
\newcommand{\ttheta}{\tilde{\theta}}

\newcommand{\curL}{{\cal L}}

\newcommand{\bk}{{\bf k}}
\newcommand{\bq}{{\bf q}}
\newcommand{\bQ}{{\bf q}}
\newcommand{\bp}{{\bf p}}

\newcommand{\grad}{{\bm{\nabla}}}
\newcommand{\zh}{\hat{z}}

\newcommand{\ch}{\hat{c}}

\newcommand{\Vt}{\tilde{V}}

\newcommand{\br}{{\bf r}}

\newcommand{\be}{\begin{equation}}
\newcommand{\ee}{\end{equation}}
\newcommand{\bea}{\begin{eqnarray}}
\newcommand{\eea}{\end{eqnarray}}
\newcommand{\bse}{\begin{subequations}}
\newcommand{\ese}{\end{subequations}}

\newcommand{\fermiint}{g}

\newcommand{\Z}{{\mathbb{Z}}}
\newcommand{\ibQ}{{\textbf{\em Q}}}
\newcommand{\ve}{\vert}

\def\rf#1{(\ref{#1})}
\def\rfs#1{Eq.~\rf{#1}}

\input{epsf}
\usepackage{bm}
\usepackage{epsfig}

\begin{document}
\title{Quantum liquid-crystal order in resonant atomic gases}
\author{Leo Radzihovsky}
\affiliation{Department of Physics, University of Colorado, 
Boulder, CO 80309}

\date{\today}

\begin{abstract}
  I review recent studies that predict a realization of quantum
  liquid-crystalline orders in resonant atomic gases. As examples of
  such putative systems I will discuss an s-wave resonant imbalanced
  Fermi gas and a p-wave resonant Bose gas.  In the former, the
  liquid-crystalline smectic, nematic and rich variety of other
  descendant states emerge from strongly quantum- and thermally-
  fluctuating Fulde-Ferrell and Larkin-Ovchinnikov states, driven by a
  competition between resonant pairing and Fermi-surface mismatch. In
  the latter, at intermediate detuning the p-wave resonant interaction
  generically drives Bose-condensation at a finite momentum, set by a
  competition between atomic kinetic energy and atom-molecule
  hybridization.  Because of the underlying rotationally-invariant
  environment of the atomic gas trapped isotropically, the putative
  striped superfluid is a realization of a quantum superfluid smectic,
  that can melt into a variety of interesting phases, such as a
  quantum nematic. I will discuss the corresponding rich phase
  diagrams and transitions, as well the low-energy properties of the
  phases and fractional topological defects generic to striped
  superfluids and their fluctuation-driven descendants.
\end{abstract}
\pacs{}

\maketitle

\section{Introduction}
\label{intro}
\subsection{Resonant atomic gases}
Experimental progress in trapping, cooling and coherently manipulating
Feshbach-resonant atomic gases opened unprecedented opportunities to
study degenerate strongly interacting quantum many-body systems in a
broad range of previously unexplored regimes
\cite{BlochReview,KetterleZwierleinReview,GRaop,GiorginiRMP,RSreview}.
These include paired fermionic superfluids (SF)
\cite{Regal2004prl,Zwierlein2004prl,Kinast2004prl,Bartenstein2004prl,
  Bourdel2004prl,ZhangPRApwave,GaeblerPRLpwave,BotelhoSdeMeloPwave,
  GRApwave,ChengYipPRLpwave}), the associated
Bardeen-Cooper-Schrieffer (BCS) to Bose-Einstein condensation (BEC)
crossover\cite{Eagles,Leggett,NSR,SdeMelo,Timmermans01,Holland,Ohashi,
  AGRswave,Stajic,GRaop}, Bose-Fermi mixtures\cite{OlsenBFmixture},
bosonic molecular
superfluids\cite{Cornish2000prl,RPWprl,RomansPRL,RWPaop}, and many
other states and regimes\cite{Reviews} under both equilibrium and
nonequilibrium conditions\cite{BLprl,AGRswave,AltmanVishwanathPRL}.

Because degenerate atomic gases are free of the underlying crystalline
matrix of the solid-state materials (though one can be imposed through
a highly tunable optical lattice potential\cite{BlochReview}), among
this rich variety of states, they admit phases that {\em
  spontaneously} partially break {\em continuous spatial} symmetries
and thereby exhibit concomitant strongly fluctuating Goldstone modes
with corresponding rich phenomenology. Resonant atomic gases are thus
uniquely suited for a realization of quantum liquid-crystalline states
of matter, that have been somewhat of a holy-grail dating back to
their studies in solid state materials such as the striped states in
high-T$_c$ superconductors, nickelates and other strongly correlated
doped Mott insulators\cite{KFE98nature,FK99prb,OKF01prb,
  Berg09prb,Berg09nature,Sachdev03aop}, heavy-fermion and organic
superconductors\cite{Agterberg08nature,Agterberg08prl}, spiral states
in helimagnets\cite{BakJensen80,BKRprb06,Green09prl}, and a
two-dimensional electron gas with a partially-filled Landau
level\cite{Fogler96,MacdonaldFisher00prb,RD02prl}.  In isotropic traps
the putative quantum liquid-crystal order is expected to exhibit all
the complexity of fluctuations and topological defects of conventional
(mesogenic) liquid crystals\cite{deGennesProst,ChaikinLubensky}, but
with the added enrichment of the accompanying quantum (off-diagonal)
order of a superfluid. Another, not insignificant virtue is that (in
contrast to other e.g., solid-state or nuclear matter systems) these
dilute gases are extremely well-characterized at the two-body level,
and are therefore described by microscopic (as opposed to effective)
Hamiltonians with well-known couplings.

\subsection{Candidate systems}

Recent theoretical studies have predicted a number of such quantum
liquid crystal realizations in degenerate atomic systems, that in
addition to internal symmetries {\em partially} break spatial
symmetries\cite{FF,LO,Sedrakian05nematic,RCprl.09,HuiZhaiSOIbosons,
  YipSOIbosons, HuiZhaiSOIfermions}. These are typically driven by
strong resonant and competing interaction that frustrates a spatially
homogeneous and isotropic superfluidity. Known examples include
bosonic and paired fermionic superfluids, where spatial order is
driven by (i) dipolar interaction\cite{dipolarStamperKurn}, (ii)
pseudo-spin-orbit interaction\cite{WangZhaiSOIbosons} (realized
through hyperfine states coupled by Raman
transitions\cite{SpielmanSOInature}), (iii) p-wave resonant
interaction\cite{RCprl.09,CRpra.11}, and (iv) a Fermi surface mismatch
(realized through species number and/or mass imbalance), that leads to
the Larkin-Ovchinnikov-Fulde-Ferrell finite-momentum
pairing\cite{FF,LO}, as well as their strongly fluctuating descendent
states. In this brief review, I will focus on the last two
realizations, and will discuss the associated microscopic models that
I believe can realize a quantum superfluid liquid-crystal order, their
phase behavior, fluctuations, topological defects, and a variety of
experimental predictions and signatures. Much of the discussed
low-energy phenomenology is shared more generally by systems
exhibiting quantum liquid crystal orders in isotropic
trap\cite{HuiZhaiSOIbosons}. For a more complete account, I refer the
reader to the original literature and the more extensive
reviews\cite{KetterleZwierleinReview,RSreview}.

I will not discuss the finite momentum states that depend on the
lattice for their realization and stability, such as the p-band and
FFLO superfluids in optical
lattices\cite{kuklov.06,liu.06,TrivediFFLOlattice09,CWuFFLOlattice11,LiuLattices11}. These
are fascinating states, but are less relevant from the liquid-crystal
perspective of this review.

\section{Imbalanced resonant Fermi gases}
\subsection{Background}
The most widely explored candidate for a realization of quantum liquid
crystal order is a species-imbalanced Feshbach-resonant Fermi gas,
\cite{KetterleZwierleinReview,GRaop,RSreview}, though it is only very
recently that it was formulated and explored in these liquid-crystal
terms\cite{RVprl,Rpra}. These studies build on well-explored
two-species Feshbach-resonant Fermi gas, that exhibits paired
superfluidity, that can be tuned between a weakly-attractive
Fermi-surface-driven BCS and a strongly-attractive molecular BEC
superfluids\cite{GiorginiRMP,GRaop}. While at $T=0$ a balanced gas
exhibits no qualitative change of state, a quantitatively accurate
description of this crossover, particularly around the strongly
interacting and universal unitary regime (where in a vacuum a
two-particle bound state first forms and the s-wave scattering length
diverges) has presented a considerable challenge with much recent
progress.

\begin{figure}[thb]
\includegraphics[width=8.5cm,scale=1]{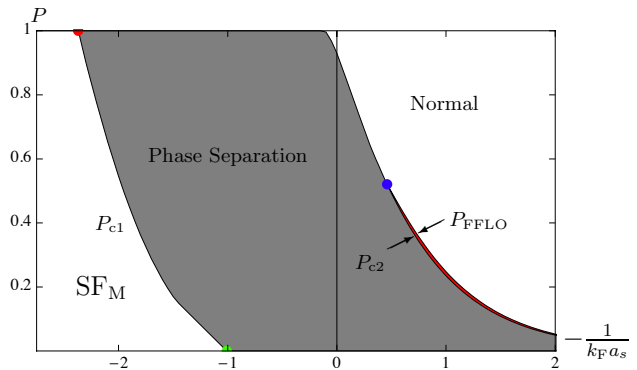}
\caption{A mean-field zero-temperature phase diagram from
  Refs.\onlinecite{SRprl,SRaop} of an imbalanced resonant Fermi gas,
  as a function of the inverse scattering length and normalized
  species imbalance $P=(N_\uparrow-N_\downarrow)/N\equiv\Delta N/N$,
  showing the magnetized (imbalanced) superfluid ($SF_{\rm M}$), the
  FFLO state (approximated as the simplest FF state) confined to a
  narrow red sliver bounded by $P_{\rm FFLO}$ and $P_{{\rm c}2}$, and
  the imbalanced normal Fermi liquid.}
\label{SRphasediagram}
\end{figure}

A species-number (and mass) imbalance in the two atomic
hyperfine-states mixture offered a new extremely fruitful experimental
knob\cite{Zwierlein06Science,Partridge06Science,Shin2006prl,Navon2009prl}.
The imbalance frustrates
pairing\cite{Combescot01,Liu03,Bedaque03,Caldas04,CarlsonReddy05,Cohen05},
driving quantum phase transitions out of the paired superfluid to a
variety of possible ground states and thermodynamic
phases\cite{Castorina05,Sedrakian05nematic,SRprl,Pao06,Son06,
  Bulgac06pwavePRL,Dukelsky}.  This rekindled considerable theoretical
activity in the context of species-imbalanced resonant Fermi
gases\cite{Mizushima,YangFFLOdetect,YangSachdev,Pieri,Torma,
  Yi,Chevy,He,DeSilva,HaqueStoof,SachdevYang,LiuHu,Chien06prl,
  Gubbels06prl,YiDuan,PaoYip,SRaop,SRcomment,Martikainen,Parish07nature,
  BulgacFFLO,Parish1dLO,Sheehy}. The corresponding imbalance versus
detuning BCS-BEC phase diagram, illustrated in
Fig.\ref{SRphasediagram} (and its extension to finite temperature) is
now well-established\cite{SRprl,SRaop,Parish07nature}, showing
qualitative agreement with
experiments\cite{Zwierlein06Science,Partridge06Science,
  Shin2006prl,Navon2009prl}. More recently, considerable progress has
been made toward establishing quantitative details of this phase
diagram through
analytical\cite{Nishida06eps,Haussmann06,Nikolic07largeN,Veillette07largeN,VeilletteRF},
numerical\cite{ProkofevSvistunovPolaron,PilatiGiorginiMCimblanced} and
experimental
approaches\cite{Shin2006prl,Navon2009prl,ZwierleinPrecision11}.

The identification of the number species imbalance with the
magnetization of an electronic system, and the chemical potential
difference with an effective Zeeman energy, connects these atomic
gases studies with a large body of research on solid state electronic
superconductors under a Zeeman
field\cite{Chandra,Clogston,Sarma,FF,LO}, as well as extensively
studied realizations in nuclear and particle
physics\cite{Alford,Bowers,Casalbuoni,Combescot}. The obvious
advantage of the newly-realized atomic system is the aforementioned
tunability, disorder-free ``samples'', and absence of the orbital part
of the magnetic field, that always accompanies a solid-state charged
superconductor in a magnetic field. In these neutral paired
superfluids the orbital field effects can be independently controlled
by a rotation of the atomic cloud\cite{DuineMacDonaldPRA}.

As illustrated in Fig.\ref{SRphasediagram}, among many interesting
features, such as the gapless imbalanced superfluid
($SF_M$)\cite{SRprl,SRcomment,Pao06,Son06}, ubiquitous phase
separation\cite{Bedaque03,SRprl,SRcomment,SRaop}, tricritical
point\cite{SRaop,Parish07nature,Sheehy}, etc., observed
experimentally\cite{Zwierlein06Science,Partridge06Science,Shin2006prl,
  Navon2009prl} and studied extensively theoretically\cite{RSreview},
the interaction--imbalance BEC-BCS phase diagram is also
predicted\cite{Mizushima,SRprl,SRaop,Son06,BulgacFFLO,Parish1dLO} to
exhibit the enigmatic Fulde-Ferrell-Larkin-Ovchinnikov state
(FFLO)\cite{FF,LO}. First predicted in the context of solid-state
superconductors over 45 years ago, the FFLO states has so
far eluded a definitive observation, though some promising solid
state\cite{evidenceFFLO} and quasi-1d atomic\cite{Hulet1dLO} candidate
systems have recently been realized.

At its most generic level the FFLO state is a fermionic superfluid,
paired at a finite center of mass momentum. It spontaneously
``breaks'' gauge and translational symmetry, a periodically-paired
superfluid (superconductor), akin to a
supersolid\cite{Andreev69,Chester70,Leggett70,KimChan}, and thus can
appropriately be called a pair-density wave
(PDW)\cite{commentNotSS,ZhangPDW}.  This state can be equivalently
thought of as a periodically ordered {\em micro}-phase separation
between the normal and paired states, that naturally replaces the {\em
  macro}-phase separation\cite{Combescot,Bedaque03} ubiquitously found
in the BCS-BEC detuning-imbalance phase
diagram\cite{SRprl,SRcomment,SRaop,Parish07nature}.

Microscopically, it is driven by Fermi surface mismatch\cite{FF,LO}
due to an imposed pairing species number (and/or
mass\cite{WuPaoYip06mass}) imbalance.  As a compromise between the
superfluid pairing and an imposed imbalance, at intermediate values of
the latter, the superconducting order parameter condenses at a set of
finite center-of-mass momenta determined by the details of the Fermi
surface mismatch and interactions. At sufficiently large imbalance, no
compromise is possible, and the resonant gas transitions to the normal
state.
%
%
%

%

\begin{figure}[thb]
\includegraphics[width=8.5cm,scale=1]{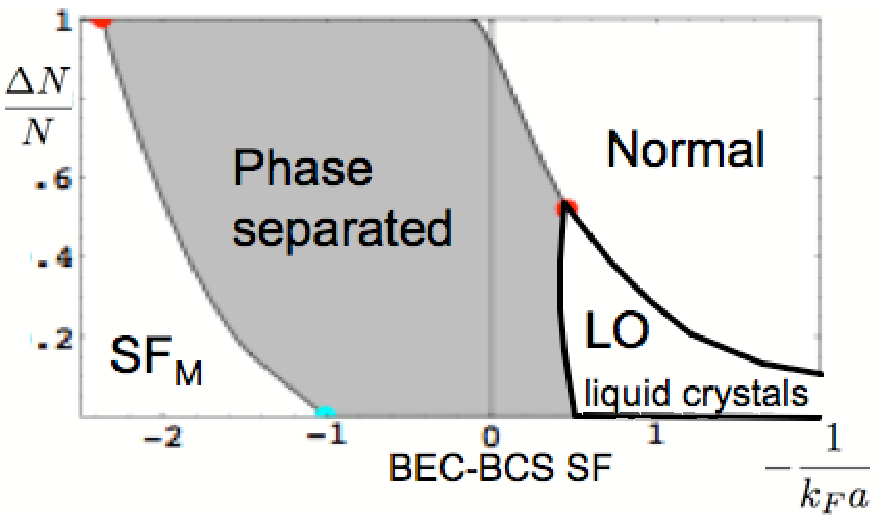}
\caption{A proposed $P=\Delta N/N$ vs. $1/(\kf\as)$ phase diagram for an
  imbalanced resonant Fermi gas, showing the more stable LO liquid
  crystal phases (discussed in the text and illustrated in detail
  Fig.\ref{fig:phasediagramLO3d}) replacing a portion of the
  phase-separated regime.}
\label{fig:phasediagramLOmft}
\end{figure}

As illustrated in Fig.\ref{SRphasediagram}, the key observation is
that, despite strong interactions, within simplest mean-field
treatments the conventional FFLO state\cite{FF,LO} remains quite
fragile, confined to a narrow sliver of polarization in the BCS
regime\cite{SRprl,SRaop,commentStabilizeFFLO,RSreview}.  I emphasize
that in fact above conclusion is only rigorously valid for the FF and
not other forms (e.g., LO) of the FFLO class of states.  Although the
upper boundary, $h_{c2}$, just below the normal state is trustworthy,
as it is shared by all FFLO states, the lower one, $h_{c1}$ can
strongly depend on the form of the FFLO state, but was determined by
Sheehy and Radzihovsky only for the FF
state\cite{SRprl,SRaop,LevinFFLO}. Furthermore, motivated by earlier studies of
the Bogoluibov-de Gennes (BdG) equation for the LO
state\cite{MachidaNakanishiLO,BurkhardtRainerLO,MatsuoLO}, combined
with finding of a negative domain-wall energy in an otherwise
fully-paired singlet BCS superfluid in Zeeman
field\cite{MatsuoLO,YoshidaYipLO}, these studies have quite
convincingly argued, that a more generic pair-density wave state (that
includes a larger set of collinear wavevectors\cite{Alford,Bowers})
may be significantly more stable. 
%
\begin{figure}[thb]
\includegraphics[width=8.5cm,scale=1]{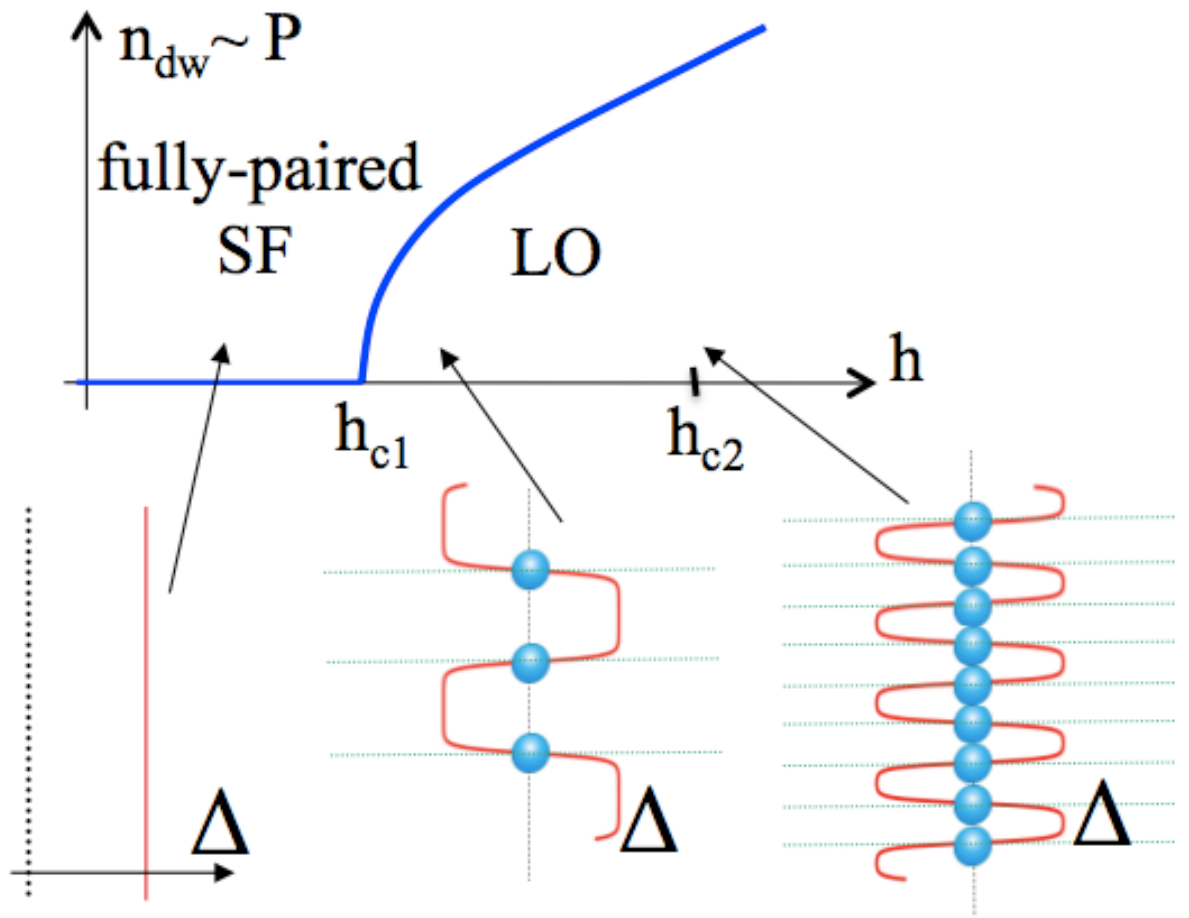}
\caption{An illustration of a continuous commensurate-incommensurate
  (CI) transition at $h_{c1}$ from a fully-gapped (balanced)
  paired-superfluid to an imbalanced Larkin-Ovchinnikov
  superfluid. The excess majority atoms are localized on the domain
  walls in (zeros of) the LO order parameter, whose number $n_{dw}(h)$
  is then proportional to the imbalance $P(h)$ and grows continuously
  with the chemical potential difference (Zeeman energy), $h-h_{c1}$.}
\label{fig:CItransition}
\end{figure}
%
The quantitative extent of the
energetic stability of a PDW states in the imbalance-detuning phase
diagram, in my view remains a widely open and urgent
question. Consistent with above arguments but in absence of controlled
quantitative analysis, I take the optimistic point of view that the LO
states can extend over significantly wider region of the phase
diagram, as schematically illustrated in
Fig.\ref{fig:phasediagramLOmft}.  Assuming it is indeed energetically
stable, its phenomenology has been explored beyond its mean-field
cartoon\cite{FF,LO} (latter only appropriate in the solid state and
optical lattices, but not in the isotropically-trapped resonant atomic
gases, where fluctuations are large).
%
%

\subsection{Summary}

Before turning to details, I summarize the salient features of the
isotropically-trapped collinear class of FFLO states, beyond their
mean-field approximation. The key observation\cite{RVprl,Rpra} is that
such striped states spontaneously break {\em continuous} rotational
symmetry, and as a result exhibit phonon-like Goldstone modes that are
{\em qualitatively} (energetically) softer than the solid-state
analogs (where only {\em discrete} rotational symmetry can be broken)
and all other superfluids. Namely, striped FF\cite{Shimahara} and LO
classes of states are characterized by highly anisotropic (with the
modulation wavevector $\qv_0=q_0\zh$ and $\perp$ transverse to it)
collective spectra, with
\begin{equation}
\omega(k_\perp,k_z)\sim \sqrt{K k_\perp^4 + B k_z^2},
\end{equation}
rather than by the usual linear-in-momentum Bogoluibov sound mode. The
more experimentally relevant striped LO state, also exhibits a
quantitatively anisotropic Bogoluibov linear-in-$k$ sound mode, but
with the superfluid stiffness ratio $\rho^s_\perp/\rho^s_z$ that
vanishes as the $h_{c2}$ transition to the normal imbalanced Fermi gas
is approached from below.

As a result, the fluctuations in such "soft" superfluid smectic states
are qualitatively stronger. Although the states are stable to quantum
fluctuations, in 3d the LO and FF long-range orders are marginally
unstable at any nonzero temperature. Consequently, (seemingly
paradoxically) inside the LO state the average LO order parameter
\begin{eqnarray}
  \langle\Delta_{LO}(\rv)\rangle_R
  &=&\langle2\Delta_{q_0}e^{i\phi(\rv)}\cos\big(\qv_0\cdot\rv +
  \theta(\rv)\big)\rangle_R,\nonumber\\
&\sim&\frac{1}{R^\eta}\cos\qv_0\cdot\rv\ \ \longrightarrow\ \ 0,
\label{LOvanishSummary}
\end{eqnarray}
vanishes in the thermodynamic limit (of a large cloud with atom number
$N$ and cloud size $R\rightarrow\infty$), suppressed to zero by
thermal phonon $\theta/q_0$ fluctuations. The LO state is therefore
strictly speaking homogeneous on long scales, exhibiting ``algebraic
topological'', but no long-ranged translational order. Namely, the
mean-field approximation fails qualitatively and the state instead is
characterized by power-law order-parameter correlations, distinguished
from the spatially short-ranged disordered phase by confined
topological defects (bound dislocations), not by a nonzero LO order
parameter. It is therefore a 3d analog of the more familiar
quasi-long-range ordered superfluid film, a 2d easy-plane ferromagnet
and a 2d
crystal\cite{Landau1dsolid,MerminWagner,Hohenberg,Berezinksii,KT}.

As a consequence, a 3d LO state is characterized by a static
structure function $S(\qv)$ and momentum distribution function
$n(\kv)$ with universal anisotropic {\em quasi}-Bragg peaks (around
$q_0$), akin to the Landau-Peierls\cite{Landau1dsolid,Peierls1dsolid}
behavior of films of a conventional superfluid and 2d
crystals\cite{MerminWagner,Hohenberg,Berezinksii,KT}.  Such novel
behavior is not, however, exhibited by 3d crystalline FFLO states with
multiple {\em non-collinear} ordering wavevectors\cite{Bowers,Alford},
that, in contrast are characterized by the long-range positional order
and a nonzero pair-condensate, that is stable to thermal fluctuations.

Another fascinating feature that arises because the LO order
parameter, $\Delta_{LO}$ is a product (rather than the sum, as it is
in a conventional supersolid) of the superfluid and density-wave
component, is the unusual topological excitation that is a half-vortex
bound to a half-dislocation -- in addition to integer vortices and
dislocations.

In 2d, at nonzero $T$ the LO state is even more strongly disordered,
characterized by short-range positional order with Lorentzian
structure function peaks, and unstable to proliferation of
dislocations\cite{TonerNelsonSm}.  The state that results from such
dislocated superfluid smectic is either a ``charge''-4 (paired Cooper
pairs) nematic superfluid\cite{RVprl,Berg09nature} or a nematic
(possibly ``fractionalized'') Fermi liquid\cite{OKF01prb,RD02prl},
latter qualitatively the same as the deformed Fermi surface state
\cite{Sedrakian05nematic}.

\begin{figure}[thb]
\vspace{0.5cm}
\includegraphics[width=9.5cm,scale=1]{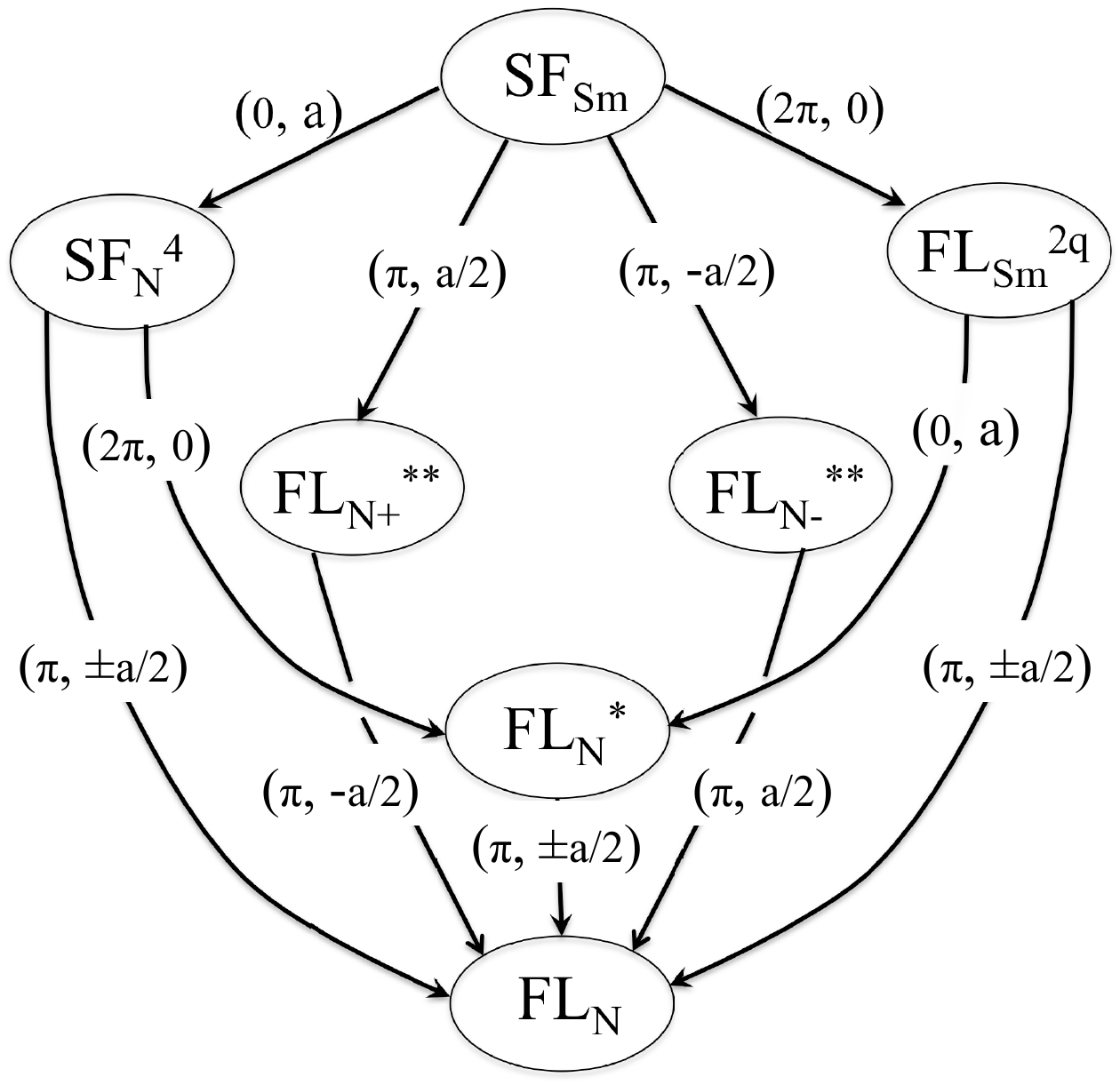}
\caption{A flowchart of superfluid ($SF$) and nonsuperfluid ($FL$)
  phases, exhibiting smectic ($Sm$) and nematic ($N$) conventional
  orders as well as topological orders (indicated by $*$ and $**$),
  induced by a proliferation of various combination of topological
  defects, $(0,a)$, $(2\pi,0)$, and $(\pi,\pm a/2)$.}
\label{fig:flowchart}
\end{figure}

\begin{widetext}

%
\begin{figure}[thb]
\vspace{0.5cm}
\includegraphics[width=15cm,scale=1]{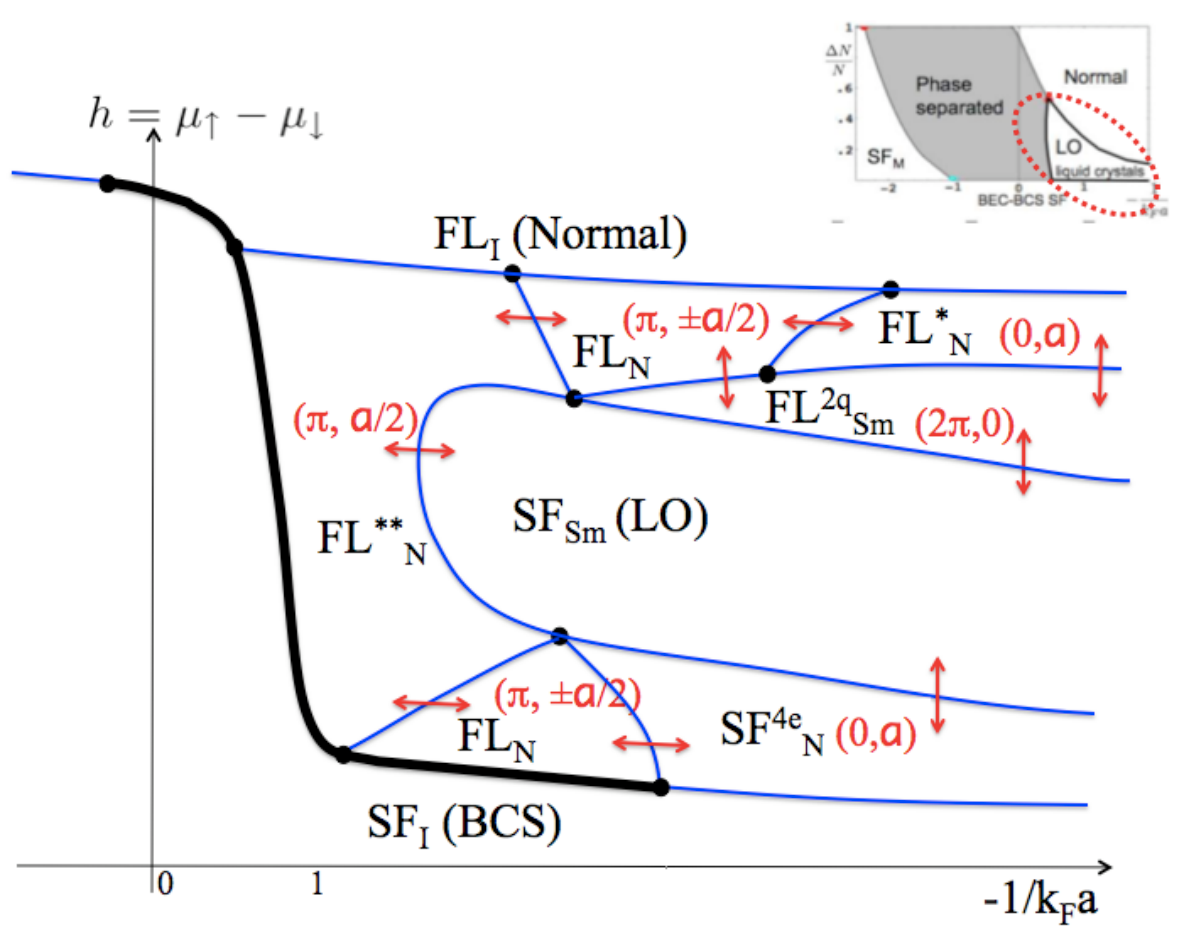}
\caption{A schematic imbalance-chemical potential (Zeeman energy),
  $h=\mu_\uparrow-\mu_\downarrow$ vs detuning (interaction strength),
  $-1/k_Fa$ phase diagram, illustrating the 3d LO smectic phase
  ($SF_{Sm}$) and its descendant (described in the text), driven by a
  proliferation of various combinations of topological defects.  The
  inset shows the global imbalance-interaction BCS-BEC phase diagram,
  illustrating the location of these putative phases.}
\label{fig:phasediagramLO3d}
\end{figure}


\end{widetext}

Furthermore, a consideration of states that arise due to unbinding of
various combination of topological defects (illustrated in the
flow-chart in Fig.\ref{fig:flowchart}) leads to a rich array of LO
descendent states, that generically must intervene between the LO
superfluid and a fully-paired conventional (isotropic and homogeneous)
superfluid and a conventional polarized Fermi liquid.  If indeed, as
argued above, the 3d LO state is energetically stable, these novel
states are expected to appear in the region collectively denoted ``LO
liquid crystals'' of the detuning-polarization phase diagram of
Fig.\ref{fig:phasediagramLOmft}. They include a nonsuperfluid smectic
($FL_{Sm}^{2q}$, driven by an unbinding of integer $2\pi$-vortices),
and a superfluid ($SF_{N}^4$, driven by a proliferation of integer
$a$-dislocations) and a nonsuperfluid ($FL_{N}$, driven by an
unbinding of both vortices and dislocations) nematics, and the
corresponding isotropic states, when disclinations also condense. In
addition, a variety of topologically-ordered isotropic and nematic
``fractionalized'' Fermi-liquid states ($FL_N^{*}$, $FL_N^{**}$,
$FL_I^{*}$, and others) were predicted\cite{RVprl,Rpra}, that are
distinguished from their more conventional fully-disordered forms by
{\em gapped} (bound) half-integer defects. These phases are summarized
by a flowchart Fig.\ref{fig:flowchart} and a schematic phase diagram
illustrated in Fig.\ref{fig:phasediagramLO3d}.
%
%

Finally, the fermionic sector of the LO gapless superconductor is also
quite unique, exhibiting a Fermi surface of Bogoliubov quasiparticles
associated with the Andreev band of states, localized on the array of
the LO domain walls. Consequences of the interplay between these
fermionic and Goldstone mode degrees of freedom remains an open
problem. 

\subsection{Microscopic model of imbalanced Fermi gas}

The physics of imbalanced atomic Fermi gases interacting through a
broad Feshbach resonance is well captured by the one-channel model
\cite{GRaop,GiorginiRMP,RSreview,Schrieffer}, characterized by a grand-canonical
Hamiltonian
\begin{eqnarray}
  H = \sum_{\bk,\sigma}(\epsilon_k - \mu_\sigma) \ch_{\bk\sigma}^\dagger  
  \ch_{\bk\sigma}^\phdag 
  + \fermiint\sum_{\bk\bq\bp} 
  \ch_{\bk\uparrow}^\dagger \ch_{\bp\downarrow}^\dagger 
  \ch_{\bk+\bq\downarrow}^\phdag\ch_{\bp-\bq\uparrow}^\phdag.
\label{eq:Hfermi}
\end{eqnarray}
with the single-particle energy $\epsilon_k =\hbar^2 k^2/2m$.  The
separately conserved number $N_\sigma = (N_\uparrow,N_\downarrow)$ of
atomic species (hyperfine states) $\sigma = \uparrow,\downarrow$) are
imposed by two chemical potentials,
$\mu_\sigma=(\mu_\uparrow,\mu_\downarrow)$. 

The key feature that distinguishes this Fermi system from those
familiar from solid state contexts is the attractive {\em resonant}
interaction parameterized by a short-range s-wave pseudopotential,
$\fermiint<0$. Through an exact T-matrix scattering
calculation\cite{GRaop}, $\fermiint$ controls the magnetic-field
tunable~\cite{RegalJinRF03} scattering length
\bse
\begin{eqnarray}
\label{eq:deltalength} 
\as(\fermiint)
&=&\frac{m}{4\pi}\frac{\fermiint}{1+\fermiint/\fermiint_c},
\end{eqnarray}
\label{as}
\ese
that diverges above a critical attraction strength,
$\fermiint_c=\frac{2\pi^2}{\Lambda m}$ ($\Lambda\sim 1/d$ is the
short-scale pseudo-potential cutoff set by the extent of the molecular
bound state), corresponding to a formation of a two-atom bound state.

The many-body thermodynamics of the resonant Fermi gas as a function
of $N,\Delta N, T, \as$, (i.e., the extension of the BEC-BCS crossover
to a finite imbalance, $\Delta N$) at large $k_F|\as|$ presents a
formidable challenge. However, much progress has been made in mapping
out its qualitative (and in some regimes quantitative) phenomenology
through a variety of approximate theoretical techniques, including
quantum Monte Carlo~\cite{CarlsonReddy05}, mean-field
theory~\cite{SRprl,SRcomment,Pao06,SRaop,Parish07nature}, the
large-$N_f$ (fermion flavor)~\cite{Nikolic07largeN,Veillette07largeN}
and $\epsilon$-expansions~\cite{Nishida06eps}.

The simplest of these is the standard mean-field 
analysis\cite{SRprl,SRaop} that gives a satisfactory qualitative
description (quantitatively valid deep in the weakly-coupled BCS
regime, $k_F|\as|\ll 1$), as a starting point of more sophisticated
treatments. To this end we assume the existence of a condensate
\begin{eqnarray}
\label{eq:Delta_q}
\Delta(\br)&=&\sum_{\bq}\Delta_{\bq} {\rm e}^{i{\bq}\cdot\br}
=\fermiint\langle \ch_\downarrow(\br) \ch_\uparrow(\br) \rangle, 
\end{eqnarray}
corresponding to pair-condensation at momenta $\bq$, with the set of
amplitudes $\Delta_\bq$ and $\bq $ to be self-consistently determined
by the minimizing the ground state energy subject to the constraints
of fixed total atom number $N=N_\uparrow + N_\downarrow$ and the atom
species number imbalance (``polarization'') $\Delta N = N_\uparrow -
N_\downarrow$, imposed by the average and difference chemical
potentials $\mu, h= \oh(\mu_\uparrow \pm \mu_\downarrow)$, latter
corresponding to the pseudo Zeeman energy.

Specializing to the simplest case of a single $\qv$ of the
Fulde-Ferrell state\cite{FF} this reduces the Hamiltonian to a
quadratic Bogoluibov form, that can be easily diagonalized. This
gives\cite{SRprl,SRaop} the ground state energy
\begin{eqnarray}
E^{FF}_{GS}&=& \sum_\bk \big(
\eps_{k} - E_{k}
+ E_{\bk\uparrow} \Theta(-E_{\bk\uparrow}) 
+ E_{\bk\downarrow} \Theta( -E_{\bk\downarrow})\big)\nonumber\\
&&-\frac{1}{\fermiint}|\Delta_\bq|^2,
\label{Egs}
\end{eqnarray}
and the excitation spectrum $E_{\bk\sigma}$
\bea
\label{eq:Esigma}
E_{\bk\uparrow/\downarrow} &=& E_k \mp h \mp \frac{\bk \cdot \bq}{2m},
\label{ekuparrow}
\eea with $\eps_k \equiv \frac{k^2}{2m} - \mu + \frac{q^2}{8m}$ and
$E_k \equiv (\eps_k^2 +\Delta_\bQ^2)^{1/2}$. $E^{FF}_{GS}$ (and its
generalizations to finite-temperature free
energy\cite{Parish07nature}), then gives all the thermodynamics,
including the phase behavior summarized by the phase diagram in
Fig.\ref{SRphasediagram}.

In particular, this analysis predicts the existence of the FF state,
stable only over a narrow sliver of imbalance, closing down for
$-1/(k_F a) > 0.5$~\cite{SRprl,SRaop}.  As mentioned in the
Introduction, there are compelling arguments suggesting that this
fragility is specific to the single $\qv$ planewave FF condensate, and
that the more generic PDW states are far more stable because they
allow energetically important amplitude
modulation~\cite{MachidaNakanishiLO,BurkhardtRainerLO,MatsuoLO,YoshidaYipLO,Rpra}.
However, general PDW states are difficult to analyze at the transition
from the fully gapped, balanced paired superfluid, near the lower
critical field $h_{c1}$ (at vanishing imbalance). In contrast, an
analytic analysis near the upper-critical chemical potential
difference $h_{c2}$ at the transition from the normal state is indeed
possible as a controlled Ginzburg-Landau expansion in the small
pair-density wave amplitude,
$\Delta_\qv$\cite{LO,RVprl,SamokhinFFLO,Rpra}. While not
quantitatively accurate away far below the $h_{c2}$ transition (where
PDW order parameter is large and is not limited to a single Fourier
component $\Delta_q$) such Landau expansion is expected to be
qualitatively correct and is a good starting point for a more complete
analysis of fluctuations and phase transitions into the PDW LO state.

\subsection{Order-parameter theory of FFLO states}

\subsubsection{Ginzburg-Landau expansion near $h_{c2}$}

The analytical treatment of the FFLO states near $h_{c2}$ relies on
the Ginzburg-Landau expansion in $\Delta_\bq$, that is small near the
(in mean-field) continuous $h_{c2}$ normal-to-FFLO transition
\cite{LO,commentMFTtrans}. This expectation is supported by the exact
1d BdG solution\cite{MachidaNakanishiLO} at high fields, where
$\Delta(x)$ is indeed well-approximated by a single sinusoid, with an
amplitude $\Delta_q$ that vanishes continuously near $h_{c2}$.

Consistent with these general arguments, by integrating out the atomic
degrees of freedom, near $h_{c2}$ the Ginzburg-Landau expansion for
the ground-state energy takes a familiar form
\begin{eqnarray}
\cH&\approx&\sum_\bq\eps_q|\Delta_\qv|^2 
+ \sum_{\{\bq_i\}}V_{\qv_1,\qv_2,\qv_3,\qv_4}
\Delta_{\qv_1}^*\Delta_{\qv_2}\Delta_{\qv_3}^*\Delta_{\qv_4},\nonumber\\
\label{HDelta_qhc2}
\end{eqnarray}
where the dispersion is given by\cite{LO,SRprl,SRaop,Rpra}
\begin{eqnarray}
\eps_q&\approx& \frac{3n}{4\epsilon_F}\left[-1
+ \frac{1}{2}\ln\frac{v_F^2q^2-4h^2}{\Delta_{BCS}^2}
+ \frac{h}{v_Fq}\ln\frac{v_F q+2h}{v_F q-2h}\right],\nonumber\\
&\approx& J(q^2-q_0^2)^2 + \eps_{q_0},
\label{epsSR}
\end{eqnarray}
whose minimum at a finite $q_0(h)\approx 1.2 h/v_F$ (near $h_{c2}$)
captures the imbalanced atomic Fermi system's energetic tendency to
pair at a finite momentum, and thereby to form a pair-density wave
characterized by a reciprocal lattice vector with magnitude $q_0$ and
a spontaneously chosen orientation. The value of $h$ at which
$\eps_{q_0}$ vanishes determines the corresponding mean-field N-FFLO
transition point. While at quadratic order, all Fourier modes with
$|\qv| = q_0$ are degenerate, becoming unstable simultaneously, the
form of the FFLO state is dictated by the interaction vertex
function, $\Vt_{\qv_1,\qv_2,\qv_3,\qv_4}$ that has been explicitly
computed.  Near the transition the physics of a {\em unidirectional}
pair-density wave (Cooper-pair stripe) order, characterized by a {\em
  collinear} set of $\qv_n$'s is well captured by focusing on
long-wavelength fluctuations of these most unstable modes, well
described by a Ginzburg-Landau Hamiltonian density
\begin{equation}
\cH = J\left[|\nabla^2\Delta|^2 - 2q_0^2|\nabla\Delta|^2\right] +
r|\Delta|^2 
+ \frac{1}{2} \lambda_1|\Delta|^4 + \frac{1}{2}\lambda_2 \jv^2,
\label{H_GL} 
\end{equation}
where deep in the BCS limit, near the $h_{c2}$ the model parameters
are given by
\bse
\begin{eqnarray}
J&\approx&\frac{0.61n}{\epsilon_Fq_0^4},\  
q_0\approx\frac{1.81\Delta_{BCS}}{\hbar v_F},\ 
r\approx\frac{3n}{4\epsilon_F}\ln\left[\frac{9h}{4h_{c2}}\right],\ \ \
\ \ \ \ 
\\
h_{c2}&\approx&\frac{3}{4}\Delta_{BCS},\ 
\lambda_1\approx\frac{3n}{4\epsilon_F\Delta_{BCS}^2},\  
\lambda_2\approx\frac{1.83n m^2}{\epsilon_F\Delta_{BCS}^2q_0^2},\ \ \
\ \ \ \ \
\end{eqnarray}
\label{Jmoduli}
\ese
and the inclusion of the current-current interaction, ${\bf
  j}=\frac{1}{m}\text{Re}
\left[-\Delta^*(\rv)i\nabla\Delta(\rv)\right]$ is necessary for a
complete description.  More generally, away from the weak-coupling BCS
limit these couplings can be taken as phenomenological parameters to
be determined experimentally, but the general form of the
Ginzburg-Landau model has broader range of applicability, capturing
all the qualitative features of the transition and the PDW state.

\subsubsection{Larkin-Ovchinnikov state near $h_{c1}$}
\label{sec:LOhc1}

However, the derivation and expressions for the associated couplings,
Eqs.\rf{Jmoduli} are limited to the weak coupling BCS regime and near
the high chemical potential imbalance (Zeeman field) normal-to-FFLO
transition at $h_{c2}$.

In a complementary, low chemical potential imbalance regime, just
above the transition from the fully-paired (BCS-BEC) superfluid to the
LO state at $h_{c1}$, a phenomenological analysis is
possible\cite{Rpra}. It treats the LO state as a periodic array of
fluctuating $\pm \Delta$ domain-walls (stripes) in $\Delta(\rv)$, akin
to the lyotropic phases in soft condensed
matter\cite{deGennesProst,ChaikinLubensky}.

However, such approach implicitly assumes that as the domain-wall
surface energy becomes
negative\cite{MachidaNakanishiLO,BurkhardtRainerLO,MatsuoLO,YoshidaYipLO}
for $h > h_{c1}$, their interaction remains {\em repulsive}, and so
the domain-walls proliferate {\em continuously} as a periodic array
inside the LO state. Under this assumption (that warrants further
study) the domain-wall density $n_{dw}$ and the associated species
imbalance $P\propto n_{dw}$ ($\approx q_0(h)$) is then set by a
balance between the negative surface energy and the domain-wall
repulsion, growing continuously as a function of $h-h_{c1}$ according
to the Pokrovsky-Talapov's commensurate-incommensurate (CI) transition
phenomenology\cite{PokrovskyTalapov}. This behavior is clearly
exhibited in 1d\cite{MachidaNakanishiLO,YangLL,ZhaoLiuLL} through an
exact solution and bosonization methods, and has been argued to
persist in higher dimensions
\cite{MachidaNakanishiLO,BurkhardtRainerLO,MatsuoLO,YoshidaYipLO}. The
CI route for a transition to the LO state contrasts sharply with the
Landau theory\cite{LO,SRprl,SRaop} of two independent order parameters
$\Delta_0$, $\Delta_{q}$, that always predicts a first-order BCS-LO
transition. The latter corresponds to the case of an {\em attractive}
domain-wall interaction, that therefore proliferate discontinuously
above $h_{c1}$, leading to the ubiquitous phase separation found in
mean-field theory\cite{SRprl,SRaop}. It is currently unclear what
dimensionless microscopic parameter, analogous to Abrikosov's $\kappa$
(distinguishing between type I and type II
superconductors)\cite{deGennes,Tinkham}, controls these two
alternatives of the macro-phase separation (a first-order transition)
and the micro-phase separated LO state (a continuous transition out of
the gapped SF state)\cite{commentMFTtrans}. A detailed analysis of
such low-imbalance approach to the SF-LO transition and the LO state
is sorely missing and is a subject of current research.

A semi-phenomenological local density approximation (LDA) model that
assembles all known ingredients is given by
\begin{eqnarray}
H[\Delta(\rv)]&\simeq& \int_\rv\left[\frac{J}{2}(|\nabla^2\Delta|^2 
-2q_0^2|\nabla\Delta|^2) + V(\Delta(\rv))\right],\nonumber\\
\label{eq:EgDelta_r}
\end{eqnarray}
where $\Delta_0 = e^{1/2}\Delta_{BCS}$ and $V(\Delta) =
-2\nu(\mu)\Delta^2\ln\left(\frac{\Delta}{\Delta_0}\right)
-\nu(\mu)\big[h\sqrt{h^2 -\Delta^2} - \Delta^2 \cosh^{-1}
(h/\Delta)\big] \Theta(h - \Delta)$ is the effective potential derived
within a BdG analysis for a uniform $\Delta$. It fully captures the
double-minimum structure and the associated 1st-order normal to
(fully-gapped BCS) superfluid transition that skips the interesting
intermediate e.g., the FFLO states. This LDA potential is supplemented
by the gradient energy functional, inherited from a microscopic GL
analysis (valid only near $h_{c2}$). Its use near $h_{c1}$ is
supported by the fact that it is the simplest form that incorporates
the underlying symmetries and encodes the expected energetics for the
system to order at a finite $h$-dependent momentum even near
$h_{c1}$. Work is under way to derive this functional through a
controlled Moyal (semi-classical) gradient expansion on the BdG
Hamiltonian.

The functional $E_G[\Delta(\rv)]$ has a double-well structure, with an
additional normal state minimum (at $\Delta=0$) developing for
$h>\Delta_{BCS}/2$. It thus allows periodic soliton structure in
$\Delta(\rv)$, corresponding to oscillations between the minima at
$\pm\Delta_{BCS}$.
\begin{figure}[tbp]
\includegraphics[scale=0.65]{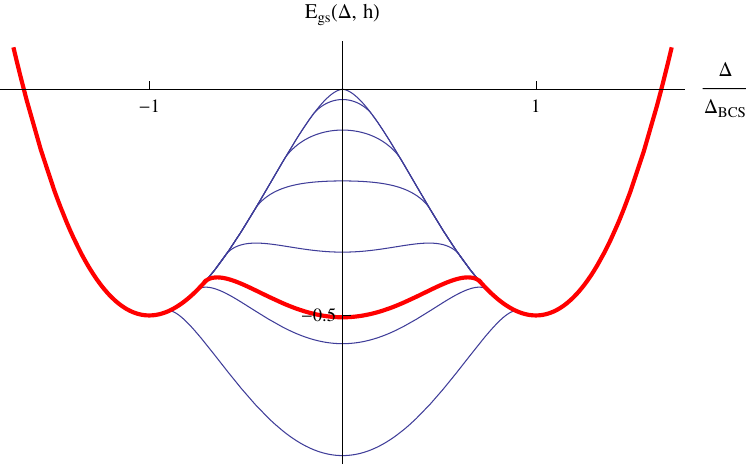}
\caption{Ground state energy $E_{gs}[\Delta,h]$ as a function of the
  order parameter $\Delta$ and Zeeman energy $h$, indicating a
  first-order transition at $h_c$ (thick red curve) between the paired
  superfluid state ($\Delta_{BCS}$) and the normal state ($\Delta=0$),
  at fixed density and/or imbalance exhibiting phase-separated
  coexistence. As argued in the text, analogously to the
  Pokrovsky-Talapov systems and type-II superconductors, this
  first-order transition can be preempted by a continuous CI-like
  transition to a striped LO superfluid and its descendent quantum
  liquid-crystal states, illustrated in Fig.\ref{fig:CItransition}.}
\end{figure}

\subsection{Goldstone modes in striped FFLO states}

Using the Ginzburg-Landau model one can develop a low-energy Goldstone
modes description of the striped paired states and use it to analyze
their stability to fluctuations. We focus on the {\em unidirectional}
(striped) pair-density wave states, with FF and LO states as simplest
representatives of two qualitatively distinct universality
classes. The corresponding order parameter is given by
\begin{equation}
  \Delta_{FFLO}(\rv) =\Delta_+(\rv) e^{i\qv\cdot\rv} + 
  \Delta_-(\rv) e^{-i\qv\cdot\rv},
\label{DeltaFFLOcollinear}
\end{equation}
where $\Delta_\pm(\rv)$ are two complex scalar order parameters, the
dominant Fourier coefficients $\Delta_\pm (\rv)= \Delta_\pm^0(\rv)
e^{i\phi_\pm(\rv)}$ and amplitudes $\Delta_\pm^0$ distinguishing
between the FF and LO states. Using this representation inside $\cH$
and minimizing the ground-state energy for $h<h_{c2}$ a simple
analysis shows that indeed it is the LO state with
$\Delta_+=\Delta_-\neq 0$ that is most stable inside the BCS regime;
the FF state is characterized by only one nonvanishing order
parameter, $\Delta_+\neq 0, \Delta_-= 0$\cite{LO}. More generally,
either state can be stabilized depending on the relative magnitudes of
$\lambda_1$ and $\lambda_2$. 

\subsubsection{Fulde-Ferrell state}

The FF state is characterized by a single (independent) nonzero
complex order parameter, 
\begin{eqnarray}
\Delta_{FF}(\rv)=\Delta_{q_0} e^{i\qv_0\cdot\rv +i\phi},
\end{eqnarray}
that is a plane-wave with the momentum $\qv_0$ and a single Goldstone
mode $\phi=\phi_+$ corresponding to the local superconducting phase.
The state carries a nonzero, uniform spontaneously-directed
supercurrent $\jv_{FF} = \frac{1}{m}|\Delta_{q_0}|^2(\qv_0+\grad\phi)$
and thereby breaks the time-reversal and rotational symmetry, chosen
spontaneously along $\qv_0$, as well as the global gauge symmetry,
corresponding to the total atom conservation.  Although the FF order
parameter itself is not translationally invariant, it is invariant
under a modified transformation of an arbitrary translation followed
by a gauge transformation, with all gauge-invariant observables thus
translationally invariant. Thus, the FF state is a uniform
orientationally-ordered polar superfluid. The underlying rotational
invariance also demands that it is invariant under a rotation of
$\qv_0=q_0\zh$ by an angle $\alpha$ that generates a nontrivial,
spatially-dependent phase $\phi^0(\rv)=z(\cos\alpha-1) +
x\sin\alpha$. Simple algebra demonstrates that the fully nonlinear
form of the longitudinal current $\delta
j_\parallel=\partial_\parallel\phi + \oh q_0^{-1}(\nabla\phi)^2$
ensures that it and the corresponding energy $\cH_{FF}$ vanish for
$\phi^0(\rv)$, as required by the underlying rotational invariance.

The analysis of the GL functional lead to a Goldstone mode $\phi$
Hamiltonian\cite{Shimahara,RVprl,Rpra}
\begin{eqnarray}
\cH_{FF} &=& \oh K (\nabla^2\phi)^2 + 
\oh\rho_s^\parallel\big(\partial_\parallel\phi 
+ \oh q_0^{-1}(\nabla\phi)^2\big)^2,\ \ \
\label{HGMff}
\end{eqnarray}
where $\partial_\parallel\equiv\hat\qv_0\cdot\grad$, $\rho^{||}_s =
8Jq^2|\Delta œôø¦_{q_0}|^2$ is the superfluid stiffness along $q_0$ and
$K = 2J|\Delta_{q_0}|$.  The Hamiltonian form, $\cH_{FF}$ is valid
beyond its weak-coupling microscopic derivation and is familiar from
studies of conventional smectic liquid
crystals\cite{deGennesProst,ChaikinLubensky,GP}, despite the fact that
FF state is a translationally-invariant polar superfluid not a
smectic. The necessity of keeping the higher order gradients and
$\delta j_\parallel$ nonlinearities in $\cH_{FF}$ is due to the
identical vanishing of the transverse superfluid stiffness,
$\rho^\perp_s = 0$ (guaranteed by the underlying rotational invariance
unique to the FFLO striped states, absent in solid state contexts)
that leads to fluctuations that are otherwise infrared-divergent in a
purely harmonic model.

\subsubsection{Larkin-Ovchinnikov state}

The LO state is instead described by {\em two} independent nonzero PDW
amplitudes $\Delta_+=\Delta_-\equiv\Delta_{q_0}$ (growing below
$h_{c2}$), that lead to a standing wave pair-density wave order
parameter, 
\bse
\begin{eqnarray}
\Delta_{LO}(\rv)
&=&2\Delta_{q_0}e^{i\oh(\phi_+ + \phi_-)}
\cos\big[\qv_0\cdot\rv + \oh(\phi_+ - \phi_-)\big],\ \ \ \ \ \ \ \\
&=&2\Delta_{q_0}e^{i\phi}
\cos\big[\qv_0\cdot\rv + \theta\big],
\end{eqnarray}
\label{DeltaLO}
\ese
that is a {\em product} of a superfluid and a unidirectional density
wave striped order parameters. These are respectively characterized by
two Goldstone modes $\phi,\theta$, corresponding to the superfluid
phase and the smectic phonon $u = -\theta/q_0$ of the striped
state. This also contrasts with the conventional
smectic\cite{deGennesProst} (e.g., in liquid crystal materials, where
one instead is dealing with a real mass density $\rho(\rv)$ not a pair
condensate wavefunction), characterized by a single phonon Goldstone
mode, $u$.

The mean-field LO order parameter, $\Delta_{LO}$ simultaneously
exhibits the ODLRO (superfluid) and the smectic (unidirectional
density wave) orders. It thus spontaneously breaks the rotational,
translational, and global gauge symmetries, and therefore realizes a
form of a paired supersolid. However, it is distinguished from a
conventional purely bosonic supersolid
\cite{Andreev69,Chester70,Leggett70,KimChan}, where homogeneous
superfluid order and periodic density wave coexist, by the vanishing
of the (``charge''-2 two-atom) zero momentum ($\qv=0$) superfluid
component in the LO condensate\cite{commentNotSS}.

Similarly to the FF state, the underlying rotational symmetry of the
LO state strongly restricts the form of the Goldstone-mode
Hamiltonian.  Namely, its $\theta=-q_0 u$ sector must be invariant
under a rotation of $\qv_0$, that defines the spontaneously-chosen
orientation of the pair-density wave, and therefore must be described
by a smectic form\cite{deGennesProst,ChaikinLubensky,GP}. On the other
hand because a rotation of the LO state leaves the superconducting
phase, $\phi$ unchanged, the superfluid phase $\phi$ sector of the
Hamiltonian is therefore expected to be of a conventional $xy$-model
type. Consistent with these symmetry-based expectations the LO
Goldstone-mode Hamiltonian was indeed found\cite{RVprl,Rpra} to be
given by
\begin{eqnarray}
\cH_{LO}&=&\oh K(\nabla^2 u)^2 + 
\oh B\big(\partial_\parallel u - \frac{1}{2}(\nabla u)^2\big)^2\nonumber\\
&&+ \frac{1}{2}\rho_s^\parallel(\partial_\parallel\phi)^2 
+ \frac{1}{2}\rho_s^\perp(\nabla_\perp\phi)^2,
\label{HgmLO}
\end{eqnarray}
with the nonlinear strain tensor $u_{qq}=\hat{\qv}\cdot\grad u
-\oh(\nabla u)^2$ ensuring the full rotational invariance.  In the
weakly-coupled BCS limit the smectic elastic moduli and the superfluid
stiffnesses are given by
\bse
\begin{eqnarray}
K&=&4J q_0^2|\Delta_{q_0}|^2
\approx\frac{0.8 n\Delta_{BCS}^2}{\epsilon_F q_0^2}\ln(h/h_{c2}),\ \ \
\\
B&=&16J q_0^4|\Delta_{q_0}|^2
\approx\frac{3.3n\Delta_{BCS}^2}{\epsilon_F}\ln(h/h_{c2}),\ \ \ \\
\rho_s^\parallel&=&B/q_0^2,\\
\rho_s^\perp &=& \frac{4\lambda_2}{m^2}|\Delta_{q_0}|^4
\approx\frac{0.8 n\Delta_{BCS}^2}{\epsilon_F q_0^2}\ln^2(h/h_{c2}).
\end{eqnarray}
\label{KBrhos_pp}
\ese

Thus, the LO state is a highly anisotropic superfluid (though less so
than the FF state, where $\rho_s^\perp=0$), with
\begin{equation}
\frac{\rho_s^\perp}{\rho_s^\parallel}=
\frac{3}{4}\left(\frac{\Delta_{q_0}}{\Delta_{BCS}}\right)^2
\approx\frac{1}{4}\ln\left(\frac{h_{c2}}{h}\right)\ll 1,
\label{derive:ratio}
\end{equation}
a ratio that vanishes for $h\rightarrow h_{c2}^-$.  

We stress that while the detailed expressions for the moduli above are
specific to the weak-coupling BCS limit near $h_{c2}$ the general form
of $\cH_{LO}$, \rf{HgmLO}, including the structure of the
symmetry-enforced nonlinearities in the $u$ (smectic) sector is valid
beyond our microscopic derivation, and holds throughout the LO
phase.

By extending the Hamiltonian to include density fluctuations, $\delta
n_\pm$, canonically conjugate to $\phi_\pm$ and integrating them out
in an imaginary time ($\tau$) coherent-state path integral, leads to a
Lagrangian density
\begin{eqnarray}
  \cL =\frac{\chi_0}{2}(\partial_\tau\phi_+)^2 
  + \frac{\chi_0}{2}(\partial_\tau\phi_-)^2 + \cH[\phi_+,\phi_-],
\label{Slo}
\end{eqnarray}
with $\cH$ given by the $\cH_{LO}$ in the LO ground state (FF state is
treated similarly using $\cH_{FF}$ and a single Goldstone mode).  For
the LO state, this analysis then predicts the existence of two
anisotropic low-frequency modes with dispersions
\bse
\begin{eqnarray}
\omega_\phi(\kv) &=&\sqrt{(\rho_s^\perp k_\perp^2 + \rho_s^\parallel
k_z^2)/\chi_0},\\
\omega_u(\kv) &=&\sqrt{(K k_\perp^4 + B k_z^2)/\chi_0},
\end{eqnarray}
\label{omegasPhiU}
\ese
where $\chi_0$ is the compressibility of the Fermi gas.  These modes
respectively correspond to the zeroth sound (the Bogoliubov mode as in
a conventional superfluid) and smectic phonon, unique to the LO
state. In cold atomic gases, these should in principle be measurable
via the Bragg spectroscopy
technique\cite{KetterleBragg,Steinhauer02prl,Papp08prl}

With the Goldstone-mode Lagrangian in hand, the effects of quantum and
thermal fluctuations as well as equilibrium correlation and response
functions can be calculated\cite{commentBerry}.

\subsubsection{Goldstone modes fluctuations}
\label{sec:GMfluctuations}

Armed with the action for the Goldstone modes, the low-energy quantum
and thermal fluctuations in the FFLO striped states are
straightforwardly computed. Despite smectic like softness of these
modes, it is easy to show that quantum fluctuations remain finite for
$d>1$ and therefore the FF and LO states remain stable at zero
temperature.

{\em Harmonic approximation:}

In contrast, at finite temperature the root-mean-squared fluctuations
of the smectic phonon modes in the LO state (and phase fluctuations in
the FF state) diverge with trap size (in a purely harmonic
Goldstone-modes theory) according to
\begin{eqnarray}
\langle u^2\rangle_0^{T}
&\approx&
\left\{\begin{array}{ll}
\frac{T}{2\sqrt{B K}}L_\perp^{3-d},& d < 3,\\
\frac{T}{4\pi\sqrt{B K}}\ln q_0L_\perp,& d = 3,\\
\end{array}\right.
\label{uuT}
\end{eqnarray}
where the fluctuations are evaluated in a trap with an aspect ratio
$L_\perp\times L_z$, with $L_z$ its extent along the ordering $\qv_0$
axis \cite{Caille,deGennesProst,ChaikinLubensky}.  These fluctuations
lead to an emergence of important crossover length scales
$\xi_z,\xi_\perp\sim (\xi_z\sqrt{K/B})^{1/2}\equiv\sqrt{\xi_z\lambda}$ that
characterize the finite-temperature LO state,
\begin{eqnarray}
\xi_\perp&\approx&
\left\{\begin{array}{ll}
\frac{a^2\sqrt{B K}}{T}\sim\frac{K}{T q_0},& d = 2,\\
a e^{4\pi a^2\sqrt{B K}/T}\sim a e^{\frac{c K}{T q_0}},& d = 3,\\
\end{array}\right.
\end{eqnarray}
\label{xiperp}
defined as scales at which LO phonon fluctuations are comparable to LO
stripe period $a=2\pi/q_0$.  

The thermal connected correlation function of LO phonons
\begin{equation}
C_{u}(\rv_\perp,z)
=\langle\left[u({\bf r_\perp},z)-u({\bf 0},0)\right]^2\rangle_0\;
\label{C_T}
\end{equation}
is also straightforwardly worked out, in 3d giving the logarithmic
Caill\'e form\cite{Caille}
\begin{eqnarray}
C^{3d}_{u}(\rv_\perp,z)
&\approx&\frac{T}{2\pi\sqrt{K  B}}\left\{\begin{array}{lr}
\ln\left(\frac{r_\perp}{a}\right),&r_\perp\gg\sqrt{\lambda|z|}\;\\
\ln\left(\frac{4\lambda z}{a^2}\right),&r_\perp\ll\sqrt{\lambda|z|}\;\\
\end{array}\right..\ \ \ \ \ \ \ 
\label{Cuu3dT0}
\end{eqnarray}
In 2d it is instead given by\cite{TonerNelsonSm}
\begin{eqnarray}
C^{2d}_{u}(x,z)
&\approx&\frac{2T}{B}\left\{\begin{array}{lr}
\left(\frac{|z|}{4\pi\lambda}\right)^{1/2},
&x\ll\sqrt{\lambda|z|}\;\\
\frac{|x|}{4\lambda}, &x\gg\sqrt{\lambda|z|}\;\\
\end{array}\right..\ \ \
\label{Cuu2dT0}
\end{eqnarray}

Above finding of the divergence of smectic phonon fluctuations at
nonzero temperature have immediate drastic implications for the
properties of the LO (and FF) states. The most important of these is
that the thermal average of the Landau's LO order parameters
\rf{DeltaLO} vanishes in thermodynamic limit
\begin{eqnarray}
\langle\Delta_{LO}(\rv)\rangle_0
&\approx&2\tilde\Delta_{q_0}(L_\perp)\cos\big(\qv_0\cdot\rv),\nonumber\\
\label{DeltaLOave2b}
\end{eqnarray}
with the thermally suppressed order parameter amplitude given by
\begin{eqnarray}
\hspace{-1cm}
\tilde\Delta_{q_0}(L_\perp)&=&
\Delta_{q_0}e^{-\oh\phi^2_{rms}}
\left\{\begin{array}{ll}
e^{-L_\perp/\xi_\perp},& d = 2,\\
\left(\frac{a}{L_\perp}\right)^{\eta/2},& d = 3,\\
\end{array}\right.\nonumber\\
&\rightarrow &0,\ \ \mbox{for $L_\perp\rightarrow\infty$},
\label{DeltaR}
\end{eqnarray}
where $\phi_{rms}^2\equiv\langle\phi^2\rangle_0$ accounts for the
finite quantum and thermal superconducting phase fluctuations, and
$\eta=\frac{q_0^2 T}{8\pi\sqrt{B K}}$ is the Caill\'e exponent\cite{Caille}.

Akin to 2d xy-model systems, the vanishing of the LO order parameter
does not imply the instability of the phase, but that the true
fluctuating state contrasts qualitatively with perfectly
periodically-ordered mean-field description. 
%
\begin{figure}[thb]
\vspace{0.5cm}
\includegraphics[width=8.5cm,scale=1]{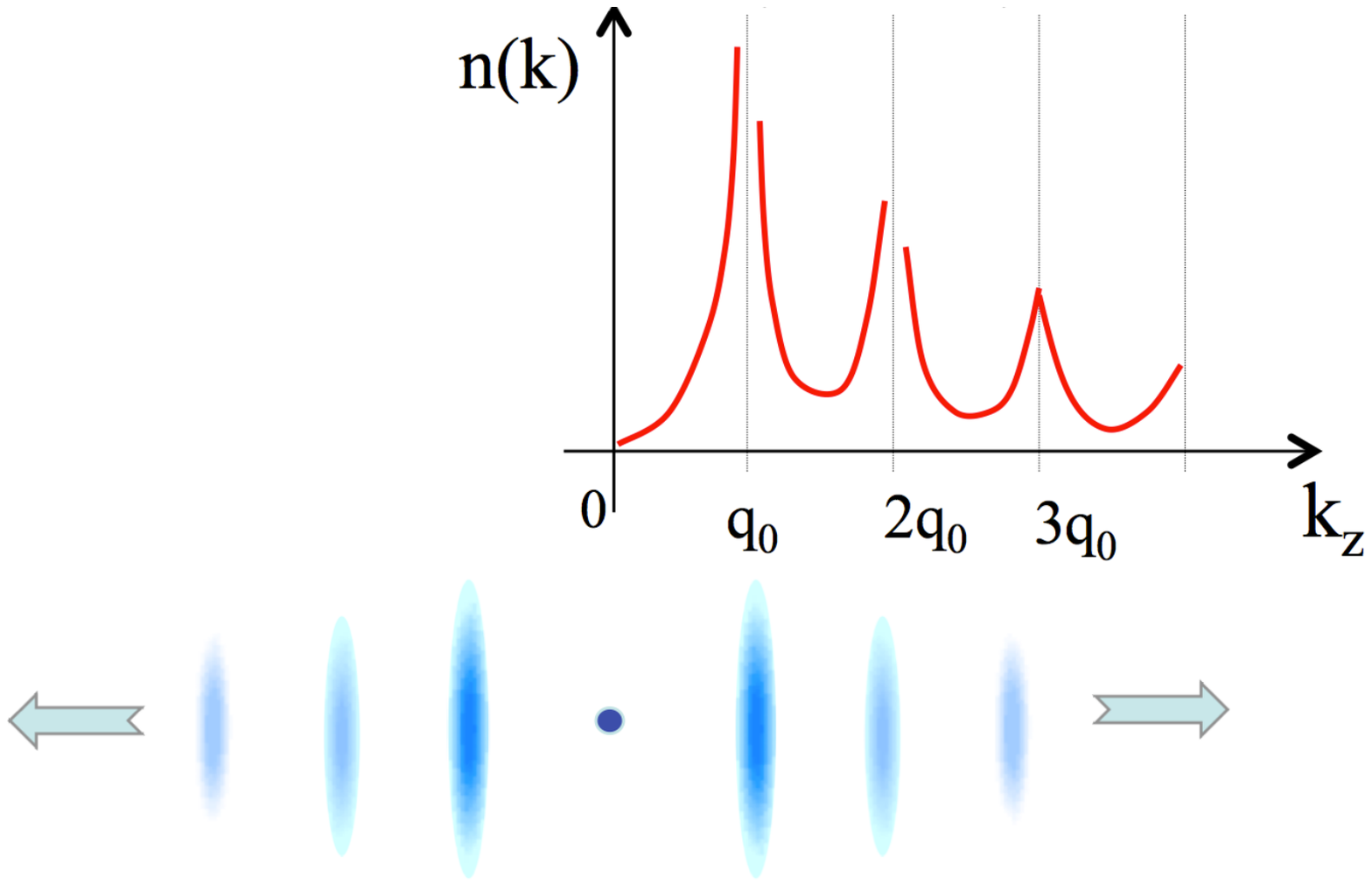}
\caption{The finite momentum pairing at $q_0$ and divergent 3d smectic
  phonon fluctuations in the LO state are reflected in the Cooper-pair
  center-of-mass momentum distribution function, $n_\kv$ (observable
  via the time-of-flight measurements), displaying power-law Bragg
  peaks, characteristic of spatial quasi-long-range order.}
\label{fig:nk}
\end{figure}
%
%
%
\begin{figure}[thb]
\vspace{0.5cm}
\includegraphics[width=7.5cm,scale=1]{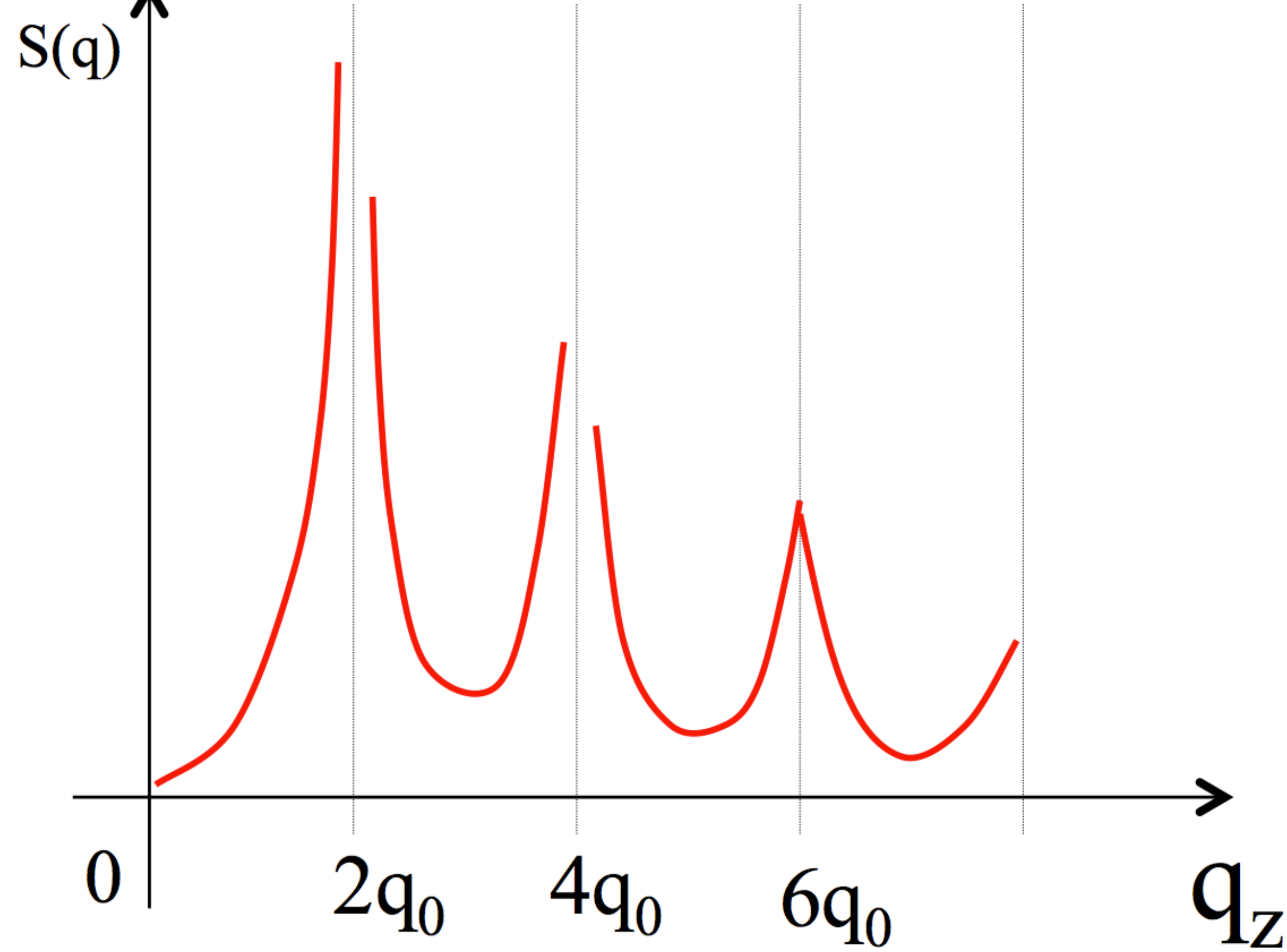}
\caption{The structure function, $S(\qv)$ for the 3d LO state,
  displaying power-law Bragg peaks, characteristic of the LO
  superfluid's spatial quasi-long-range order.}
\label{fig:Sq}
\end{figure}
%

These divergent LO phonon fluctuations also qualitatively modify the
Cooper-pair momentum distribution function $n_\kv=\langle
\Delta^\dagger_\kv \Delta_\kv\rangle$ and structure function $S(\qv)$
(respectively measurable via time-of-flight and Bragg
spectroscopy\cite{KetterleBragg,Steinhauer02prl,Papp08prl} imaging of
pair-condensate) from their true Bragg ($\delta$-function peaks) form
characteristic of the mean-field long-range periodic order. They are
highly anisotropic ($q_z\sim q_\perp^2/\lambda$) and exhibit
quasi-Bragg peaks (see Fig.\ref{fig:Sq}) around the ordering
wavevector $\qv_0$ (and its harmonics, $\qv_n$), reminiscent of (1+1)d
Luttinger liquids and two-dimensional
crystals\cite{Landau1dsolid,Peierls1dsolid,MerminWagner,Hohenberg,Berezinksii,KT}
\bse
\begin{eqnarray}
  n^{LO}_{k_z} &\approx& \sum_{n}\frac{n_{q_n}}{|k_z - n
    q_0|^{2-n^2\eta}},
\label{nkresult}\\
S^{LO}(q_z)
&\approx&\sum_{n}\frac{|\Delta_{q_n}|^4}{|q_z - 2n q_0|^{2-4n^2\eta}}, 
\label{Sq}
\end{eqnarray}
\ese
These predictions are a reflection of the
well-known\cite{deGennesProst,ChaikinLubensky} and experimentally
tested\cite{SqSmExp} behavior of conventional smectic liquid crystals.
In two dimensions, the LO order is even more strongly suppressed by
thermal fluctuations down to short-ranged correlations, strongly
suggesting true instability of the LO order. Because it is the
``soft'' smectic Goldstone mode that is responsible for these
interesting properties they are necessarily also shared by the FF
state\cite{Shimahara}.

I emphasize that analogous Kosterlitz-Thouless phase fluctuation
physics has been observed in conventional 2d trapped superfluids,
despite the finite trap size\cite{Hadzibabic,ChengChinKT}. Thus, I am
hopeful that they can be similarly seen in the 3d LO state. A more
detailed analysis of the trap effects is necessary for direct
comparison with experiments.

Another fascinating consequence of the vanishing of
$\langle\Delta_{LO}\rangle$, \rf{DeltaR} is that the leading nonzero
Landau order parameter characterizing the LO state is the
translationally-invariant ``charge''-4 (4-atom pairing)
superconducting order parameter,
\begin{eqnarray}
  \Delta^{(4)}_{sc}&\sim&\int_{\rv,\rv'}
  \langle\ch_\downarrow(\br) \ch_\uparrow(\br)
\ch_\downarrow(\br') \ch_\uparrow(\br')\rangle
e^{i\qv\cdot(\rv-\rv')},\nonumber\\
&\sim&\sum_{\kv,\kv'}
  \langle\ch_{\downarrow,-\kv}\ch_{\uparrow,\kv+\qv}
\ch_{\downarrow,-\kv'}\ch_{\uparrow,\kv'-\qv}\rangle.
\end{eqnarray}
Thus in the presence of thermal fluctuations the LO phase corresponds
to an exotic state in which the off-diagonal order is exhibited by
pairs of Cooper pairs, i.e., a bound quartet of atoms, rather than by
the conventional 2-atom Cooper pairs\cite{Berg09nature,RVprl}. In 2d
and 3d this higher order pairing is driven by arbitrarily low-$T$
fluctuation, rather than by a fine-tuned attractive interaction
between Cooper pairs, and therefore has no simple mean-field
description.  A microscopic formulation of such a state and its
detailed properties remain an open problem.

{\em Nonlinear elasticity beyond Gaussian fluctuations:}

As discussed in the context of conventional smectics\cite{GP} and more
recently for the FFLO state\cite{RVprl,Rpra}, above predictions
neglect the effects of Goldstone mode nonlinearities
$\cH_{\text{nonlinear}}=-\oh B (\partial_z u)(\nabla u)^2
+\frac{1}{8} B (\nabla u)^4$, in Eq.\rf{HGMff},\rf{HgmLO}, that modify
the asymptotics on scales longer than the crossover scales
$\xi^{NL}_{\perp}=\sqrt{\xi^{NL}_{z}\lambda}$,
\begin{eqnarray}
\xi^{NL}_\perp&\approx&
\left\{\begin{array}{ll}
\frac{1}{T}\left(\frac{K^3}{B}\right)^{1/2},& d = 2,\\
a e^{\frac{c}{T}\left(\frac{K^3}{B}\right)^{1/2}},& d = 3,
\end{array}\right.
\label{xiNL}
\end{eqnarray}
The behavior on scales beyond $\xi^{NL}_{\perp,z}$ can be obtained
using renormalization-group analysis for $d\leq 3$\cite{GP,Rpra}, with
an exact solution in 2d\cite{GW}. The finite-temperature asymptotics
is well-approximated by a correlation function
\begin{eqnarray}
G_{u}(\kv)
&\approx&\frac{T}{B(\kv) k_z^2 + K(\kv) k_\perp^4},
\label{Guu}
\end{eqnarray}
with moduli $B(\kv)$ and $K(\kv)$ that display a universal singular
wavevector-dependence, that is asymptotically {\em exact}
logarithmic\cite{GP} in 3d
\bse
\begin{eqnarray}
  K({\kv_\perp,k_z=0})&\sim&K|1+
\frac{5g}{64\pi}\ln(1/k_\perp a)|^{2/5}\;,\label{K3d}\\
  B({\kv_\perp=0,k_z})&\sim&B|1+\frac{5g}{128\pi}
\ln(\lambda/k_za^2)|^{-4/5}.\ \ \ \ \ \ \ \ \ \ \label{B3d}
\end{eqnarray}
\label{KB3d}
\ese
and power-law in 2d
\bse
\begin{eqnarray}
K({\bf k})&=&K\left(k_\perp\xi^{NL}_\perp\right)^{-\eta_K}
f_K(k_z\xi^{NL}_{z}/(k_\perp\xi^{NL}_\perp)^\zeta)\;,\label{Kg}
\ \ \ \ \ \ \ \ \ \\
&\sim& k_\perp^{-\eta_K},\nonumber\\
B({\bf k})&=&B\left(k_\perp\xi^{NL}_\perp\right)^{\eta_B}
f_B(k_z\xi^{NL}_{z}/(k_\perp\xi^{NL}_\perp)^\zeta)\;,\label{Bg}\\
&\sim& k_\perp^{\eta_B}.\nonumber
\end{eqnarray}
\label{KgBg}
\ese
with $f_B(x),f_K(x)$ universal scaling functions and 
$\eta_B^{2d}=1/2,\eta_K^{2d}=1/2,\zeta^{2d}=3/2$ exact\cite{GW}.
In 3d this translates into an equal-time LO order parameter
correlations given by\cite{GP}
\bse
\begin{eqnarray}
n(z,\rv_\perp=0)&=& \langle\Delta^*_{LO}(z)\Delta_{LO}(0)\rangle,\\
&\sim&e^{-c_1(\ln z)^{6/5}}\cos(q_0 z).
\end{eqnarray}
\label{SrGP}
\ese
Although these 3d anomalous effects are less dramatic and likely to be
difficult to observe in practice, theoretically they are quite
significant as they represent a qualitative breakdown of the
mean-field and harmonic descriptions of the FFLO striped states.

\subsection{Phases and transitions}
\label{sec:DefectsTransitions}

\subsubsection{Topological defects}
Associated with its two compact Goldstone modes,$\phi_\pm$
(equivalently $\phi(\rv)=\oh(\phi_++\phi_-)$, $u(\rv)=\oh
a(\phi_+-\phi_-)$) the LO state admits two types of topological
defects, characterized by integers $\Nv_v=(n_+,n_-)$ defined by $\oint
d\vec{\ell}\cdot\vec{\nabla}\phi_\pm = 2\pi n_\pm$. These equivalently
correspond to superfluid vortices and edge dislocations in the striped
PDW\cite{comment2op}. These are characterized by multiples of {\em
  half}-integers $n_v, n_d = (n_+ \pm n_-)/2$ and therefore allow four
types of elementary defects: integer $(2\pi,0)$ vortex, integer
dislocation $(0,a)$ and two half-integer vortex-dislocation composites
$(\pi,\pm a/2)$. The latter composite fractional defects are allowed
because a sign change in $\Delta_{LO}$ due to a $a/2$-dislocation in
$u$ is compensated by a $\pi$-vortex in $\phi$ (thereby preserving a
single-valuedness of
$\Delta_{LO}$)\cite{Agterberg08prl,Berg09nature,RVprl,Rpra}.  In terms
of the two coupled $\phi_+,\phi_-$ Goldstone modes, these correspond
to an integer vortex in one and no vortex in the other superfluid
phase.

Their thermodynamics and correlations can be treated via a mapping on
a multi-component Coulomb gas 
\begin{eqnarray}
H^{xy-sm}_{CG}&=&
\oh\int_{\bf  q}\left[\frac{\sqrt{\rho^\perp_s\rho^\parallel_s}}
{\Gamma_\qv^{xy}}|m_{\qv,v}|^2
+\frac{K q_\perp^2}{\Gamma_\qv^{sm}}|m_{\qv,d}|^2\right]\nonumber\\
&&+\sum_{\rv_i}\left(E^v_c n_{\rv_i,v}^2 + E^d_c n_{\rv_i,d}^2\right),
\end{eqnarray}
where 
\bse
\begin{eqnarray}
\Gamma_\qv^{xy}&=& q_\perp^2\sqrt{\rho_s^\perp/\rho_s^\parallel}
+  q_z^2\sqrt{\rho_s^\parallel/\rho_s^\perp},\\
\Gamma_\qv^{sm}&=& q_z^2 + \lambda^2 q_\perp^4.
\end{eqnarray}
\ese
with $m_{\qv,v}, m_{\qv,d}$ the Fourier transforms of the vortex and
dislocation densities. Equivalently, defects thermodynamics can be
analyzed via duality transformation. In two dimensions it leads to a
sine-Gordon-like model for dual $\tphi,\ttheta$ fields characterizing
fractional defects, with the dual Hamiltonian\cite{Rpra,Berg09nature,LiuSineGordon}
\begin{widetext}
\begin{eqnarray}
\tilde H_{SG}&=&
\oh\int_{\bf q}\left[
\frac{\Gamma^{xy}_\qv}{\sqrt{\rho^\perp_s\rho^\parallel_s}}|\tphi_\qv|^2
+\frac{\Gamma^{sm}_\qv q_0^2}{K q_\perp^2}|\ttheta_\qv|^2\right]
-\int_\rv\left[g_{\pi,a/2}\cos(\pi\tphi)\cos(\pi\ttheta)
+ g_{2\pi,0}\cos(2\pi\tphi) + g_{0,a}\cos(2\pi\ttheta)\right],
\label{HdualSG}
\end{eqnarray}
\end{widetext}
It is convenient for analyzing the effects of defects on the LO state,
particularly for a computation of their screening on long scales,
unbinding, and for the analysis of the resulting disordered state.
From the form \rf{HdualSG} it is clear that (aside from an
inconsequential anisotropy) the dual vortex sector described by
$\tphi$ has a standard sine-Gordon form.  In contrast, the dual
dislocation sector, described by $\ttheta$ is qualitatively modified
by the highly nonlocal and qualitatively anisotropic smectic kernel,
$\Gamma^{sm}_\qv$.

A standard analysis gives the relative energetics of these defects, in
the thermodynamic limit ($L_{\perp,z}\rightarrow\infty$) given by
\begin{widetext}
\begin{eqnarray}
  &&E^d_{(0,a)}\sim K L \ll 
E^{v-d}_{(\pi,a/2)}\sim\frac{\rho_s}{4}L\ln L + \frac{K}{4}L
\ll E^{v}_{(2\pi,0)}\sim \rho_s L\ln L,\ \ \ \mbox{for
    $L_{\perp,z}\sim L\rightarrow\infty$},
\label{eq:defectsE}
\end{eqnarray}
\end{widetext}
Based on this energetics one may be tempted to conclude that in this
limit (unless preempted by a first-order transition) it is the integer
dislocation loop defects that proliferate first and the LO smectic
preferentially disorders into a nematic superfluid, $SF_{N}$. However,
in contrast to the 2d KT mechanism\cite{KT}, the 3d disordering
transitions take place when the relevant stiffness, renormalized by
quantum and thermal fluctuations is continuously driven to zero at the
transition, or takes place at a finite (rather than a vanishing)
defects fugacity. For a thermal transition this roughly corresponds to
a transition temperature set by the corresponding stiffnesses,
$\overline{\rho}_s=\sqrt{\rho_s^\parallel\rho_s^\perp}$ and $K,
B$. Thus, in principle by tuning these stiffnesses via imbalance and
resonant interaction, a variety of phases can be accessed.

\subsubsection{Conventional phases and transitions}

By considering all possible basic combinations of spontaneously
``broken'' subset of spatial and gauge symmetries leads to an array of
partially spatially-ordered paired superfluids and Fermi-liquid
states, that are descendants of the smectic LO (SF$_{Sm}$)
state. These isotropic (I), nematic (N) and smectic (Sm) SF and FL
states are summarized in Table I.  It is notable that the isotropic
superfluid, $SF_I$ exhibits a finite species imbalance and
off-diagonal long-range order, symmetry-wise isomorphic to the
polarized superfluid, $SF_M$\cite{SRprl,SRaop}, latter confined to the
BEC side of the BCS-BEC crossover. In contrast, (as a descendant of
the LO state expected to be stabilized by Fermi surfaces imbalance)
the $SF_I$ state is realized in the BCS regime, something that has
been searched for dating back to Sarma\cite{Sarma}, but has not been
possible within mean-field treatments, that instead predict an
instability to phase separation\cite{Bedaque03,SRprl,SRaop}.  The
isotropic Fermi liquid, $FL_I$ is isomorphic to the conventional
normal state.  Together these intermediate fluctuation-induced phases
naturally interpolate between the fully gapped singlet (homogeneously
and isotropic) BCS superconductor at zero imbalance and low
temperature, and a polarized Fermi liquid at large imbalance and/or
high temperature.

\begin{table}[t]
\begin{tabular}{ c  c  c  c  c  } 
$FL_{Sm}$&$\rightarrow$ &$FL_N$ &$\rightarrow$ & $FL_I$\\[0.2cm]
$\ \ \ \ \big\uparrow$\small{U(1)}&    
&$\ \ \ \ \ \big\uparrow$\small{U(1)}& 
&$\ \ \ \ \ \big\uparrow$\small{U(1)} \\[0.2cm]
$SF_{Sm}$&$\rightarrow$ &$SF_N$ &$\rightarrow$ & $SF_I$ \\[0.5cm]
\end{tabular}
\caption{Five phases that naturally emerge as disordered descendants
  of the LO (superfluid smectic, $SF_{Sm}$) state.}
\label{table:phasesRelate}
\end{table}

\begin{table}[t]
\begin{tabular}{| c | c | c | c | } 
\hline
phase/symmetry & $U(1)$ & $T_{\qv_0}$ & $R$ \\
\hline
$FL_I$ &   $\surd$       &  $\surd$        &   $\surd$      \\
\hline
$FL_N$ &  $\surd$        &  $\surd$        & X \\
\hline
$FL_{Sm}$ & $\surd$      & X & X \\
\hline
$SF_I$ & X & $\surd$       &  $\surd$      \\
\hline
$SF_N$ & X &  $\surd$       & X \\
\hline
$SF_{Sm}$ & X & X & X \\
\hline
\end{tabular}
\caption{A summary of LO liquid crystal Fermi-liquid (FL) and
  superfluid (SF) phases, and corresponding order parameters and
  broken symmetries, indicated by X's. Unbroken
  symmetries (gauge $U(1)$, translational $T_{q_0}$, rotational $R$)
  are marked by check marks. The subscripts $I, N, Sm$
  respectively indicate the Isotropic, Nematic and Smectic orders.}
\label{table:phases}
\end{table}

\subsubsection{Topological phases via defects unbinding}

The phases discussed above can be complementarily characterized
through unbinding of different combinations of topological defects.
The smectic (whether SF LO state or FL smectic) to nematic transition
is driven by unbinding of defects with edge dislocation charge,
followed by transition into the isotropic state driven by
proliferation of disclinations. The superfluid and Fermi-liquid
version of these liquid crystal states are distinguished by unbinding
defects with superfluid vortex charge.

However, a characterization in terms of topological defects also
allows a distinction between topologically distinct phases with the
same conventional order, where a Landau order parameter is
insufficient to distinguish them.  In fact because of the vanishing LO
order parameter, description in terms of topological order is
necessary even for the smectic LO state, distinguished from its more
disordered descendants by the absence of unbound topological defects,
in direct analogy with the quasi-long-range ordered state of the 2d xy
model.

A rich variety of possible phases and transitions is displayed in a
schematic imbalance-detuning $P-1/k_Fa$ phase diagram,
Fig.\ref{fig:phasediagramLO3d}.  Increasing the imbalance suppresses
the superfluid stiffness and drives the system toward a conventional
Fermi liquid state, $FL_I$ at $h_{c2}$. Conversely, a reduction in
species imbalance primarily reduces the elastic moduli of the smectic
pair-density wave by increasing its period $1/q_0$ and thereby
weakening the interaction between the LO domain-walls, driving the
system toward a conventional isotropic and homogeneous superfluid
$SF_I$ at $h_{c1}$.

Thus, starting with the LO $SF_{Sm}$ state and {\em decreasing} $h$
leads to the unbinding of the integer dislocations $(0,a)$, and a
transition to an orientationally-ordered, i.e., a nematic ``charge''-4
superfluid, $SF^{4}_N$. The later ``charge''-4 feature of $SF^{4}_N$
naturally appears as the remaining secondary order parameter
$\Delta_{sc}^{(4)}=\Delta_{LO}^2$ once the LO positional order
$\Delta_{LO}$ is destroyed by unbinding of integer dislocation loops.
Since in contrast, the ``charge''-2 SF nematic order vanishes in the
LO state, a direct transition to it from the LO state can generically
only proceed through a first-order transition.

Conversely, {\em increasing} $h$ starting with the LO $SF_{Sm}$ is
expected to lead to a suppression of $\rho_s$, a proliferation $(2\pi,
0)$ vortices, and a transition to a $2q$-smectic Fermi liquid,
$FL^{2q}_{Sm}$, a non-superfluid periodic state with a wavevector that
is twice the LO state. Alternatively, as suggested by the energetics
in \rfs{eq:defectsE}, if instead the lower-energy half-vortex
dislocation defects $(\pi, a/2)$ (or the $(\pi, -a/2)$, but not both)
unbinds first, a transition to a nematic Fermi liquid, $FL^{**}_N$
(with the restored translational and U(1) charge symmetries) will take
place. The resulting state is qualitatively distinct from the more
conventional nematic (orientationally ordered) $FL_N$ phase in which
{\em both} $(\pi, a/2)$ and $(\pi, -a/2)$ are proliferated. Both are
also distinct from the nematically ordered $FL^{*}_N$ state, in which
only integer dislocations, $(0,a)$ and integer vortices, $(2\pi, 0)$
are unbound.  One can envision a number of other states and phase
transitions at low $h$ by further considering the disordering of the
nematic superfluid, $SF_N^{4}$ by unbinding various patterns of
disclinations and $\pi$-vortices. Many open questions remain about the
relative energetics and detailed properties of these phases.

Above considerations lead to at least three topologically distinct
Fermi liquid phases that naturally emerge from disordering of the LO
($SF_{Sm}$) phase by unbinding different combinations of allowed
defects. Because the conventional vortex $(2\pi, 0)$ and the
conventional dislocation $(0, a)$ are composites of the fundamental
defects $(\pi, \pm a/2),$ the nonsuperfluid states $FL^*_N$,
$FL^{**}_N$ and their isotropic cousins $FL_I^*$, $FL_I^{**}$ (in
which disclinations are also unbound) are expected to be
``fractionalized''\cite{SenthilFisher}, topologically distinct from
their conventional Fermi liquid analogs, where $(\pi, \pm a/2)$ are
also unbound.

These novel phases are analogous to the putative phase-disordered
fractionalized states obtained by unbinding double ($hc/e$) vortices,
studied extensively by Sachdev, and by Balents, Senthil, Fisher, and
collaborators\cite{SachdevDoubleVortex,Balents,SenthilFisher} in the
context of high temperature superconductors.  The resulting
nonsuperfluid phase is distinguished from a conventional Fermi liquid
by a gapped ``vison'', a $\Z_2$ defect that is a remnant of the
fundamental $hc/2e$ vortex after the composite $hc/e$ (double)
vortices proliferate.

The states $FL^*_N$, $FL^{**}_N$ also bare a close relation to the
collective mode fractionalization discussed by Sachdev, et
al. \cite{SachdevBook,SachdevMI,NussinovZaanen} in the context of
quantum paramagnetic phases, emerging from disordering a collinear
spin-density wave.  As with the $\left(U(1)\otimes U(1)\right)/\Z_2$
LO state, where the order parameter is a product of the superfluid and
smectic order parameters, Eq.\rf{DeltaLO}, there too the order
parameter is of $\left(S_2\otimes U(1)\right)/\Z_2$ product form,
encoding spatial modulation of the spin density, and therefore admits
half-integer (vison-like) defects. Correspondingly, the phases and
transitions can equivalently be captured via an effective Ising gauge
theory. Its nonzero $\Z_2$ flux through a plaquette encodes the
presence of a half-integer $(\pi,a/2)$ defect. The Ising gauge field
encodes the local $\Z_2$ redundancy of splitting the LO order
parameter \rf{DeltaLO} into a ``charge''-2 boson,
$b_\rv^\dagger=e^{-i\phi_\rv}$, that creates a zero-momentum
Cooper-pair (diatomic molecule) and a neutral boson,
$\rho_{q,\rv}^\dagger=e^{-i\theta_\rv}$, that creates a density wave
at the LO wavevector $q$.  In the above notation this is the
nonsuperfluid periodic state dubbed $FL_{Sm}^{2q}$, in which
$(2\pi,0)$ vortices have proliferated, but dislocations remain
bound. For a vison remaining gapped, the resulting nonsuperfluid
nematic state is the topologically ordered $FL_N^*$, qualitatively
distinct from a conventional $FL_N$ in which the vison is gapless. The
two FL phases are separated by a deconfinement transition of vison
condensation, $FL_N^*$-$FL_N$, corresponding to a proliferation of the
$(\pi,a/2)$ fractional defects, that is expected to be in the inverted
Ising universality class.

The variety of phases and transitions between them are summarized in
the phase diagram, Fig.\ref{fig:phasediagramLO3d} and a flow-chart,
Fig.\ref{fig:flowchart}.  This rich fluctuations-driven phase behavior
contrasts sharply with a direct LO-N transition (described by
$U(1)\times U(1)$ Landau theory $H_{mft}=r(|\Delta_+|^2+|\Delta_-|^2)
+ \lambda_1 (|\Delta_+|^4+|\Delta_-|^4)
+\lambda_2|\Delta_+|^2|\Delta_-|^2$) found in mean-field theory.

\subsection{Fermionic excitations}
\label{sec:fermions}

The subtlety of the LO and its descendent states is in addition to
strongly fluctuating Goldstone modes, they exhibit gapless fermionic
excitations that can be strongly coupled to the superfluid phase
$\phi$ and smectic phonon $u$ Goldstone modes.  A complete
comprehensive treatment remains an open problem. However, a number of
approximate numerical and analytical treatments has lead to a
consistent
picture\cite{MachidaNakanishiLO,BurkhardtRainerLO,MatsuoLO,YoshidaYipLO,Rpra}.

At large imbalance, near $h_{c2}$ where $\Delta$ modulation is weak,
the fermionic spectrum can be approximated by various $\qv$ branches
of the form Eq.\rf{eq:Esigma} for the FF state. These lead to a novel
paired superfluid with a Fermi surface of Bogoliubov quasi-particles.

\begin{figure}[thb]
\vspace{0.5cm}
\includegraphics[width=7.5cm,scale=1]{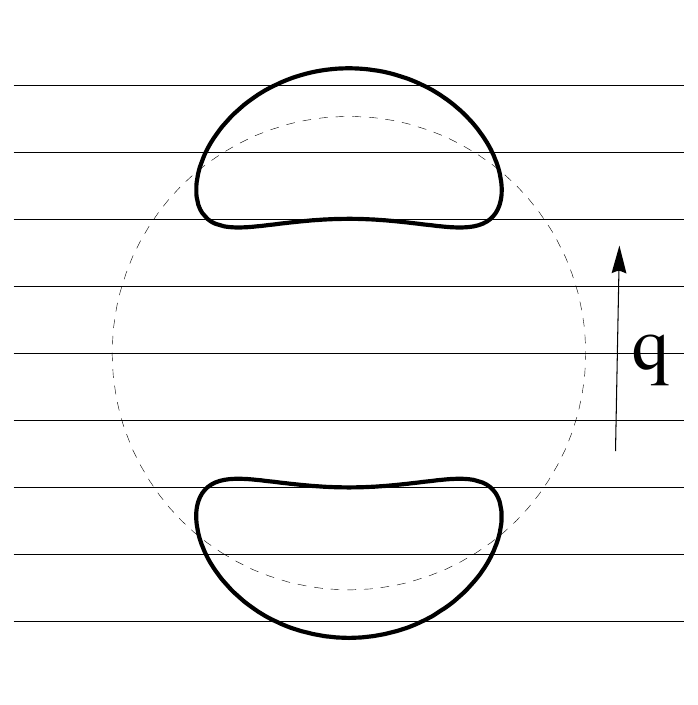}
\caption{An illustration of Fermi pockets (full curve) of the gapless
  Bogoliubov quasi-particles characteristic of the Larkin-Ovchinnikov
  ground state. The periodic array of domain-walls in
  $\Delta_{LO}(z)$, the associated wavevector $\qv$, and the
  Fermi-surface of the underlying normal state (dashed circle) are
  also indicated.}
\label{fig:fermiPockets}
\end{figure}

Near $h_{c1}$, a periodic form of $\Delta(\rv)$ is more appropriately
treated as an array of domain-walls of width
$\xi$\cite{MachidaNakanishiLO,BurkhardtRainerLO,MatsuoLO,YoshidaYipLO},
rather than a single harmonic.  Diagonalizing the BdG equations in
their presence leads to a band of midgap states as in 1d where this
can be done exactly\cite{SamokhinFFLO,ZagoskinBook,Rpra} corresponding
to imbalanced fermionic atoms residing on the nodes of the solitons in
$\Delta$ akin to the
polyacetylene\cite{SuSchriefferHeeger}. Consistent with the large
imbalance regime this leads to a Fermi surface
pockets\cite{comment1dPockets} illustrated in
Fig.\ref{fig:fermiPockets}.  Despite considerable progress a fully
self-consistent quantum-mechanical treatment is still missing.

%

\subsection{Open questions}
\label{sec:open}

While a broad range of phenomena associated with the
Larkin-Ovchinnikov state has been explored, many interesting and
difficult questions remain open. Probably the most urgent of these is
the long-standing question of the range of energetic stability of the
crystalline superconductor. If the state is indeed stable over a
sufficiently broad range of detuning and imbalance to be
experimentally accessible, is its lowest energy form indeed the
striped collinear LO type? While for large atomic clouds and shallow
traps (such that LDA remains valid) only a small deformation of the LO
state near the boundaries is expected, for tighter traps a more
detailed treatment of the trap is necessary, and may lead to a
distinct global form of the LO state, such as the ``onion'' and
``radial'' structures. To address such questions undoubtedly requires
numerical solutions in experiment-specific geometries.

Furthermore, the nature of the (2d and 3d) transition into the LO
state at the lower-critical Zeeman field
$h_{c1}$,\cite{BurkhardtRainerLO}, and the extent to which it
resembles a commensurate-incommensurate transition (as in
1d\cite{MachidaNakanishiLO,YangLL}) remains an open question. More
broadly, while a number of LO descendent states have been proposed
their detailed phenomenology, stability to quantum and thermal
fluctuations, as well the nature of the associated phase transitions
remains wide open. Similarly to the LO state, these phases are
expected to exhibit gapless fermionic excitations coupled to their
Goldstone modes. Understanding the effects of these fermionic modes on
the phases and the associated transitions remain an extremely
interesting and challenging problem.

To summarize, I reviewed a wide range of fluctuation phenomena in a LO
state, expected to be realizable in an imbalanced resonant Fermi
gas. Combining a microscopic analysis with robust model-independent
symmetry arguments predicts that the LO state in an isotropic trap is
a gapless superfluid smectic liquid crystal. Consequently, the state
is extremely sensitive to thermal fluctuations that destroy its
long-range positional order even in three dimensions, replacing it by
a quasi-long range order, characterized by power-law correlations
akin to a system tuned to a critical point or two-dimensional xy-model
systems. This exotic state also exhibits vortex fractionalization,
where the basic superfluid vortex is half the strength of a vortex in
a regular paired condensate, and is accompanied by half-dislocations
in the LO smectic (layered) structure.

Analysis of the fluctuation-driven disordering of the LO smectic
predicts a rich variety of descendant quantum liquid states, such as
the superfluid ($SF_N$) and Fermi liquid ($FL_N$) nematics and the
fractionalized nonsuperfluid states ($FL^*$), that generically
intervene between the LO state and the conventional BCS superfluid (at
low population imbalance) and a conventional Fermi liquid (at high
population imbalance). This phenomenology has a rich variety of
experimental implications, many of which await detailed analysis.

\section{p-wave resonant Bose gas}
\subsection{Background}
A $p$-wave resonant Bose gas is another system that was recently
predicted to exhibit quantum liquid crystal order. Its study was in
part motivated by the interesting phenomenology found in $s$-wave
resonant bosonic systems (e.g., in $^{85}$Rb~\cite{Cornish2000prl}),
that was demonstrated to exhibit a magnetic field-driven quantum Ising
transition between molecular and atomic superfluids
~\cite{RPWprl,RomansPRL,RWPaop,lee.lee.04}, contrasting with a smooth
BEC-BCS crossover in balanced fermionic isotopes.

Although instabilities intrinsic to resonant
bosons~\cite{Cornish2000prl,Papp08prl} challenge realization of such bosonic
molecular condensates, many features of the phase diagram are expected
to survive away from the resonance and/or reflected in the
nonequilibrium phenomenology (before the onset of the instability) of
a resonant Bose gas. Furthermore, recent extension of an $s$-wave
resonant Bose gas to an optical lattice~\cite{zoller1.10,zoller2.10}
demonstrated the stabilization through a quantum Zeno mechanism
proposed by Rempe~\cite{rempe.08}, that dates back to
Bethe's~\cite{bethe} analysis of the triplet linewidth in hydrogen.

The predictions~\cite{RPWprl,RomansPRL,RWPaop,zoller1.10,
  zoller2.10} for the $s$-wave case have been supported by recent
density matrix renormalization group~\cite{Ejima.Simons.11}, exact
diagonalization~\cite{Bhaseen.Simons.09}, and quantum Monte
Carlo~\cite{Bonnes.Wessel.11} studies. I am enthusiastic about similar
progress in the substantially richer $p$-wave case, that I discuss
below.

Experimental realization of two-species degenerate Bose gas with a
$p$-wave Feshbach resonant inter-species interaction in $^{85}$Rb -
$^{87}$Rb mixtures~\cite{Papp08prl} provides a direct motivation for
the theoretical studies discussed
below\cite{RCprl.09,CRpra.11}. Related studies of Bose condensation in
$p$-(and higher) bands in optical lattices have also been carried
out\cite{kuklov.06, liu.06}, but are only tangentially relevant to the
present focus of quantum liquid crystals.

\subsection{Summary}

\begin{figure}[thb]
\includegraphics[width=8.5cm]{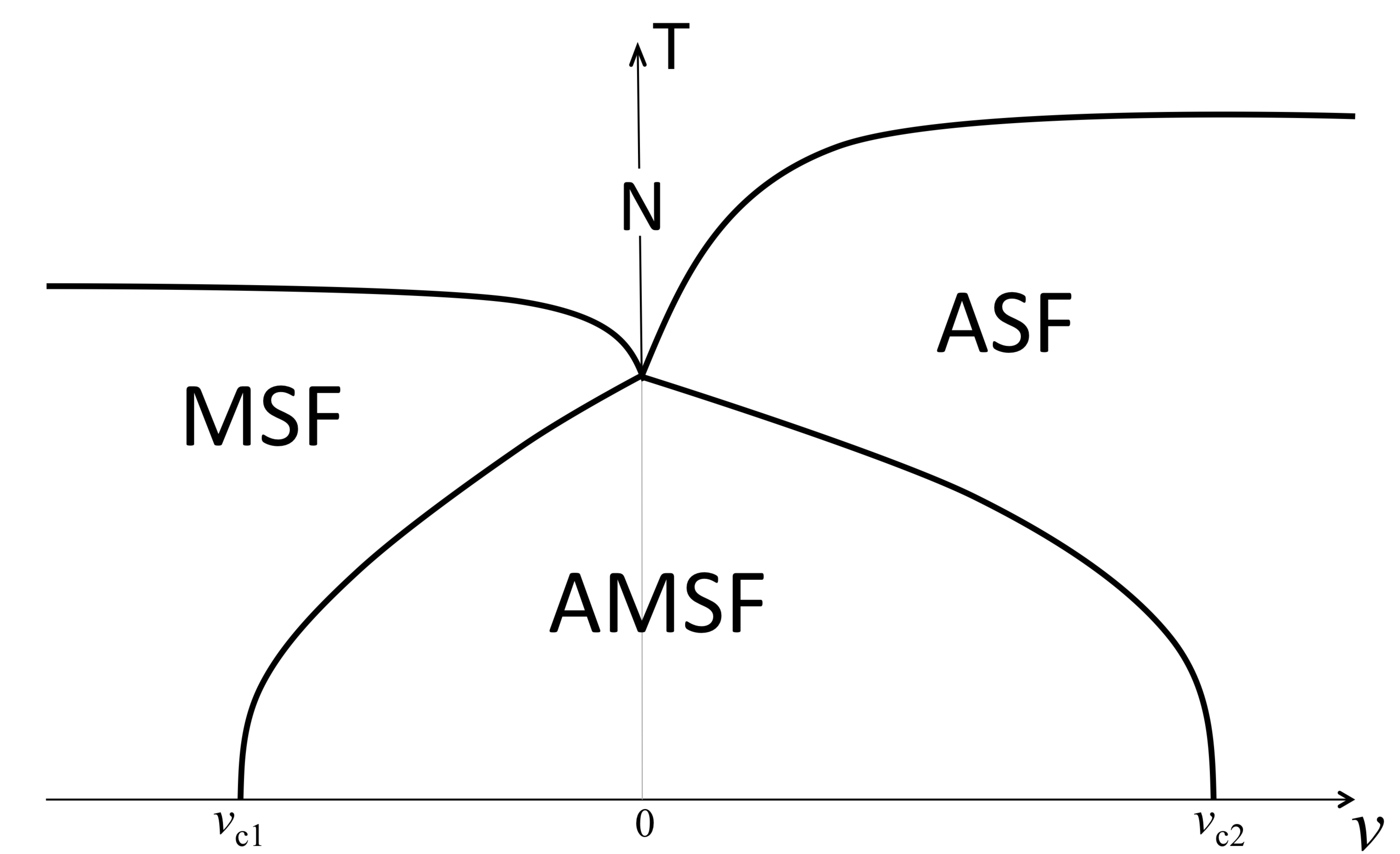} 
\caption{Schematic temperature-detuning phase diagram for a balanced
  two-species $p$-wave resonant Bose gas. As illustrated, it exhibits
  atomic (ASF), molecular (MSF), and atomic-molecular (AMSF)
  superfluid phases. The novel AMSF state is characterized by a
  $p$-wave, molecular and a finite-momentum $Q$ (see Fig.~\ref{fig:Q})
  atomic superfluidity.}
\label{fig:phasediagramAMSF}
\end{figure}

The phenomenology of a balanced two-component $p$-wave resonant Bose
gas is summarized by a temperature-detuning phase diagram. As
illustrated in Fig.~\ref{fig:phasediagramAMSF}
at low temperature such two-component Bose gas generically exhibits
three classes of superfluid phases, atomic (ASF), molecular (MSF) and
atomic-molecular (AMSF) condensates.
\begin{figure}[thb]
\includegraphics[width=6.2cm]{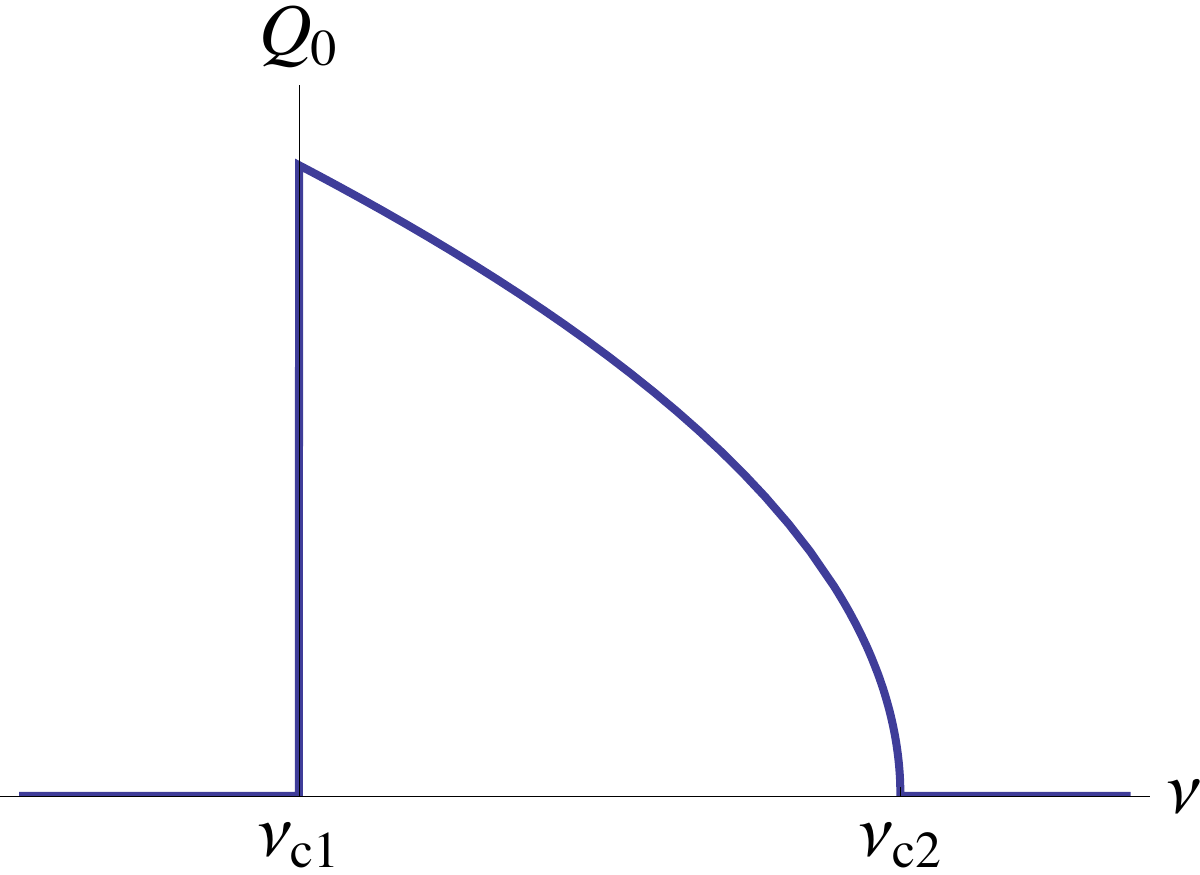} 
\caption{The momentum $Q(\nu)$ characteristic of the AMSF (polar) state,
  ranging between zero and the $p$-wave FR width-dependent value.}
\label{fig:Q}
\end{figure} 

The ASF appears at a large {\em positive} detuning (weak FR attraction) and
low temperature, where one of the three combinations (ASF$_1$,
ASF$_2$, ASF$_{12}$) of the $^{85}$Rb and $^{87}$Rb atoms are
Bose-condensed into a conventional, uniform superfluid, and the
$p$-wave $^{85}$Rb-$^{87}$Rb molecules are energetically costly and
therefore appear only as gapped excitations.
\begin{figure}[b]
\includegraphics[width=7cm]{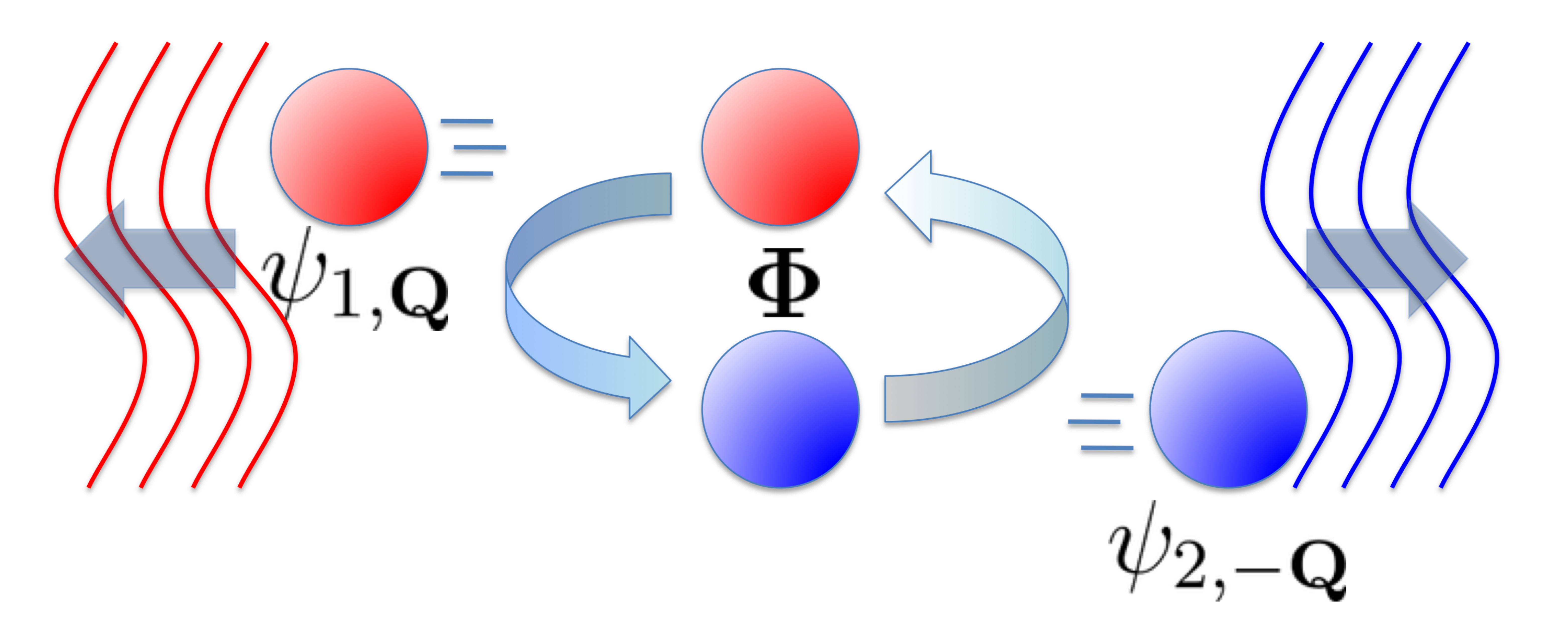} 
\caption{A cartoon of a $p$-wave molecule decaying into two oppositely
  moving two species of atoms, illustrating a resonant mechanism for a
  finite momentum $\ibQ$ atomic superfluidity (indicated by wavy
  lines) in the AMSF phase.}
\label{fig:cartoonAMSF}
\end{figure}

In the complementary regime of a large {\em negative} detuning, the
attraction between two flavors of atoms is sufficiently strong so as
to bind them into a tight $p$-wave hetero-molecules (e.g.,
$^{85}$Rb-$^{87}$Rb molecule), which at low temperature condense into
a $p$-wave molecular superfluid isomorphic to a spinor-$1$
condensate~\cite{matthews.cornell.98,stenger.ketterle.98,
stamperkurn.98,dipolarStamperKurn,
  ho.spinor.98,ohmi.machida.98,ho.yip.00,zhou.spinor.01,demler.zhou.02,
  barnett.demler.06,mukerjee.moore.06}.  The latter is known to come
in two forms, thereby predicting the $\ell_z=0$ ``polar'' (MSF$_{\rm
  p}$) and $\ell_z=\pm 1$ ``ferromagnetic'' (MSF$_{\rm fm}$) molecular
$p$-wave superfluid phases, with their relative stability determined
by the ratio $a_0/a_2$ of molecular spin-0 ($a_0$) to molecular spin-2
($a_2$) scattering lengths.

Besides these fairly conventional {\em uniform} atomic and molecular
BEC's, for intermediate detuning around a unitary point the gas is
predicted to exhibits novel AMSF$_{\rm p}$ and AMSF$_{\rm fm}$ phases,
characterized by a {\em non-zero momentum} $\hbar Q$ atomic
condensate~\cite{kuklov.06, RCprl.09,CRpra.11}, that is a
superposition of two atomic species. The momentum
\begin{equation}
Q=\alpha m \sqrt{n_m} \sim \sqrt{\gamma_p\ell n_m}
\lesssim\sqrt{\gamma_p}/\ell
\label{Q}
\end{equation}
is tunable via the FR detuning, $\nu$, primarily entering through the
molecular condensate $n_m(\nu)\lesssim 1/\ell^3$ density, and sensitive
to the FR width $\gamma_p$~\cite{GRaop}.  In addition to exhibiting an
off-diagonal long-range order (ODLRO) of an ordinary superfluid the
two AMSF$_{\rm p,fm}$ states (distinguished by the polar versus
ferromagnetic nature of their $p$-wave molecular condensates)
spontaneously partially break orientational and translational
symmetries, akin to polar and smectic liquid
crystals~\cite{deGennesProst} and the putative
Fulde-Ferrell-Larkin-Ovchinnikov states of imbalanced paired
fermions~\cite{FF,LO,KetterleZwierleinReview,Zwierlein06Science,
  Partridge06Science,Navon2009prl,Mizushima,SRprl,SRaop,RSreview}
discussed in the first part of this review.  This state is a {\em
  finite momentum}, $\hbar Q$ spinor superfluid, akin to (but distinct
from) a supersolid~\cite{Andreev69,Chester70,Leggett70,KimChan}.

The physical picture behind such a finite-momentum AMSF formation is
quite clear and is illustrated as a cartoon in AMSF phase of
Fig.~\ref{fig:phasediagramAMSF}. At intermediate detuning, where atomic gap
closes within the MSF state, $p$-wave molecules decay via FR into a
pair of atoms, which (due to the $p$-wave nature of the molecules) are
necessarily created at finite and opposite momenta, $\pm\bk$, and
therefore at low temperature form a finite momentum atomic condensate,
AMSF. The energetic cost ($\sim Q^2/2m$) of a finite momentum atomic
condensation is balanced by the lowering of the energy ($\sim\alpha
Q\sqrt{n_m}$) through FR hybridization between closed-channel $p$-wave
molecule and open-channel pair of atoms that is only possible at
finite atomic momentum, giving $Q$ in \rfs{Q}.

\begin{figure}[thb]
\includegraphics[width=8.5cm]{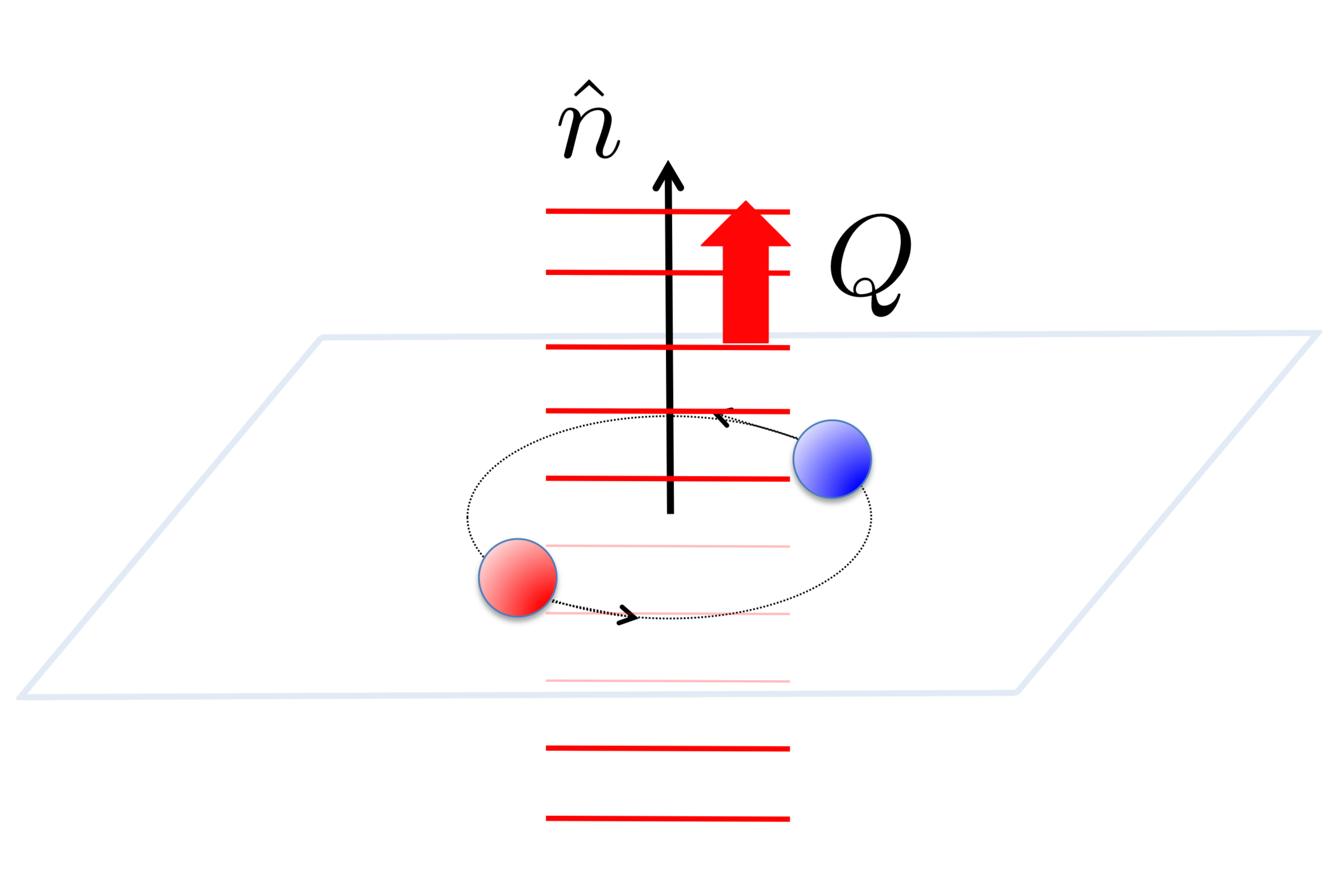}  \\
\caption{Schematic of the AMSF$_{\rm p}$ polar state.  The thick arrow
  indicates the atomic condensate momentum $\ibQ$ and the $\hat n$
  arrow denotes the quantization axis along which the projection of
  molecular internal orbital angular momentum vanishes.}
\label{fig:AMSFp}
\end{figure}

\begin{figure}[thb]
\includegraphics[width=8.5cm]{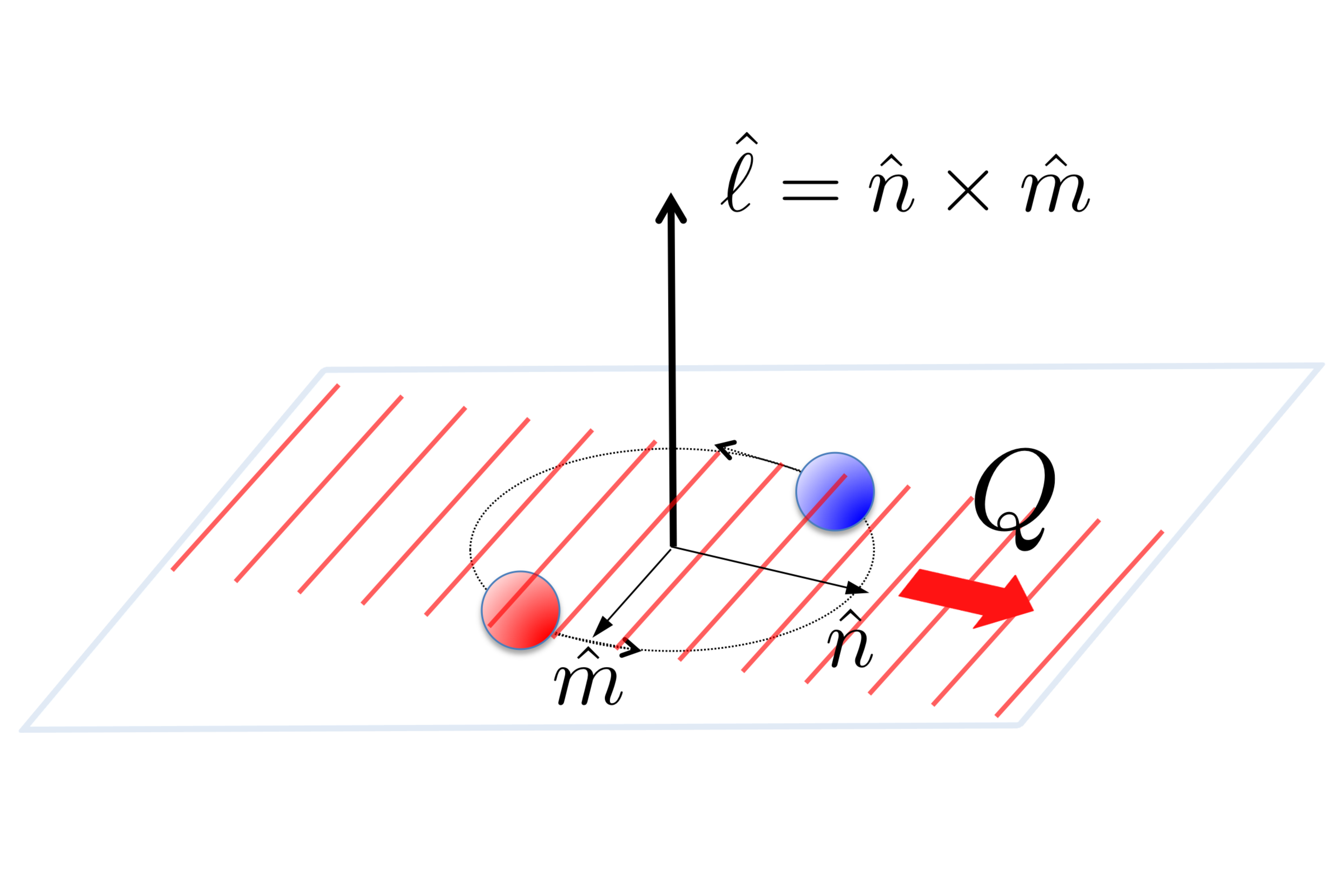}  \\\
\caption{Schematic of the AMSF$_{\rm fm}$ ferromagnetic state.  The
  thick arrow indicates the atomic condensate momentum $\ibQ$, lying
  in the plane transverse to the quantization axis $\hat{\bm\ell}$,
  along which the projection of the molecular internal orbital angular
  momentum is $\ell_z=+1$.}
\label{fig:AMSFfm}
\end{figure}

As illustrated in Fig.~\ref{fig:AMSFp}, in the polar AMSF$_{\rm p}$ state,
$\ibQ$ aligns along the quantization axis along which the molecular
condensate has a zero projection of its internal $\ell=1$ angular
momentum. 
For the case of the ferromagnetic AMSF$_{\rm fm}$ state, $\ibQ$ lies
in the otherwise isotropic plane, transverse to the $p$-wave molecular
condensate axis, as illustrated in Fig.~\ref{fig:AMSFfm}.

As any neutral superfluid, ASF, MSF, and AMSF are each characterized
by Bogoliubov modes, with long wavelength acoustic ``sound''
dispersions
\be
E^{\rm B}_\sigma({\bf k}) \approx c_\sigma \hbar k, 
\ee 
where $c_\sigma$ (with $\sigma =$ASF$_{1,2,12}$, MSF$_{\rm p,fm}$,
AMSF$_{\rm p,fm}$) are the associated sound speeds with standard
Bogoliubov form $c_\sigma\approx\sqrt{g_\sigma n_\sigma/2m}$.  In each
of these SF states one Bogoliubov mode (and only one in the ASF$_i$
states) corresponds to the overall condensate phase fluctuations. In
addition, the MSF$_{\rm p}$ exhibits two degenerate ``transverse''
Bogoliubov orientational acoustic modes. The MSF$_{\rm fm}$ is also
additionally characterized by one ``ferromagnetic'' spin-wave mode,
$E_k^{\rm MSF_{fm}}\sim k^2$ and one gapped mode, consistent with the
characteristics of a conventional spinor-1
condensate~\cite{ho.spinor.98, ohmi.machida.98}.

Because MSF$_{\rm p,fm}$ are paired molecular superfluids, they also
exhibit gapped single atom-like quasiparticles (akin to Bogoliubov
excitations in a fermionic paired BCS state), that do not carry a
definite atom number.  These single-particle excitations are
``squeezed'' by the presence of the molecular condensate, offering a
mechanism to realize atomic squeezed states~\cite{kimble.90}, that can
be measured by interference experiments, similar to those reported in
Ref.~\onlinecite{kasevich.01}.  The low-energy nature of these
single-atom excitations is guaranteed by the vanishing of the gap at
the MSF-AMSF transition at $\nu_c^{\rm MSF_{p,fm}-AMSF_{p,fm}}$, with
$E^{\rm gap}_{\rm MSF}(\nu_c) = 0$.

In addition to conventional Bogoliubov modes the AMSF exhibits a
Goldstone mode corresponding to the fluctuations of a {\em relative}
phase between the two atomic condensate components. A spatially
periodic collinear AMSF state exhibits a smectic-like anisotropic
phonon mode, akin to striped FFLO states~\cite{LO,Shimahara,RVprl,Rpra}
discussed in the first part of this review.
\bse 
\bea
\omega_{\rm AMSF_p}(\bk)&=& \sqrt{(B k_z^2+K k^4_\perp)/\chi_-},
\label{omegaAMSF_p}\\
\omega_{\rm AMSF_{fm}}(\bk)&=& \sqrt{(B
  k_z^2+k^2(K_x k_x^2+K_y k_y^2))/\chi_-},
\label{omegaAMSF_fm}\nonumber\\
\eea 
\ese
and an orientational mode $\omega_{\rm fm}^\gamma$, associated with
orientational symmetry breaking in AMSF$_{\rm fm}$
\bse
\begin{align}
\omega_{+p}(\bk) =&\sqrt{\frac{2\rho_s}{\chi_+m}}k, \\
\omega_{\rm fm}^\gamma(\bk) =& \sqrt{\frac{Jk^2\left[Bk^2_z+k^2(K_xk_x^2+K_yk_y^2)\right]}{J\chi_-k^2+\kappa^2k_y^2}},
\end{align}
\ese
where $B$, $K$'s, $J$, $\kappa$, and $\chi$ are compressional and
bending moduli characterizing the phases.

\subsection{Microscopic model of a balanced $p$-wave resonant Bose
  gas}

As required by bosonic statistics a $p$-wave resonance takes place
between two distinguishable bosonic atoms created by
$\psi_\sigma^\dagger(\br)=\left(\psi_1^\dagger(\br),\psi_2^\dagger(\br)\right)$,
(e.g., $^{85}$Rb, $^{87}$Rb)~\cite{Papp08prl}, interacting through a
closed-channel $\ell = 1$ hetero-molecule created by a vector field
operator
${\bm\phi}^\dagger(\br)=(\phi_x^\dagger,\phi_y^\dagger,\phi^\dagger_z)$. The
corresponding Hamiltonian is given by,
\bea
\curH &=&  
\sum_{\sigma=1,2}\hat\psi_{\sigma}^{\dag} \hat{\varepsilon}_{\sigma}
\hat\psi_\sigma 
+\hat{\bm\phi}^{\dag}\cdot\hat{\omega}\cdot\hat{\bm\phi} 
+ \curH_{bg} 
\label{Hmain} \\
&&+\frac{\alpha}{2}
\left(\hat{\bm\phi}^{\dag}\cdot
\left[\hat\psi_{1}(-i\grad)\hat\psi_{2} -
\hat\psi_{2}(-i\grad)\hat\psi_{1}\right]+h.c. \right),\nonumber
\eea
where $\hat{\varepsilon}_{\sigma} = -\frac{1}{2m}\grad^2-\mu_\sigma$,
$\hat{\omega}=-\frac{1}{4m}\grad^2-\mu_m$, with molecular chemical
potential $\mu_m=\mu_1+\mu_2-\nu$, adjustable by a magnetic field
dependent detuning $\nu$, latter the rest energy of the closed-channel
molecule relative to a pair of open-channel atoms. A generalization to
number and mass imbalanced mixtures (as studied for fermionic
atoms\cite{KetterleZwierleinReview,Zwierlein06Science,
  Partridge06Science,Navon2009prl,Mizushima,SRprl,SRaop,RSreview}
remain to be explored\cite{commentFRequalmass}.

For simplicity, above form is specialized to a rotationally invariant
FR interaction, with $\hat{\omega}$ and $\alpha$ independent of the
molecular component $i$. This is an approximation for the system of
interest, $^{85}$Rb-$^{87}$Rb mixture, where indeed the $p$-wave FR
around $B = 257.8$ Gauss~\cite{Papp08prl} is split into a
doublet by approximately $\Delta B = 0.6$ Gauss, similar to the
fermionic case of
$^{40}$K~\cite{RegalPwave03,GaeblerPRLpwave,GRApwave,GRaop}. A more
realistic, richer case remains to be explored.

The background (non-resonant) interaction density $\curH_{bg}$ is the
short-scale two-body interaction between spin and number densities
characterized by atomic scattering lengths $a_1,a_2, a_{12}$ as well
as molecular scattering lengths. A miscibility of a two-component
atomic gas requires a condition on the corresponding atomic $s$-wave
scattering lengths $a_1 a_2 > a_{12}^2$~\cite{greene.97}, which may be
problematic for the case of $^{85}$Rb-$^{87}$Rb due to the negative
background scattering length of $^{85}$Rb.

The corresponding imaginary time ($\tau$) coherent state Lagrangian
density is given by
\begin{widetext}
\begin{eqnarray}
\curL &=& 
\psi^*_\sigma(\partial_\tau-\frac{\nabla^2}{2m}-\mu_\sigma)\psi_\sigma +
\bm\phi^*\cdot(\partial_\tau-\frac{\nabla^2}{4m}-\mu_m)\cdot\bm\phi 
+\frac{\lambda_\sigma}{2}\ve\psi_\sigma\ve^4 \nonumber \\
&+&\lambda_{12}\ve\psi_{1}\ve^2\ve\psi_{2}\ve^2
+g_{am}\left(\ve\psi_{1}\ve^2+\ve\psi_{2}\ve^2\right)\ve{\bm\phi}\ve^2 
+\frac{g_1}{2}\ve{\bm\phi}^*\cdot{\bm\phi}\ve^2
+\frac{g_2}{2}\ve{\bm\phi}\cdot{\bm\phi}\ve^2 \nonumber \\
&+&\frac{\alpha}{2}
\left({\bm\phi}^*\cdot \left[\psi_1(-i\grad)\psi_2-\psi_2(-i\grad)\psi_1\right]+c.c.\right),
\label{curlL}
\end{eqnarray}
\end{widetext}
where $\lambda_i$s and $g_i$s are the atomic and molecular $s$-wave
pseudopotentials.  Closely related models also arise in completely
distinct physically contexts. These include quantum magnets that
exhibit incommensurate spin liquids states~\cite{SachdevMI} and
bosonic atoms in the presence of spin-orbit
interactions~\cite{HuiZhaiSOIbosons}.

The low-energy two-atom vacuum scattering in the above two-channel
model can be computed exactly. It faithfully captures all the features
of the low-energy $p$-wave resonant and $s$-wave nonresonant
scattering phenomenology of the $^{85}$Rb-$^{87}$Rb $p$-wave
Feshbach-resonant mixture~\cite{Papp08prl}. By matching experiments in
the dilute limit one can fix the model's key
parameters\cite{GRaop,CRpra.11}. The analysis at nonzero, balanced
atomic densities leads to the predictions summarized above. All
predictions are qualitatively robust, and can furthermore be made
quantitatively accurate in a narrow-resonance limit.

\subsection{Phases, symmetries and Goldstone modes}

Qualitative features of the phase diagram for this system can be
mapped out through a mean-field treatment of the Lagrangian,
\rf{curlL}, supplemented by symmetry arguments and the analysis of
fluctuations about the ordered state.

In the absence of periodic or disorder potential at sufficiently low
temperatures a {\em bosonic} gas is always a superfluid, that in three
dimensions exhibits Bose-Einstein condensation, characterized by
scalar atomic, $\Psi_\sigma$ and/or $3$-vector molecular, ${\bm\Phi}$
complex order parameters. Thus, at low T the gas exhibits three
classes of SF phases:
\begin{enumerate}
\item Atomic Superfluid (ASF), $\Psi_\sigma \neq 0$ and ${\bm\Phi}=0$
\item Molecular Superfluid (MSF), $\Psi_\sigma = 0$ and ${\bm\Phi}\neq 0$
\item Atomic Molecular Superfluid (AMSF), $\Psi_\sigma \neq 0$ and
  ${\bm\Phi}\neq 0$
\end{enumerate}
with the transition between them driven by the magnetic
field-dependent detuning, $\nu$.

\subsubsection{Atomic superfluids: ASF$_1$, ASF$_2$, ASF$_{12}$}
For large {\em positive} detuning $\nu$, closed-channel molecules are
gapped and the ground state is a molecular vacuum, and a zero-momentum
atomic condensate ASF.  The latter itself comes in three forms: (i)
ASF$_1$ with $\Psi_1\neq 0,\Psi_2 = 0$, (ii) ASF$_2$ with $\Psi_1=
0,\Psi_2 \neq 0$, (iii) ASF$_{12}$ with $\Psi_1\neq 0,\Psi_2 \neq 0$,
separated by continuous phase transitions.  For a balanced mixture
$\tilde{\mu}_1 = \tilde{\mu}_2$, the system exhibits a direct
N-ASF$_{12}$ transition through a tetracritical point, $\tilde{\mu}_1
= \tilde{\mu}_2 = 0$, that is believed to be in the decoupled
universality class\cite{LiuFisher,KNF,Vicari}.  All other transitions
(N-ASF$_1$, N-ASF$_2$, and ASF$_i$-ASF$_{12}$) are in the XY
universality class, breaking associated $U(1)$ symmetries. The phase
boundaries and the values of the atomic condensate order parameters
can be straightforwardly computed within mean-field theory
\cite{RCprl.09,CRpra.11}, but are modified by
fluctuations\cite{KNF,Vicari}.

Within ASF phases, the spectrum of fluctuations can be
straightforwardly computed by a Bogoliubov diagonalization of coupled
atomic and molecular excitations, with details depending on which of
the three possible ASF phases is studied. In general these states
exhibit one Bogoliubov sound mode per broken atomic $U(1)$ symmetry,
with one Goldstone mode in ASF$_1$ and ASF$_2$ phases and two in
ASF$_{12}$\cite{CRpra.11}.

\subsubsection{Molecular superfluids: MSF$_{p}$ and MSF$_{fm}$}
In the opposite limit of a large {\em negative} detuning, atoms are
gapped, tightly bound into heteromolecules, that at low temperature
condense into a $p$-wave molecular superfluid, MSF. In this regime of
atomic vacuum, the gas reduces to that of interacting $p$-wave
molecules, isomorphic to that of the extensively studied $F=1$ spinor
condensate~\cite{matthews.cornell.98,stenger.ketterle.98,stamperkurn.98,
  dipolarStamperKurn,
  ho.spinor.98,ohmi.machida.98,ho.yip.00,zhou.spinor.01,demler.zhou.02,
  barnett.demler.06,mukerjee.moore.06}, with the hyperfine spin $F$
here replaced by the orbital $\ell=1$ angular momentum of two
constituent atoms.

Like $F=1$ spinor condensates, the $p$-wave molecular superfluid, MSF
exhibits two distinct phases depending on the sign of the renormalized
interaction coupling $g_2$ in \rfs{curlL}, or equivalently the sign of
the difference $a_0^{(m)}-a_2^{(m)}$ of the molecular $L=0$ and $L=2$
channels $s$-wave scattering lengths.

{\em Polar molecular superfluid, MSF$_{p}$:}

For $g_2 < 0$ the ground state is the so-called ``polar'' molecular
superfluid, MSF$_{\rm p}$, characterized by a (collinear) order
parameter ${\bm\Phi}=\Phi_{\rm p}e^{i\varphi}{\hat{\bm n}}$, with
${\hat{\bm n}}$ a real unit vector, $\varphi$ a (real) phase, and
$\Phi_{\rm p}$ a (real) order-parameter amplitude, with the state
corresponding to $\ell_z=0$ projection of the internal molecular
orbital angular momentum along ${\hat{\bm n}}$. MSF$_{\rm p}$ clearly
spontaneously breaks rotational symmetry by its choice of the
$\ell_z=0$ quantization axis ${\hat{\bm n}}$, and the global gauge
symmetry, corresponding to a total atom number conservation. The
low-energy order parameter manifold that characterizes MSF$_{\rm p}$
is given by the coset space $(U(1)\otimes S_2) /\mathbb{Z}_2$,
admitting half-integer ``charge'' vortices~\cite{mukerjee.moore.06}
akin to (but distinct from) the $s$-wave
MSF~\cite{RPWprl,RomansPRL,RWPaop}.

The polar MSF$_{\rm p}$ state exhibits three gapless Bogoliubov-like
modes that are calculated in a standard way\cite{CRpra.11}.  One
corresponds to breaking of the global atom number conservation and two
associated with breaking of rotational $O(3)$
symmetry~\cite{ho.spinor.98,CRpra.11}.

{\em Ferromagnetic molecular superfluid, MSF$_{fm}$:}

Alternatively, for $g_2>0$ the ground state is the ``ferromagnetic''
molecular superfluid, MSF$_{\rm fm}$, characterized by an (coplanar)
order parameter ${\bm\Phi}=\frac{\Phi_{\rm fm}}{\sqrt{2}}({\hat{\bm
    n}} \pm i{\hat{\bm m}})$, with ${\hat{\bm n}}$, ${\hat{\bm m}}$,
$\hat{\bm\ell}\equiv{\hat{\bm n}}\times{\hat{\bm m}}$ a real
orthonormal triad, $\Phi_{\rm fm}$ a real amplitude, and the state
corresponds to $\ell_z=\pm 1$ projection of the internal molecular
orbital angular momentum along the $\hat{\bm\ell}$ axis. MSF$_{\rm
  fm}$ spontaneously breaks the time reversal, the $O(3)$ rotational
and the global gauge $U_N(1)$ symmetries, latter corresponding to a
total atom number $N$ conservation. Inside MSF$_{\rm fm}$ the
low-energy order parameter manifold is that of the
$O(3)=SU(2)/\mathbb{Z}_2$ group, corresponding to orientations of the
orthonormal triad ${\hat{\bm n}},{\hat{\bm m}}, \hat{\bm\ell}$.

As its hyperfine spinor-condensate cousin, the ferromagnetic MSF$_{\rm
  fm}$ exhibits (only) {\em two} gapless Goldstone modes, one linear
($\propto k$) conventional Bogoliubov mode associates with the broken
global $U(1)$ gauge symmetry, and another quadratic ($\propto k^2$)
corresponding to the ferromagnetic order, with associated
spin-waves~\cite{ho.spinor.98} reflecting the precessional FM
dynamics. This is despite the three-dimensionality of its $SO(3)$
coset space and can be traced back to the fact that the two components
of the spinor are canonically conjugate and as a result combine into a
single low-frequency $k^2$ mode.

\begin{figure}[thb]
\includegraphics[width=7cm]{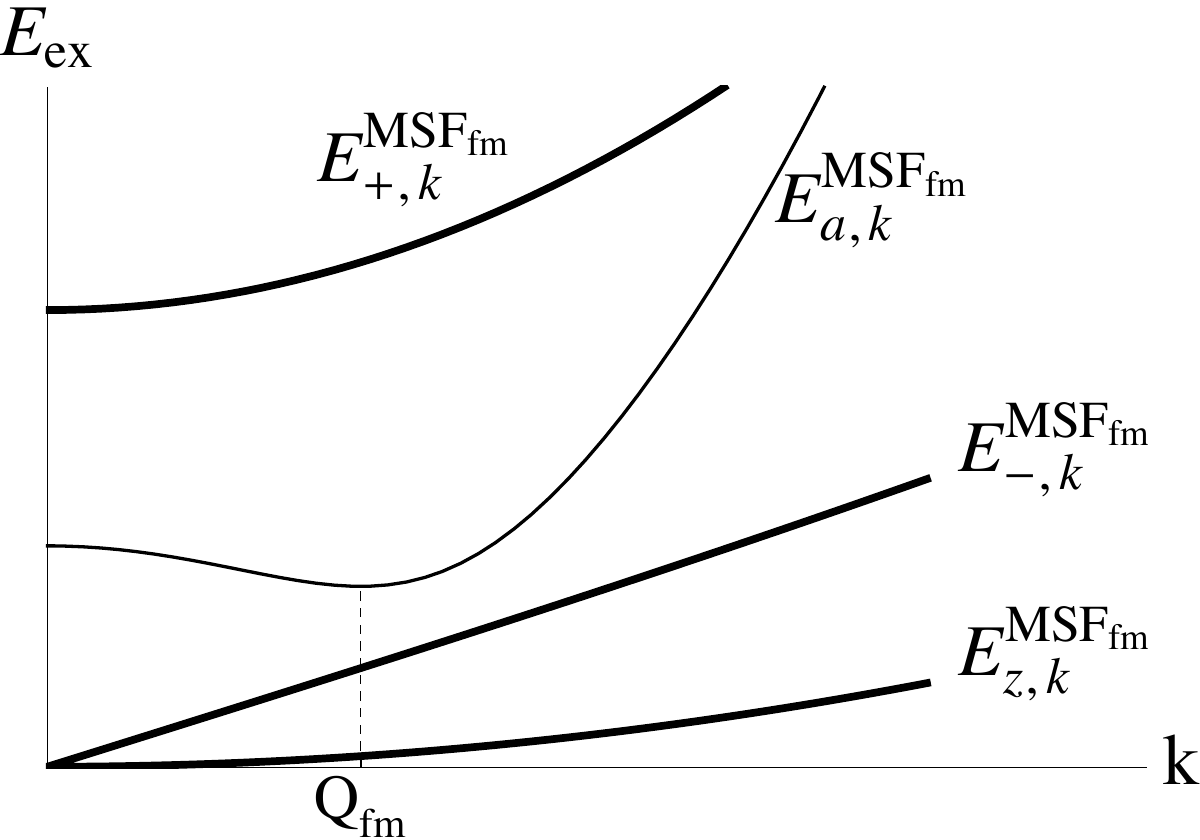} 
\caption{Excitation spectrum for the ferromagnetic molecular
  superfluid, MSF$_{\rm fm}$. The doubly-degenerate atomic spectrum
  (thin curves) exhibits a minimum gap at nonzero $k$, a precursor of
  finite momentum atomic condensation into the AMSF$_{\rm fm}$. The
  molecular spectrum (thick curves), consists of a longitudinal
  gapless quadratic ferromagnetic spin-wave mode (lowest), a
  Bogoliubov sound mode and a quadratic gapped mode.}
\label{msfexcFer}
\end{figure} 
As illustrated in Fig.\rf{msfexcFer}, while atomic excitations are
gapped (confined into molecules), because of the $p$-wave FR coupling
in \rf{curlL} the minimum in the atomic spectrum occurs at a nonzero
momentum, $Q_p$, a precursor of the Bose-condensation transition into
a nonzero momentum AMSF state, when this atomic gap closes.

\subsubsection{Atomic-molecular superfluids: AMSF$_{p}$ and
  AMSF$_{fm}$}

As detuning is further increased from a large negative value of the
MSF$_{\rm p,fm}$ phases, for intermediate $\nu$ the gap to atomic
excitations decreases, closing at a critical value of $\nu_c^{\rm
  MSF-AMSF}$ and leading to atomic Bose-condensation into the
corresponding AMSF$_{\rm p}$ and AMSF$_{fm}$ states. It is clear from
the {\em linear} momentum dependence of the $p$-wave FR coupling in
the Hamiltonian \rf{Hmain}, that, quite generally, the MSF-AMSF
transition is robustly into a {\em nonzero}-momentum atomic
condensate, with $k = Q$, set by a balance of the $p$-wave FR
hybridization and the atomic kinetic energies.

As with other (partially) crystalline states of
matter~\cite{ChaikinLubensky,FF,LO}, the nature of the resulting AMSF
states depends on the symmetry of the crystalline order, encoded into
the atomic condensate order parameter
\bea
\hspace{-1cm}
\Psi_{\sigma}(\br)&=&
\begin{pmatrix}
\Psi_1(\br) \\ 
\Psi_2(\br)
\end{pmatrix}
=\sum_{\ibQ_n}
\begin{pmatrix}
\Psi_{1,\ibQ_n}e^{i\ibQ_n\cdot\br} \\
\Psi_{2,-\ibQ_n}e^{-i\ibQ_n\cdot\br}
\end{pmatrix},
\eea
via the associated set of the reciprocal lattice vectors, $\ibQ_n$ at
which the condensation takes place. Determined by a detailed nature of
interactions and fluctuations, typically the nature of crystalline
order is challenging to deduce.

The nature and symmetries of these AMSF states furthermore
qualitatively depends on the parent MSF, with the polar AMSF$_{\rm p}$
and the ferromagnetic AMSF$_{\rm fm}$ as two possibilities. In
addition to the symmetries already broken in its MSF parent, by virtue
of atomic condensation the AMSF state breaks the remaining $U_{\Delta
  N}(1)$ global gauge symmetry associated with the conservation of the
difference in atom species number, $\Delta N$. Other symmetries that
it breaks depend on the detailed structure of the AMSF$_{\rm fm,p}$
states.

{\em Polar atomic-molecular superfluid, AMSF$_{p}$}

\begin{figure}[bht]
\begin{tabular}{c}
\includegraphics[width=8.5cm]{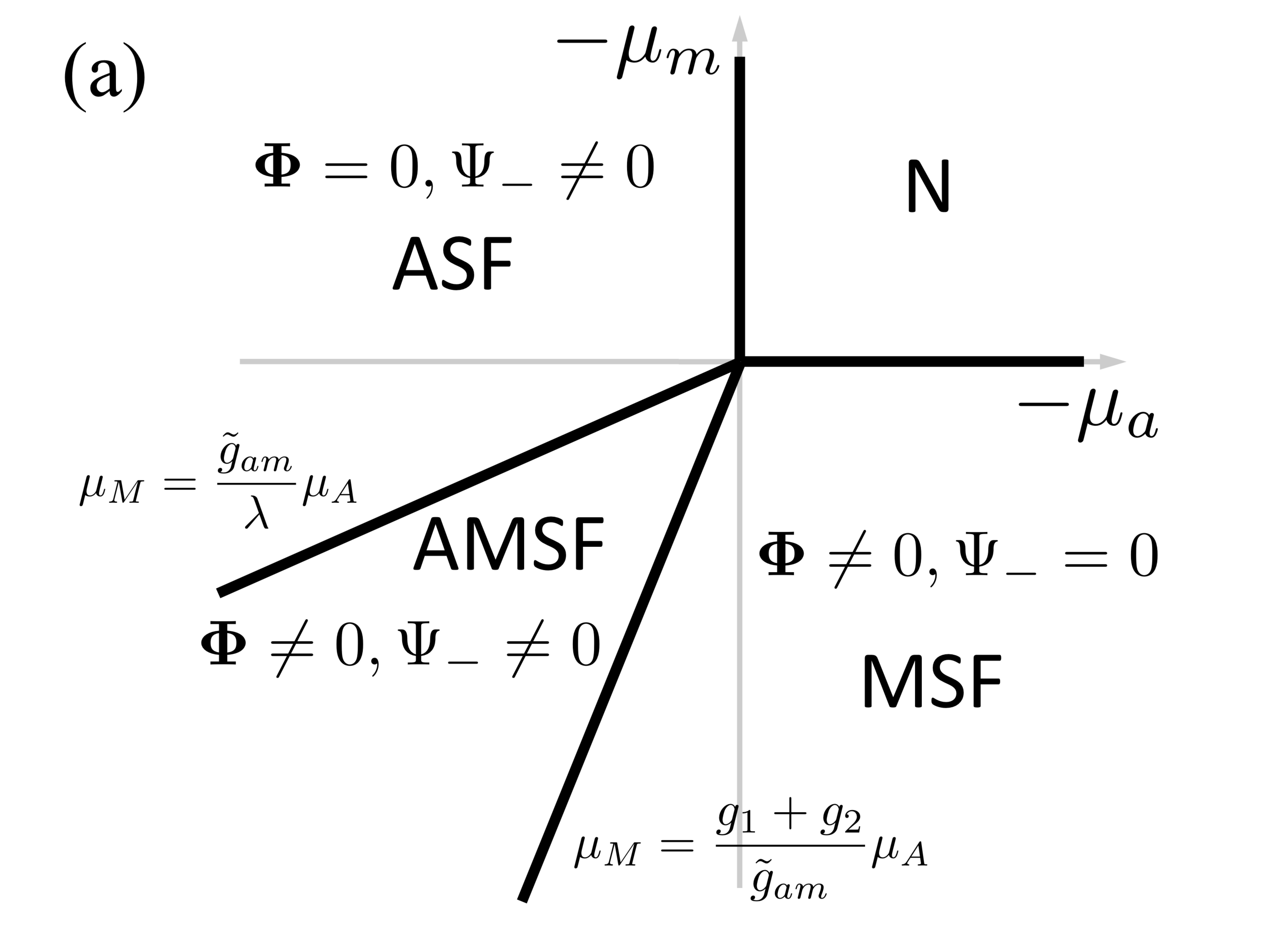} \\
\includegraphics[width=8.5cm]{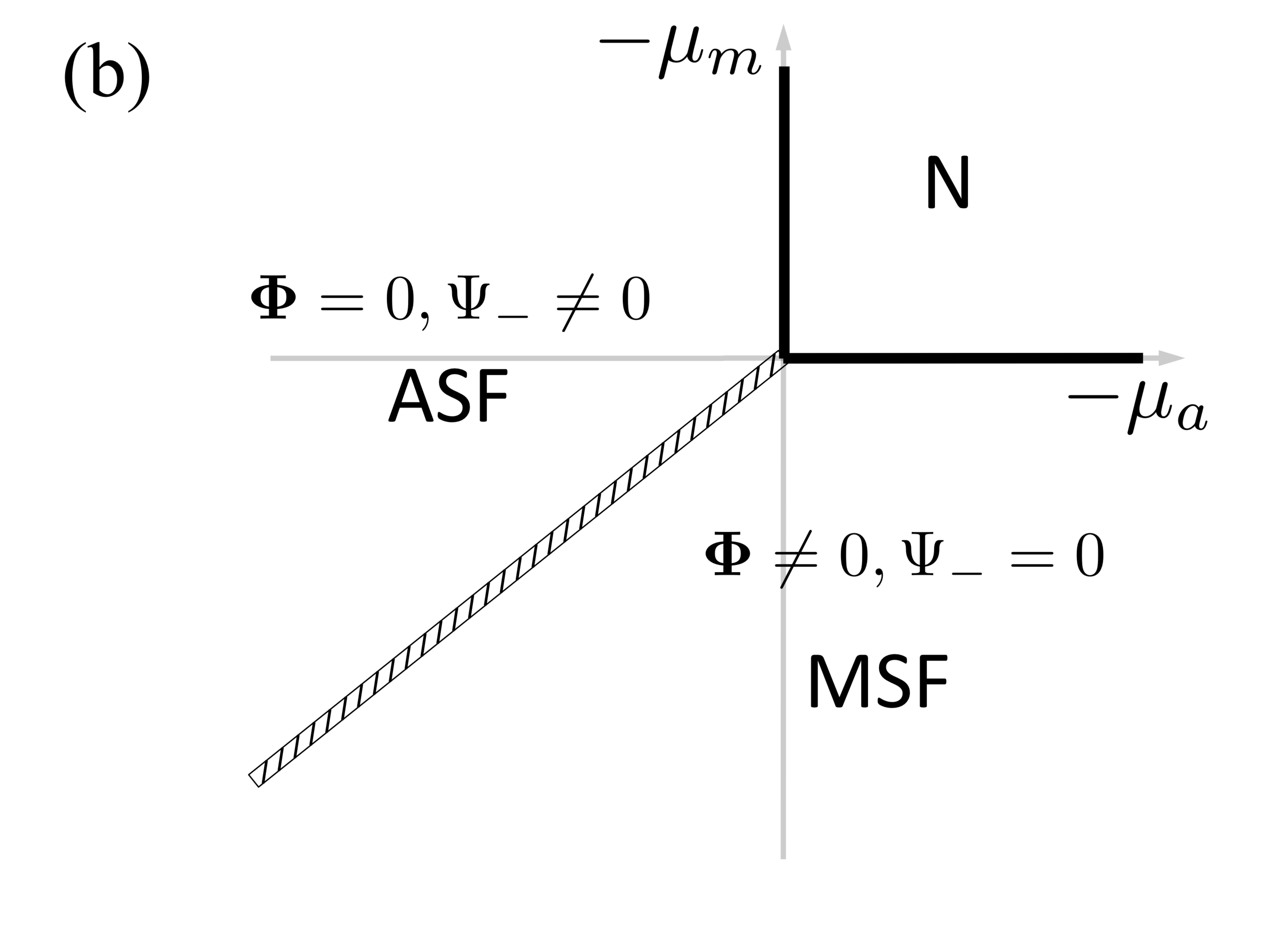}
\end{tabular}
\caption{Mean field phase diagrams for polar phase as a function of atomic and
  molecular chemical potentials, $\mu_a$, $\mu_m$, respectively.  
  Ferromagnetic phase is similar but with different parameters. 
  (a) For $\lambda(g_1+g_2)-{\tilde g_{am}}^{2}>0$, all three superfluid
  phases, ASF, AMSF, and MSF appear and are separated by continuous
  phase transitions (thick black lines), (b) For $\lambda(
  g_1+g_2)-{\tilde g_{am}}^{2}<0$, AMSF is unstable, and the ASF and
  MSF are separated by a first-order transition (hatched double
  line).}
\label{fig:PolarPhasediagram}
\end{figure}

\begin{figure}[thb]
\includegraphics[width=7cm]{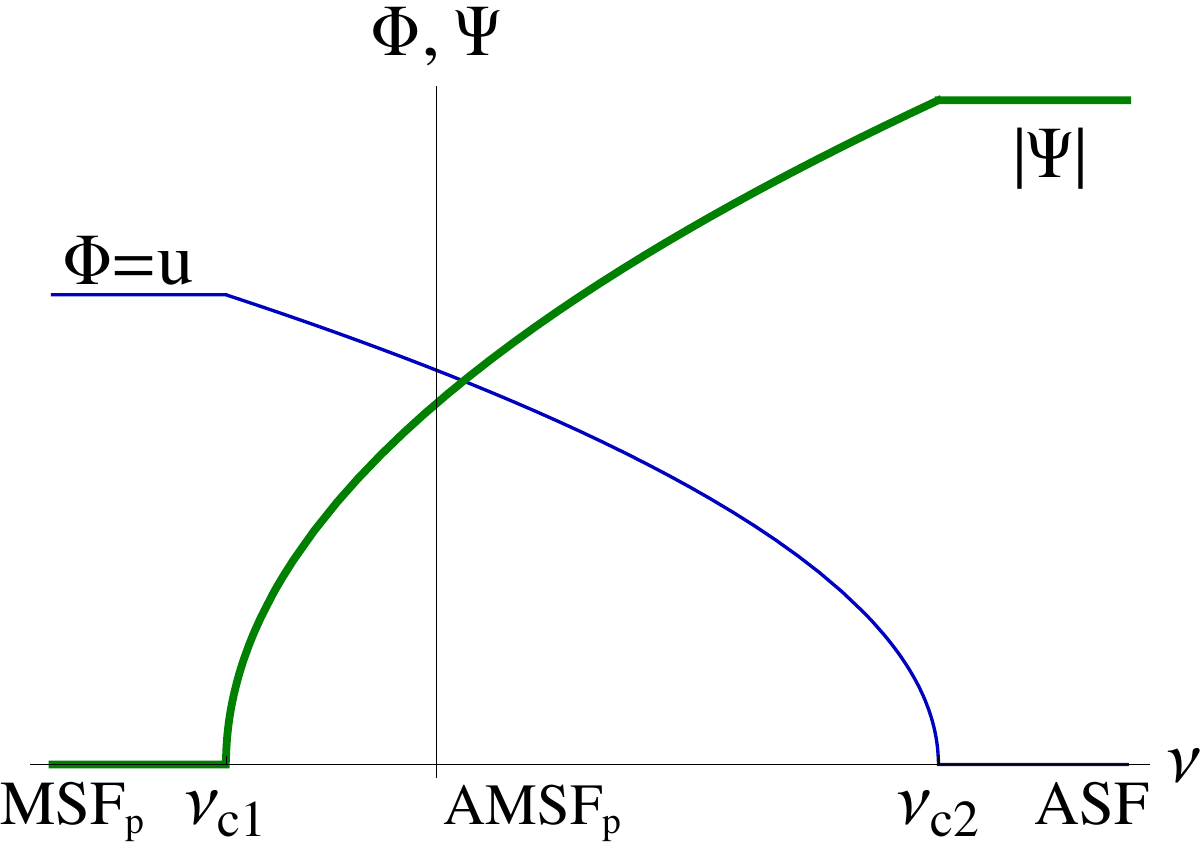} 
\caption{Atomic (thick) and molecular (thin) order
  parameters versus the FR detuning $\nu$ for the polar phase, with
  $\nu_{\rm c1}= \nu_{\rm c}^{\rm MSF_{\rm p}-AMSF_{\rm p}}$ and
  $\nu_{\rm c2}= \nu_{\rm c}^{\rm AMSF_{\rm p}-ASF}$.}
\label{polarOPplot}
\end{figure}

The polar atomic-molecular superfluid AMSF$_{\rm p}$ emerges from the
polar molecular state, MSF$_{\rm p}$.  As illustrated in
Fig.\ref{fig:AMSFp} and is clear from the form of \rf{curlL} in the
AMSF$_{\rm p}$ the finite momentum atomic condensate orders with
$\ibQ$ locked along the molecular condensate field
${\bm\Phi}=\Phi_{\rm p}e^{i\varphi}{\hat{\bm n}}$. Hence, quite
generically it is a Fulde-Ferrell like~\cite{FF} {\em collinear}
plane-wave state, that spontaneously breaks the time reversal
symmetry, but not any additional spatial symmetries.  A schematic
phase diagram for this polar class of states is illustrated in
Fig.\ref{fig:PolarPhasediagram}
The phase boundaries corresponding to the MSF$_{\rm p}$ - AMSF$_{\rm
  p}$ and the AMSF$_{\rm p}$ - ASF transitions are given by
\bse
\begin{align}
\nu_{\rm c}^{\rm MSF_p-AMSF_p} &= -\left(g_1+ g_2-2\tilde g_{am}\right)n_m, 
\label{pMSF-AMSF}\\ 
&\approx -\frac{1}{2}\left(g_1+ g_2-2\tilde g_{am}\right)n,\\ 
\nu_{\rm c}^{\rm AMSF_p-ASF} &= \left(2\lambda-\tilde g_{am}\right)n_a, 
\label{pASF-AMSF}\\
&\approx\left(2\lambda-\tilde g_{am}\right)n, 
\end{align}
\ese
The evolution of the order parameters across the phase diagram is
illustrated in Fig.\ref{polarOPplot}.

Despite a nonzero momentum ($Q$) of its condensate, in its minimal
form such AMSF$_p$ state is a {\em nematic} (not smectic) superfluid
that is anisotropic but spatially {\em homogeneous}. A detailed
analysis\cite{CRpra.11} shows that Goldstone modes in this phase are
governed by a Lagrangian density
\begin{eqnarray}
\delta\curL_{\rm p} &=& 
\oh\chi_+(\partial_\tau\theta_+)^2 +
\frac{\rho_{s0}}{m}(\grad\theta_+)^2\\
&&+\oh\chi_-(\partial_\tau\theta_-)^2 +
\frac{\rho_0}{m}(\partial_z\theta_-)^2+
\frac{K}{2}(\grad^2_\perp\theta_-)^2,\nonumber
\label{deltaLpolarFinal}
\end{eqnarray}
where the compressibilities $\chi_\pm$ and other couplings are
functions of the microscopic parameters that have been computed in
weakly interacting limit\cite{CRpra.11}. Thus, despite a
translationally invariant nature of the AMSF$_p$ phase, in addition to
the conventional (linear in $k$) Bogoluibov mode, $\theta_+$, akin to
the FF state\cite{FF,Shimahara,RVprl,Rpra} it exhibits a {\em
  smectic}-like Goldstone mode, $\theta_-$, with dispersion given in
Eq.\rf{omegaAMSF_p}\cite{CRpra.11}.

{\em Ferromagnetic atomic-molecular superfluid, AMSF$_{fm}$}

Similarly, upon increase of the detuning, a finite-momentum atomic
condensation from the MSF$_{\rm fm}$ leads to the ferromagnetic
atomic-molecular superfluid, AMSF$_{\rm fm}$. In this state, a
$p$-wave Feshbach resonant interaction leads to the energetic
preference for a {\em transverse} orientation of the atomic condensate
momentum $\ibQ$ to the molecular quantization axis, $\hat{\bm
  \ell}=\hat{\bm n}\times\hat{\bm m}$. Consequently, as illustrated in
Fig.\ref{fig:AMSFfm}, the AMSF$_{\rm fm}$ state spontaneously breaks
additional rotational symmetry of the uniaxial molecular state in the
plane perpendicular to the molecular quantization axis
$\hat{\bm\ell}$. In fact the nature of spatial order inside the
$\hat{\bm n}-\hat{\bm m}$ plane is not necessarily of co-linear
striped order, and generically admits a 2d crystalline condensate,
nontrivially determined by interactions and fluctuations. Motivated by
the polar version of AMSF, so far only a striped order has been
investigated, with general form remaining an open problem. For such
collinear order, a mean-field analysis predicts a FF (rather than
LO~\cite{LO}) type AMSF$_{fm}$ state in the $\hat{\bm n}-\hat{\bm m}$
plane. Thus, the AMSF$_{\rm fm}$ state is a {\em biaxial nematic}
superfluid, defined by $\ibQ$ and $\hat{\bm\ell}$ axes.

The corresponding Goldstone modes Lagrangian density is given by:
\begin{widetext}
\begin{eqnarray}
\delta\curL_{\rm fm} &\approx& 
\oh\chi_+(\partial_\tau\theta_+)^2 +
\frac{\rho_{s0}}{m}(\grad\theta_+)^2 +
\oh\chi_-(\partial_\tau\theta_-)^2 +
\oh B(\partial_z\theta_-)^2+
\oh K_x (\grad\partial_x\theta_-)^2+
\oh K_y (\grad\partial_y\theta_-)^2\nonumber\\
&+&
i\kappa\partial_y\theta_-\partial_\tau\gamma 
+ \oh J (\grad\gamma)^2,
\label{deltaLferro}
\end{eqnarray}
\end{widetext}
where the Goldstone mode $\gamma$ describes one additional fluctuating
angle of the $\hat{\bm m}$ axis outside of the ${\bm\nh}$-${\bm\mh}$
plane, into the $\hat{\bm \ell}$ axis.

The biaxiality is expected and arises due to the vector $p$-wave
order, characterized by a spinor $\bm\Phi_{fm}$, with the quantization
axis, ${\bm\lh}$. The finite angular momentum, $\ell_z=\pm 1$ along
${\bm\lh}$ distinguishes AMSF$_{\rm fm}$ from AMSF$_{\rm p}$ and leads
to this additional Goldstone mode $\gamma$.

A straightforward diagonalization of the above Lagrangian leads to
dispersions for three Goldstone modes inside the AMSF$_{\rm fm}$ state:
\bse
\begin{eqnarray}
  \omega^+_{fm}(\bk)&=&c_+ k,\label{omega+AMSFfm}\\
  \omega^-_{fm}(\bk)&=&\sqrt{[B k_z^2 + k^2(K_x k_x^2 +
K_yk_y^2)]/\chi_-},\hspace{1cm}\label{omega-AMSFfm}\\
  \omega^\gamma_{fm}(\bk)&=&
\sqrt{\frac{J k^2[B k_z^2 + k^2(K_x k_x^2 + K_yk_y^2)]}
{J\chi_-k^2 +\kappa^2 k_y^2}}.\,\label{omega_gammaAMSFfm}
\end{eqnarray}
\label{omegasAMSFfm}
\ese
The anisotropic $\omega^\gamma_{fm}(\bk)$ dispersion corresponds to
the ferromagnetic spin-waves in the plane of atomic condensate
phase-fronts (``smectic layers'') of the $p$-wave atomic-molecular
condensate, AMSF$_{\rm fm}$, reducing to the dispersion of MSF$_{\rm
  fm}$ for a vanishing smectic order, with $B=0$.

\subsection{Phase transitions}
Phase transitions from the high temperature normal state into the ASF
and MSF states have been well-explored in the context of conventional
and spinor superfluids.

The quantum MSF - AMSF phase transitions are more interesting and
require a beyond mean-field treatment. In large part the complexity is 
associated with the nontrivial coupling of the Goldstone modes of the
MSF phase to the finite-momentum order parameter condensing in the
AMSF state. 

Starting with the coherent-state microscopic Lagrangian, specializing
to the MSF$_p$ ordered state, and integrating out massive modes, one
obtains the effective Lagrangian density 
\begin{align}
\curL_{\rm p}=&
\eps_+^{-1}\ve\partial_\tau\psi_-\ve^2
+\frac{1}{2m}\ve\left(-i\grad - \alpha m \sqrt{\rho_{m0}}\delta{\bm\nh}\right) \psi_-\ve^2 \nonumber \\
&+\epsilon_-\ve\psi_-\ve^2
+\frac{\lambda}{2}\ve\psi_-\ve^4
+\frac{1}{2g_2}(\partial_\tau\nh)^2
+\frac{\rho_{m0}}{4m}(\grad\nh)^2 \nonumber \\
&+\frac{1}{2g}(\partial_\tau\varphi)^2
+\frac{\rho_{m0}}{4m}(\grad\varphi)^2,
\end{align}
that describes the quantum MSF$_p$-AMSF$_p$ transition.  The critical
field $\psi_-=\frac{1}{\sqrt{2}}(\psi_{1,\ibQ}+\psi_{2,-\ibQ}^*)$ is
the finite-momentum atomic condensate order parameter, $\bm\nh$ the
$\ell_z=0$ quantization axis and $\phi$ the superfluid phase, defining
the MSF$_p$ state.

Thus, as can also be argued on symmetry grounds, the zero-temperature
transition is indeed described by a quantum ($(d+1)$-dimensional) de
Gennes model (also known as the Ginzburg-Landau or Abelian-Higg's
models)~\cite{deGennesProst}, akin to that for a
normal-to-superconductor and nematic-to-smectic-A transitions. The
role of the nematic director (gauge-field) is played by the $\ell_z=0$
quantization axis of the $p$-wave molecular condensate.  Based on the
extensive work for these systems \cite{HLM74,Coleman73}, in three
(spatial) dimensions ($d=3$) the effective gauge-field fluctuations
drive this transition first-order.

Similarly, the quantum MSF$_{\rm fm}$-AMSF$_{\rm fm}$ transition is
described by the effective low-energy Lagrangian density
\begin{align}
\curL_{\rm fm}=&
\eps_+^{-1}\ve\partial_\tau\psi_-\ve^2
+\frac{1}{2m}\Bigg\ve\left(-i\grad - \frac{\alpha m \sqrt{\rho_{m0}}}{\sqrt{2}}\delta{\bm\nh}\right)\psi_-\Bigg\ve^2 
\nonumber \\
&+\epsilon_-\ve\psi_-\ve^2
+\frac{\lambda}{2}\ve\psi_-\ve^4
+\frac{\rho_{m0}}{8m}(\grad\bm\nh)^2
+\frac{\rho_{m0}}{8m}(\grad\bm\mh)^2 \nonumber \\
&+\frac{1}{2g}(\bm\nh\cdot\partial_\tau\bm\mh)^2.
\end{align}
It is distinguished from the polar case by the additional biaxial
order whose fluctuations are characterized by $\bm\mh$.

\subsection{Topological defects}

In addition the smooth low-energy configurations of these Goldstone
modes, the compact nature of the order parameters characterizing the
ASF, MSF and AMSF states admits a rich variety of topological
defects. These singular excitations are crucial to a complete
characterization of the states and their disordering, particularly in
the case of non-meanfield (e.g., partially disordered) states that are
not uniquely characterized by a Landau order parameter. In addition to
the conventional vortices in the ASF$_i$ states and well explored
defects in the spin-1 AMSF$_{p,fm}$ condensates, the AMSF$_{p,fm}$
exhibit interesting combination of vortex, dislocation and domain-wall
defects similar to those discussed in the context of FFLO and other
striped superfluids. While some preliminary studies have been carried
out\cite{CRpra.11}, their detailed analysis, response to rotation, and
realization remain subjects of future research.

\section{Summary and Conclusion}

In this review I discussed two recently studied quantum liquid crystal
orders, proposed to be realizable in $s$-wave resonant fermionic and
$p$-wave resonant bosonic atomic gases. The key ingredient is the
interaction-driven superfluid condensation at a nonzero momentum, that
in an isotropic trap spontaneously partially breaks orientational and
in some cases translational symmetry.  The main state that generically
appears at low temperature is a quantum superfluid smectic. A rich
variety of other phases results from partially disordering this state
by unbinding various combinations of topological defects associated
with compact nature of the broken symmetries.  An interesting
interplay ensues between Goldstone modes, associated spontaneously
broken gauge and spatial symmetries and the fermionic and bosonic
single-particle excitations. Although a general formulation of such
states is credibly established many interesting open questions remain.

\section{Acknowledgments}

I thank A. Vishwanath and Sungsoo Choi for collaborations on which
much of this review is based\cite{RVprl,RCprl.09,CRpra.11}. I acknowledge
fruitful discussions with V. Gurarie, M. Hermele, D. Huse, and
M. Levin. This work was supported by the National Science Foundation
through grants DMR-1001240 and MRSEC DMR-0820579.


\begin{thebibliography}{99}
\bibitem{BlochReview}
I. Bloch, J. Dalibard, W. Zwerger,
Rev. Mod. Phys. {\bf 80}, 885 (2008).
\bibitem{KetterleZwierleinReview} W. Ketterle and M. Zwierlein, ``Making,
  probing and understanding ultracold Fermi gases'', in {\it
    Ultracold Fermi Gases\/}, Proceedings of the International School
  of Physics ``Enrico Fermi", Course CLXIV, Varenna, 20 - 30 June
  2006, edited by M. Inguscio, W. Ketterle, and C. Salomon.
\bibitem{GRaop}
V. Gurarie and L. Radzihovsky, Ann. of Phys. {\bf 322\/}, 2  (2007).
\bibitem{GiorginiRMP}
S. Giorgini, L.P. Pitaevskii, 
and S. Stringari, Rev. Mod. Phys. {\bf 80}, 1215 (2008).
\bibitem{RSreview}
L. Radzihovsky, D E. Sheehy, Rep. Prog. Phys. {\bf 73}, 076501 (2010)
\bibitem{Regal2004prl}
C.A. Regal, M. Greiner, and D.S. Jin,
%
Phys. Rev. Lett. {\bf 92}, 040403 (2004).
%
%
\bibitem{Zwierlein2004prl}
M.W. Zwierlein, C.A. Stan, C.H. Schunck, S.M.F. Raupach, A.J. Kerman,
and W. Ketterle, Phys. Rev. Lett. {\bf 92}, 120403 (2004).
%
\bibitem{Kinast2004prl}
J. Kinast, S.L. Hemmer, M.E. Gehm, A. Turlapov, and J.E. Thomas,
 Phys. Rev. Lett. {\bf 92}, 150402 (2004).
%
\bibitem{Bartenstein2004prl}
    M. Bartenstein, A. Altmeyer, S. Riedl, S. Jochim, C. Chin, 
J.H. Denschlag, and R. Grimm, Phys. Rev. Lett. {\bf 92}, 120401 (2004).
%
\bibitem{Bourdel2004prl}
T. Bourdel, L. Khaykovich, J. Cubizolles, J. Zhang, F. Chevy,
M. Teichmann, L. Tarruell, S.J.J.M.F. Kokkelmans, and C. Salomon,
Phys. Rev. Lett. {\bf 93}, 050401 (2004).
%
\bibitem{ZhangPRApwave} J. Zhang, E. G. M. van Kempen, T. Bourdel,
  L. Khaykovich, J. Cubizolles, F. Chevy, M. Teichmann, L. Tarruell,
  S. J. J. M. F. Kokkelmans, and C. Salomon, Phys. Rev. A {\bf 70},
  030702(R) (2004).
\bibitem{GaeblerPRLpwave} J. P. Gaebler, J. T. Stewart,
  J. L. Bohn, D. S. Jin, Phys. Rev. Lett. {\bf 98}, 200403 (2007).
\bibitem{BotelhoSdeMeloPwave} S. S. Botelho and C. A. R. S. de Melo,
  J. Low Temp. Phys. {\bf 140}, 409 (2005).
\bibitem{GRApwave}
V. Gurarie, L. Radzihovsky and A.V. Andreev,
Phys. Rev. Lett. {\bf 94\/}, 230403 (2005).
%
\bibitem{ChengYipPRLpwave}
C.-H. Cheng and S.-K. Yip, Phys. Rev. Lett. {\bf 95}, 070404 (2005).
\bibitem{Eagles}
D.M. Eagles, Phys. Rev. {\bf 186\/}, 456 (1969).
%
\bibitem{Leggett}
A.J. Leggett, 
 in {\it Modern Trends in the Theory of Condensed Matter\/}, 
edited by A. Pekalski and R. Przystawa, (Springer-Verlag, Berlin, 1980).
\bibitem{NSR}
 P. Nozi\`eres and S. Schmitt-Rink, J. Low Temp. Phys. {\bf 59}, 195 (1985).
\bibitem{SdeMelo}
C.A.R. S\'a de Melo, M. Randeria, and J.R. Engelbrecht,
Phys. Rev. Lett. {\bf 71}, 3202 (1993).
%
\bibitem{Timmermans01}
E. Timmermans, K. Furuya, P.W. Milonni, and A.K. Kerman,
Physics Letters A {\bf 285}, 228 (2001).
\bibitem{Holland}
M. Holland, S.J.J.M.F. Kokkelmans, M.L. Chiofalo, and R. Walser,
 Phys. Rev. Lett. {\bf 87}, 120406 (2001).
%
\bibitem{Ohashi}
Y. Ohashi and A. Griffin, 
Phys. Rev. Lett. {\bf 89}, 130402 (2002);
%
%
%
%
\bibitem{AGRswave}
A.V. Andreev, V. Gurarie, and L. Radzihovsky,
Phys. Rev. Lett. {\bf 93}, 130402 (2004).
%
\bibitem{Stajic}
J. Stajic, J.N. Milstein, Q. Chen, M.L. Chiofalo, M.J. Holland, and K. Levin,  
Phys. Rev. A {\bf 69}, 063610 (2004).
%
\bibitem{OlsenBFmixture} M. L. Olsen, J. D. Perreault, T. D. Cumby, D. S. Jin
Phys. Rev. A {\bf 80}, 030701(R) (2009).
\bibitem{Cornish2000prl} S. L. Cornish,
N. R. Claussen, J. L. Roberts, E. A. Cornell, and C. E. Wieman,
Phys. Rev. Lett. {\bf 85}, 1795 (2000).
\bibitem{RPWprl} L. Radzihovsky, J. Park, and P. Weichman,
Phys. Rev. Lett. {\bf 92}, 160402 (2004).
\bibitem{RomansPRL} M. W. J. Romans, R. A. Duine, S. Sachdev, and 
H. T. C. Stoof, Phys. Rev. Lett. {\bf 93}, 020405 (2004).
\bibitem{RWPaop} L. Radzihovsky, P. Weichman, and J. Park,
Annals of Physics {\bf 323}, 2376 (2008).
\bibitem{Reviews} see e.g.,
  Refs.\onlinecite{BlochReview,KetterleZwierleinReview,
GRaop,GiorginiRMP,RSreview} and references therein.
\bibitem{BLprl} R. A. Barankov, L. S. Levitov, B. Z. Spivak,
Phys. Rev. Lett. {\bf 93}, 160401 (2004).
\bibitem{AltmanVishwanathPRL} E. Altman and A. Vishwanath,
Phys. Rev. Lett. {\bf 95}, 110404 (2005).
\bibitem{KFE98nature} S. A. Kivelson, 
E. Fradkin, V. J. Emery, Nature {\bf 393}, 550 (1998).
\bibitem{FK99prb} E. Fradkin, S. A. Kivelson, 
Phys. Rev. B {\bf 59}, 8065 (1999). 
\bibitem{OKF01prb} V. Oganesyan, S. Kivelson, E. Fradkin, 
Phys. Rev. B {\bf 64}, 195109 (2001).
\bibitem{Berg09prb} E. Berg, E. Fradkin, S. A. Kivelson,
Phys. Rev. B {\bf 79}, 064515 (2009).
\bibitem{Berg09nature} E. Berg, E. Fradkin, S. A. Kivelson,
Nature Phys. {\bf 5}, 830 (2009).
\bibitem{Sachdev03aop}
S. Sachdev, Annals Phys. {\bf 303} 226 (2003).
\bibitem{Agterberg08nature} D. F. Agterberg and H. Tsunetsugu, Nature
  Physics {\bf 4} 639 (2008).  
\bibitem{Agterberg08prl}
  D. F. Agterberg, S. Mukherjee, and Z. Zheng, Phys. Rev. Lett. {\bf
    100}, 017001 (2008).
\bibitem{BakJensen80} P. Bak, M.H. Jensen, J. Phys. C (Solid State)
  {\bf 13} L88 (1980).
\bibitem{BKRprb06} D. Belitz, T. R. Kirkpatrick, and A. Rosch,
  Phys. Rev. B {\bf 73}, 054431 (2006).
\bibitem{Green09prl} A. M. Berridge, A. G. Green, S. A. Grigera,
  B. D. Simons Phys. Rev. Lett. {\bf 102}, 136404 (2009).
\bibitem{Fogler96} A. A. Koulakov, M. M. Fogler, and B. I. Schklovskii, Phys.
Rev. Lett. {\bf 76}, 499 (1996); Phys. Rev. B {\bf 54}, 1853 (1996).
\bibitem{EisensteinPRL99} M. P. Lilly, K. B. Cooper, J. P. Eisenstein, 
L. N. Pfeiffer, and K. W. West, Phys. Rev. Lett. {\bf 82}, 394 (1999).
\bibitem{MacdonaldFisher00prb} A. H. MacDonald and M. P. A. Fisher,
  Phys. Rev. B {\bf 61}, 5724 (2000).
\bibitem{RD02prl} L. Radzihovsky, A. T. Dorsey,
  Phys. Rev. Lett. {\bf 88}, 216802 (2002). 
\bibitem{deGennesProst} P. deGennes and J. Prost,  {\em The Physics of Liquid
  Crystals}, Clarendon Press, Oxford 1993.
\bibitem{ChaikinLubensky}
P. Chaikin and T.C. Lubensky, {\it Principles of Condensed Matter Physics},
Cambridge University Press, Cambridge 1995.
\bibitem{FF}
P. Fulde and R.A. Ferrell,
Phys. Rev. {\bf 135}, A550 (1964).
\bibitem{LO}
A.I. Larkin and Yu.N. Ovchinnikov,
Zh. Eksp. Teor. Fiz. {\bf 47}, 1136 (1964) 
[Sov. Phys. JETP {\bf 20}, 762 (1965)].
\bibitem{Sedrakian05nematic}
A. Sedrakian, J. Mur-Petit, A. Polls, and H. M\"uther,
Phys. Rev. A {\bf 72},  013613 (2005).
%
%
\bibitem{RCprl.09}
L. Radzihovsky and S. Choi, Phys. Rev. Lett. {\bf 103}, 095302 (2009).
\bibitem{HuiZhaiSOIbosons}
  C.-M. Jian, H. Zhai, Phys. Rev B {\bf 84}, 060508 (2011).
\bibitem{YipSOIbosons} S. K. Yip, Phys. Rev. A {\bf 83}, 043616
  (2011).
\bibitem{HuiZhaiSOIfermions}
  Z.-Q. Yu and H. Zhai, Phys. Rev. Lett. {\bf 107}, 195305 (2011).
\bibitem{WangZhaiSOIbosons} C. J. Wang, G. Chao, C. M. Jian, and
  H. Zhai, Phys.  Rev. Lett. {\bf 105}, 160403 (2010).
\bibitem{SpielmanSOInature} Y. J. Lin, K. Jimenez-Garcia, and
  I. B. Spielman, Nature {\bf 471}, 83 (2011).
\bibitem{CRpra.11} S. Choi and L. Radzihovsky, Phys. Rev. A {\bf 84},
  043612 (2011).
\bibitem{dipolarStamperKurn}
M. Vengalattore, S. R. Leslie, J. Guzman, D. M. Stamper-Kurn,
Phys. Rev. Lett. {\bf 100}, 170403 (2008).
\bibitem{kuklov.06}
A. B. Kuklov, Phys. Rev. Lett. {\bf 97}, 110405 (2006).
\bibitem{liu.06}
W. V. Liu and C. Wu, Phys. Rev. A {\bf 74}, 013607 (2006).
\bibitem{TrivediFFLOlattice09} 
Y. L. Loh, N. Trivedi,  arXiv:0907.0679.
\bibitem{CWuFFLOlattice11} Z. Cai, Y. Wang, C. Wu, Phys. Rev. A {\bf
    83}, 063621 (2011).
\bibitem{LiuLattices11}
Z. Zhang, X. Li, W. V. Liu, arXiv:1105.3387.
\bibitem{RVprl} L. Radzihovsky and A. Vishwanath, Phys. Rev. Lett. {\bf
    103}, 010404, (2009).
\bibitem{Rpra} L. Radzihovsky, Phys. Rev. A {\bf 84}, 023611 (2011).
\bibitem{Zwierlein06Science}
M.W. Zwierlein, A. Schirotzek, C.H. Schunck, and 
W. Ketterle, Science {\bf 311}, 492 (2006).
%
\bibitem{Partridge06Science} 
G.B. Partridge, W. Li, R.I. Kamar, Y. Liao, and R.G. Hulet,
Science {\bf 311\/}, 503 (2006).
%
\bibitem{Shin2006prl}
Y. Shin, M.W. Zwierlein, C.H. Schunck, A. Schirotzek, and W. Ketterle, 
Phys. Rev. Lett. {\bf 97}, 030401 (2006).
%
\bibitem{Navon2009prl} S. Nascimb`ene, N. Navon, K.J. Jiang,
  L. Tarruell, M. Teichmann, J. McKeever, F. Chevy, and C. Salomon, 
Phys. Rev. Lett. {\bf 103}, 170402 (2009).
\bibitem{Combescot01}
R. Combescot, Europhys. Lett. {\bf 55}, 150 (2001).
%
\bibitem{Liu03}
W.V. Liu and F. Wilczek,
Phys. Rev. Lett. {\bf 90}, 047002 (2003).
%
\bibitem{Bedaque03}
P.F. Bedaque, H. Caldas, and G. Rupak,
Phys. Rev. Lett. {\bf 91}, 247002 (2003).
%
\bibitem{Caldas04}
H. Caldas,
 Phys. Rev. A {\bf 69}, 063602 (2004).
%
\bibitem{CarlsonReddy05}
J. Carlson and S. Reddy, 
Phys. Rev. Lett. {\bf 95}, 060401 (2005).
%
\bibitem{Cohen05}
T.D. Cohen, Phys. Rev. Lett. {\bf 95}, 120403 (2005).
%
\bibitem{Castorina05}
P. Castorina, M. Grasso, M. Oertel, M. Urban and D. Zappal\`a,
Phys. Rev. A {\bf 72}, 025601 (2005).
%
\bibitem{SRprl}
D.E. Sheehy and L. Radzihovsky, Phys. Rev. Lett. {\bf 96}, 060401 (2006).
%
\bibitem{Pao06}  C.-H. Pao, S.-T. Wu, and S.-K. Yip, Phys. Rev. B {\bf 73}, 132506 (2006).
\bibitem{Son06}
D.T. Son and M.A. Stephanov, Phys. Rev. A {\bf 74}, 013614 (2006).
%
%
\bibitem{Bulgac06pwavePRL} A. Bulgac, M. Forbes, and A. Schwenk,
  Phys. Rev. Lett {\bf 97}, 020402 (2006).
\bibitem{Dukelsky}
J. Dukelsky, G. Ortiz, and S.M.A. Rombouts, and K. van Houcke,
Phys. Rev. Lett. {\bf 96}, 180404 (2006).
%
%
%
\bibitem{Mizushima}
T. Mizushima, K. Machida, and M. Ichioka,
Phys. Rev. Lett. {\bf 94}, 060404 (2005).
%
\bibitem{YangFFLOdetect}
K. Yang,  Phys. Rev. Lett. {\bf 95}, 218903 (2005).
%
\bibitem{YangSachdev}
K. Yang and S. Sachdev, 
Phys. Rev. Lett. {\bf 96}, 187001 (2006).
%
\bibitem{Pieri}
 P. Pieri and G.C. Strinati, Phys. Rev. Lett.
 {\bf 96}, 150404 (2006).
%
\bibitem{Torma}
 J. Kinnunen, L. M. Jensen, and P. T\"orm\"a,
Phys. Rev. Lett. {\bf 96}, 110403 (2006).
%
\bibitem{Yi}
 W. Yi and L.-M. Duan, Phys. Rev. A {\bf 73}, 031604 (2006).
%
\bibitem{Chevy}
F. Chevy, Phys. Rev. Lett. {\bf 96}, 130401 (2006).
%
\bibitem{He}
L. He, M. Jin, and P. Zhuang, Phys. Rev. B {\bf 73}, 214527 (2006).
%
\bibitem{DeSilva}
T.N. De Silva and E.J. Mueller,
Phys. Rev. A {\bf 73}, 051602 (2006).
%
%
\bibitem{HaqueStoof}
M. Haque and H.T.C. Stoof, Phys. Rev. A {\bf 74}, 011602 (2006). 
%
%
\bibitem{SachdevYang}
S. Sachdev and K. Yang, Phys. Rev. B {\bf 73}, 174504  (2006).
%
\bibitem{LiuHu}
X.-J. Liu and H. Hu, 
Europhys. Lett. {\bf 75}, 364 (2006).
%
\bibitem{Chien06prl} C. Chien, Q. Chen, Y. He, and K. Levin,
  Phys. Rev. Lett. {\bf 97}, 090402 (2006).
\bibitem{Gubbels06prl} K.B. Gubbels, M.W.J. Romans, and H.T.C. Stoof, Phys. Rev. Lett. 97, 210402 (2006).
%
%
\bibitem{YiDuan}
W. Yi and  L.-M. Duan, Phys. Rev. A {\bf 74}, 013610 (2006).
%
%
\bibitem{PaoYip}
C.-H. Pao and  S.-K. Yip, J. Phys. Cond. Mat. {\bf 18}, 5567 (2006).
\bibitem{SRaop} 
D.E. Sheehy and L. Radzihovsky, Ann. of Phys. {\bf 322}, 1790 (2007).
\bibitem{SRcomment} D.E. Sheehy and L. Radzihovsky, 
Phys. Rev. B {\bf 75}, 136501 (2007).
\bibitem{Martikainen}
J.-P. Martikainen, Phys. Rev. A {\bf 74\/}, 013602 (2006). 
%
%
\bibitem{Parish07nature}
M. M. Parish, F.M. Marchetti, A. Lamacraft, and B.D. Simons,
Nature Phys. {\bf 3}, 124 (2007).
%
\bibitem{BulgacFFLO} A. Bulgac, M. Forbes,
Phys. Rev. Lett. {\bf 101}, 215301, 2008.
\bibitem{Parish1dLO} M. M. Parish, S. K. Baur, E. J. Mueller, D.
  A. Huse, Phys. Rev. Lett. {\bf 99}, 250403 (2007).
\bibitem{Sheehy} D. E. Sheehy, Phys. Rev. A {\bf 79}, 033606 (2009).
\bibitem{Nishida06eps} Y. Nishida and D. T. Son, Phys. Rev. Lett. 97,
  050403 (2006);
\bibitem{Haussmann06} R. Haussmann, W. Rantner, S. Cerrito, and W. Zwerger, Phys.
Rev. A {\bf 75}, 023610 (2007).
\bibitem{Nikolic07largeN} P. Nikoli\'c and S. Sachdev, Phys. Rev. A {\bf
    75}, 033608 (2007).
\bibitem{Veillette07largeN} M. Y. Veillette, D. E. Sheehy and
  L. Radzihovsky, Phys. Rev. A {\bf 75}, 043614 (2007).
\bibitem{VeilletteRF}
M. Veillette, E. G. Moon, A. Lamacraft, L. Radzihovsky,
S. Sachdev, D.E. Sheehy, Phys. Rev. A {\bf 78}, 033614 (2008).
\bibitem{ProkofevSvistunovPolaron} N. Prokof'ev, B. Svistunov,
  Phys. Rev. B {\bf 77}, 020408(R) (2008).
\bibitem{PilatiGiorginiMCimblanced} S. Pilati, S. Giorgini,
  Phys. Rev. Lett. {\bf 100} 030401 (2008).
\bibitem{ZwierleinPrecision11}
K. Van Houcke, F. Werner, E. Kozik, N. Prokofev, B. Svistunov, M. Ku, 
A. Sommer, L. W. Cheuk, A. Schirotzek, M. W. Zwierlein,  arXiv:1110.3747.
\bibitem{Chandra} B. S. Chandrasekhar, Appl. Phys. Lett. {\bf 1}, 7
  (1962).
\bibitem{Clogston} A.M. Clogston, Phys. Rev. Lett. {\bf 9}, 266
  (1962).
%
%
\bibitem{Sarma}
G. Sarma, J. Phys. Chem. Solids {\bf 24}, 1029 (1963).
%
\bibitem{Alford}
M. Alford, J.A. Bowers, and K. Rajagopal,
Phys. Rev. D {\bf 63}, 074016 (2001).
%
\bibitem{Bowers}
J.A. Bowers and K. Rajagopal,
Phys. Rev. D {\bf 66}, 065002 (2002).
%
\bibitem{Casalbuoni}
R. Casalbuoni and G. Nardulli, Rev. Mod. Phys. {\bf 76}, 263 (2004). 
\bibitem{Combescot}
R. Combescot and C. Mora,
Europhys. Lett. {\bf 68}, 79 (2004).
%
\bibitem{DuineMacDonaldPRA} Y.-P. Shim, R.A. Duine, and
  A.H. MacDonald, Phys. Rev. A {\bf 74}, 053602 (2006).
\bibitem{evidenceFFLO} Recent experiments found evidence of a FFLO
  state in heavy fermion, CeCoIn$_5$ and in organic
  $\kappa$-(BEDT-TTF)$_2$Cu(NCS)$_2$ superconductors. See e.g.,
  A. Bianchi, R. Movshovich, C. Capan, P.G. Pagliuso, and J.L. Sarrao,
  Phys. Rev. Lett. {\bf 91}, 187004 (2003); H.A. Radovan,
  N.A. Fortune, T.P. Murphy, S.T. Hannahs, E.C. Palm, S.W. Tozer, and
  D. Hall, Nature {\bf 425\/}, 51 (2003), and R. Lortz, Y. Wang,
  A. Demuer, P. H. M. Bo\"ttger, B. Bergk, G. Zwicknagl, Y. Nakazawa,
  and J. Wosnitza, Phys. Rev. Lett.  {\bf 99}, 187002 (2007).
%
%
\bibitem{Hulet1dLO} Y. Liao, A. S C. Rittner, T.
  Paprotta, W. Li,
  G. B. Partridge, R. G. Hulet, S. K. Baur, E. J. Mueller, Nature {\bf
    467}, 567 (2010).
\bibitem{Andreev69} A.F. Andreev and I.M. Lifshitz,
  Zh. Eksp. Teor. Fiz. {\bf 56}, 2057 (1969) [Sov. Phys. JETP {\bf
    29}, 1107 (1969)].
\bibitem{Chester70}
G.V. Chester, Phys. Rev. A {\bf 2}, 256 (1970).
\bibitem{Leggett70}
A.J. Leggett, Phys. Rev. Lett. {\bf 25}, 1543 (1970).
\bibitem{KimChan}
E. Kim and M.H.W. Chan, Nature {\bf 427}, 225 (2004); Science {\bf 305}, 1941 (2004).
\bibitem{commentNotSS}A LO state is qualitatively distinguished from a
  supersolid by the absence of a zero momentum condensate. While a
  conventional supersolid also breaks translational and gauge
  symmetry, it is characterized by two {\em independent} order
  parameters, in contrast to a LO state's single product order
  parameters. This puts LO state into a distinct univesality class,
  that is more appropriately referred to as a ``pair-density wave''
  (PDW)\cite{ZhangPDW}.

\bibitem{ZhangPDW} 
H.-D. Chen, O. Vafek, A. Yazdani, S.-C. Zhang,
Phys. Rev. Lett. {\bf 93}, 187002 (2004).
\bibitem{commentStabilizeFFLO} The stability of the FFLO is
  significantly extended in optical
  lattices\cite{CWuFFLOlattice11,TrivediFFLOlattice09} and in one
  dimension\cite{Parish1dLO}. However, our focus here is on a
  lattice-free realization, where rich liquid-crystal phenomenology is
  realized.
\bibitem{WuPaoYip06mass} S.-T. Wu, C.-H. Pao, S.-K. Yip,
Phys. Rev. B {\bf 74}, 224504 (2006).
\bibitem{LevinFFLO} Y. He, C.-C. Chien, Q. Chen, K. Levin
Phys. Rev. A {\bf 75}, 021602(R) (2007).
\bibitem{MachidaNakanishiLO} K. Machida and H. Nakanishi, Phys. Rev. B
  {\bf 30}, 122 (1984).
\bibitem{BurkhardtRainerLO} H. Burkhardt and D. Rainer, Ann. Physik
  {\bf 3}, 181-194 (1994).
\bibitem{MatsuoLO} 
S. Matsuo, S. Higashitani, Y. Nagato, and K. Nagai, 
J. Phys. Soc. Japan {\bf 67}, 280 (1998).
\bibitem{YoshidaYipLO}
N. Yoshida, and S.-K. Yip, 
Phys. Rev. A {\bf 75}, 063601 (2007). 
\bibitem{Shimahara} 
H. Shimahara, J. Phys. Soc. Jpn. {\bf 67}, 1872 (1998).
\bibitem{Landau1dsolid} L. D. Landau, in {\em Collected Papers of
    L. D. Landau}, edited by D. ter Haar (Gordon and Breach, New York,
  1965), p. 209; L. D. Landau and E. M. Lifshitz, {\em Statistical
    Physics} (Pergamon, London, 1969), p. 403.
\bibitem{Peierls1dsolid} R. E. Peierls, Helv. Phys. Acta Suppl. {\bf
    7}, 81 (1934).
\bibitem{MerminWagner} N.D. Mermin, H. Wagner,
Phys. Rev. Lett. {\bf 17}, 1133 (1966).
\bibitem{Hohenberg} P.C. Hohenberg, Phys. Rev. {\bf 158}, 383 (1967).
\bibitem{Berezinksii}
V.L. Berezinskii, Zh. Eksp. Teor. Fiz.
{\bf 59}, 907 (1970) [Sov. Phys. JETP 32, 493 (1971)]; Zh.
Eksp. Teor. Fiz. {\bf 61}, 1144 (1971) [Sov. Phys. JETP {\bf 34}, 610 (1972)].
\bibitem{KT} J.M. Kosterlitz and D.J. Thouless, J. Phys. C {\bf 6},
  1181 (1973).
\bibitem{TonerNelsonSm} J. Toner and D. R. Nelson, Phys. Rev. B {\bf
    23}, 316 (1981).
\bibitem{Schrieffer}
J.R. Schrieffer, {\em Theory of Superconductivity}, Perseus, 1999.
\bibitem{SamokhinFFLO} K. Samokhin, Phys. Rev. B {\bf 81}, 224507
  (2010).
\bibitem{RegalJinRF03}
C.A. Regal and D.S. Jin, Phys. Rev. Lett. {\bf 90}, 230404 (2003).
%
\bibitem{commentMFTtrans} The FFLO-N transition is only continuous
  within mean-field theory, and can be argued to be either generically
  driven first-order by fluctuations.~\cite{Landau1dsolid,Brazovskii}
  or split into intermediate phases as illustrated in
  Fig.\ref{fig:phasediagramLO3d}. This observation is quite generic
  and is due to the fact that LO state spontaneously breaks {\em both}
  translational and rotational symmetries, while the conventional
  paired SF breaks neither. Consequently, at least two continuous
  transitions are required.
\bibitem{PokrovskyTalapov} For a review see e.g.: V.L. Pokrovsky,
  A. L. Talapov and P. Bak in {\em Solitons}, edited by 
S. E. Trullinger, V. E. Zakharov and V. L. Pokrovsky 
(North Holland, Amsterdam, 1986), Chap. 3, pp. 71œôòó127.
\bibitem{Brazovskii} S. A. Brazovskii, Zh. Eksp. Teor. Fiz. {\bf 68},
  175 (1975) [Sov.  Phys. œôòó JETP {\bf 41}, 85 (1975)].
\bibitem{YangLL} K. Yang, 
Phys. Rev. B {\bf 63}, 140511 (2001).
\bibitem{ZhaoLiuLL}
E. Zhao, W. V. Liu
Phys. Rev. A {\bf 78}, 063605 (2008).
\bibitem{deGennes} 
P.-G. de Gennes, 
{\em Superconductivity of Metals and Alloys}, Benjamin, New York, 1966.
\bibitem{Tinkham}
M. Tinkham, {\em Introduction to Superconductivity},
McGraw-Hill, 1996.
%
\bibitem{GP} G. Grinstein and R. A. Pelcovits, Phys. Rev. Lett. {\bf
    47}, 856 (1981).
\bibitem{KetterleBragg} 
D. M. Stamper-Kurn, A. P. Chikkatur, A. G¨orlitz, S. Inouye,
S. Gupta, D. E. Pritchard, and W. Ketterle, Phys. Rev. Lett. {\bf 83}, 
2876 (1999).
\bibitem{Steinhauer02prl} J. Steinhauer, R. Ozeri, N. Katz, and
  N. Davidson, Phys. Rev. Lett. {\bf 88}, 120407 (2002)
\bibitem{Papp08prl} S. B Papp, J. M Pino, R. J Wild, S Ronen, C. E
  Wieman, D. S Jin, E. A Cornell, Phys. Rev. Lett. {\bf 101}, 135301 (2008).
\bibitem{commentBerry} Additional Berry phase effects may arise and
  qualitatively modify the dynamics, as argued in e.g.,
  Ref.\onlinecite{FujimotoFFLOberry}. 
\bibitem{FujimotoFFLOberry} S. Fujimoto, arXiv:1008.5183.
\bibitem{Caille} A. Caille', C. R. Acad. Sci. Ser. B {\bf 274}, 891
  (1972). 
\bibitem{SqSmExp} J. Als-Nielsen, J. D. Litster, R. J. Brigeneau,
  M. Kaplan, C. R. Safinya, A. Lindegaard-Andersen, and S. Mathiesen,
  Phys. Rev. B {\bf 22}, 312 (1980).
\bibitem{Hadzibabic} Z. Hadzibabic, P. Kr\"uger, M.
  Cheneau,
  B. Battelier, and J. Dalibard, Nature {\bf 441}, 1118 (2006).
\bibitem{ChengChinKT} C.-L. Hung, X. Zhang, N. Gemelke, and C. Chin,
  Nature {\bf 470}, 236 (2011).
\bibitem{GW} L. Golubovic, Z. Wang, Phys. Rev. Lett. {\bf
    69} 2535 (1992).
\bibitem{comment2op} These integer vector defects, $\Nv_v$ are
  associated with the fundamental group $\Pi_1$ of the torus
  $U(1)\otimes U(1)$,\cite{commentZ2} that characterizes the
  low-energy manifold of Goldstone modes of the LO state. In this
  respect the LO superfluid has similarities to other $U(1)\otimes
  U(1)$ systems, such as easy-plane spinor-1
  condensates\cite{PodolskyS1prb} and two-gap superconductors, e.g.,
  MgB$_2$\cite{Babaev02prl}.
\bibitem{commentZ2} In terms of the ``rotated'' Goldstone modes
  $\phi=\oh(\phi_++\phi_-)$ and $\theta=\oh(\phi_+-\phi_-)$, the
  low-energy LO manifold is described as a half-twisted torus,
  $\big(U(1)\otimes U(1)\big)/\Z_2$. Because of the structure of the
  Hamiltonian the energetics is best analyzed in terms of
  $\phi,\theta$, while topological constraints on defects are most
  transparently implemented in terms $\phi_\pm$.
\bibitem{PodolskyS1prb} D. Podolsky, S. Chandrasekharan,
  A. Vishwanath, Phys. Rev. B {\bf 80}, 214513 (2009).
\bibitem{Babaev02prl} E. Babaev, Phys. Rev. Lett. {\bf 89}, 067001 (2002).
\bibitem{LiuSineGordon} C. Lin, X. Li, and W. Vincent Liu, Phys. Rev.
  B 83, 092501 (2011).
\bibitem{SenthilFisher} T. Senthil and M. P. A. Fisher, 
Phys. Rev. B. {\bf 62}, 7850 (2000); ibid, {\bf 63}, 134521 (2001).
\bibitem{SachdevDoubleVortex} S. Sachdev, Phys. Rev. B {\bf 45},
  389 (1992).
\bibitem{Balents} L. Balents, M. P. A. Fisher, and C. Nayak,
  Phys. Rev. B {\bf 60}, 1654 (1999).
%
\bibitem{SachdevBook}
S. Sachdev, {\it Quantum Phase Transitions}, Cambridge University Press, 
Cambridge, 1999. 
\bibitem{SachdevMI} S. Sachdev, Annals Phys. {\bf 303}, 226
  (2003).
%
\bibitem{NussinovZaanen} Z. Nussinov, J. Zaanen, J. Phys. IV France
  {\bf 12}, 245 (2002); Physica Status Solidi B {\bf 236}, 332 (2003).
\bibitem{ZagoskinBook} {\em Quantum Theory of Many-Body Systems
Techniques and Applications}, A. M. Zagoskin, New York, Springer, 1998.
\bibitem{comment1dPockets} In contrast, in 1d the effective chemical
  potential generically lies in the gap between Andreev
  bands\cite{SamokhinFFLO}. Thus, a 1d LO state exhibits only
  collective low-energy excitations $\phi,\theta$, consistent with its
  Luttinger-liquid description\cite{YangLL,ZhaoLiuLL}.
\bibitem{SuSchriefferHeeger} W. P. Su, J. P. Schrieffer, A. J. Heeger,
  Phys. Rev. Lett. {\bf 42}, 1698 (1979); Phys. Rev. B {\bf 22}, 2099
  (1980).
%
\bibitem{Mukerjee} S. Mukerjee, C. Xu, and J. E. Moore,
  Phys. Rev. Lett. {\bf 97}, 120406 (2006).
\bibitem{commentU_N} We distinguish two independent $U(1)$ symmetries,
  $U_N(1)$ and $U_{\Delta N}(1)$, associated with independent total
  atom number $N=N_1+N_2 + 2N_m$ conservation and the difference
  $\Delta N = N_1-N_2$ conservation, respectively.
\bibitem{lee.lee.04}
Y. W. Lee and Y. L. Lee, Phys. Rev. B. {\bf 70}, 224506 (2004).
\bibitem{zoller1.10}
S. Diehl, M. Baranov, A. J. Daley, and P. Zoller,
Phys. Rev. B {\bf 82}, 064509 (2010).
\bibitem{zoller2.10}
S. Diehl, M. Baranov, A. J. Daley, and P. Zoller,
Phys. Rev. B {\bf 82}, 064510 (2010).
\bibitem{rempe.08} 
N. Syassen, D. M. Bauer, M. Lettner, T. Volz,
  D. Dietze, J. J. Garcia-Ripoll, J. I. Cirac, G. Rempe, and
  S. D\"urr, Science {\bf 320}, 1329 (2008).
\bibitem{bethe}
H. A. Bethe {\it Handbuch der Physik} (1933), second edition, Vol. 24/1, pp. 452-462.
\bibitem{Ejima.Simons.11}
S. Ejima, M. J. Bhaseen, M. Hohenadler, F. H. L. Essler, H. Fehske, and B. D. Simons,
Phys. Rev. Lett. {\bf 106}, 015303 (2011).
\bibitem{Bhaseen.Simons.09}
M. J. Bhaseen, A. O. Silver, M. Hohenadler, and B. D. Simons,
Phys. Rev. Lett. {\bf 103}, 265302 (2009).
\bibitem{Bonnes.Wessel.11}
L. Bonnes and S. Wessel,
Phys. Rev. Lett. {\bf 106}, 185302 (2011).
\bibitem{matthews.cornell.98} M. R. Matthews, D. S. Hall, D. S. Jin,
  J. R. Ensher, C. E. Wieman, E. A. Cornell, F. Dalfovo, C. Minniti,
  and S. Stringari, Phys. Rev. Lett. {\bf 81}, 243 (1998).
\bibitem{stenger.ketterle.98} J. Stenger, S. Inouye,
  D. M. Stamper-Kurn, H.-J. Miesner, A. P. Chikkatur, and W. Ketterle,
  Nature {\bf 396}, 345 (1998).
\bibitem{stamperkurn.98} D. M. Stamper-Kurn, M. R. Andrews,
  A. P. Chikkatur, S. Inouye, H.-J Miesner, J. Stenger, and
  W. Ketterle, Phys. Rev. Lett. {\bf 80}, 2027 (1998).
\bibitem{ho.spinor.98}
T.-L. Ho, Phys. Rev. Lett. {\bf 81}, 742 (1998).
\bibitem{ohmi.machida.98}
T. Ohmi and K. Machida, J. Phys. Soc. Japan {\bf 67}, 1822 (1998).
\bibitem{ho.yip.00}
T.-L. Ho and S. K. Yip, Phys. Rev. Lett. {\bf 84}, 4031 (2000).
\bibitem{zhou.spinor.01}
F. Zhou, Phys. Rev. Lett. {\bf 87}, 080401 (2001).
\bibitem{demler.zhou.02}
E. Demler and F. Zhou, Phys. Rev. Lett. {\bf 88}, 163001 (2002).
\bibitem{barnett.demler.06}
R. Barnett, A. Turner, and E. Demler, Phys. Rev. Lett. {\bf 97}, 180412 (2006).
\bibitem{mukerjee.moore.06}
S. Mukerjee, C. Xu, and J. E. Moore, Phys. Rev. Lett. {\bf 97}, 120406 (2006).
\bibitem{kimble.90}
H. J. Kimble, {\it Quantum fluctuations in quantum optics - squeezing and related phenomena}, les Houches,
Session LIII, 1990 (Elsvier, 1992) and references therein.
\bibitem{kasevich.01} C. Orzel, A. K. Tuchman, M. L. Fenselau,
  M. Yasuda, and M. A. Kasevich, Science {\bf 291}, 2386 (2001).
\bibitem{RegalPwave03} C. A. Regal, C. Ticknor, J. L. Bohn, and
  D. S. Jin, Phys. Rev. Lett. {\bf 90}, 053201 (2003).
\bibitem{commentFRequalmass} 
For unequal masses FR interaction is given by
$H_{FR}=\alpha\hat{\bm\phi}^{\dag}\cdot
\left[\frac{m_1}{m_1+m_2}\hat\psi_{1}(-i\grad)\hat\psi_{2} 
-\frac{m_2}{m_1+m_2}\hat\psi_{2}(-i\grad)\hat\psi_{1}
\right]+h.c.$
as required to preserve Galilean invariance.
\bibitem{greene.97} B. D. Esry, C. H. Greene, J. P. Burke, Jr., and
  J. L. Bohn, {\it et al.}, Phys. Rev. Lett. {\bf 78}, 3594 (1997).
\bibitem{LiuFisher} K.-S. Liu and M. E. Fisher, J. Low
  Temp. Phys. {\bf 10}, 655 (1973).
\bibitem{KNF} J.M. Kosterlitz, D.R. Nelson, and M.E. Fisher, 
Phys. Rev. B 13, 412 (1976).
\bibitem{Vicari} P. Calabrese, A. Pelissetto, and E. Vicari,
Phys. Rev. B {\bf 67}, 054505 (2003)
\bibitem{HLM74}
B. I. Halperin, T. C. Lubensky, and S. K. Ma, Phys. Rev. Lett. {\bf 32}, 292 (1974).
\bibitem{Coleman73}
S. Coleman and E. Weinberg, Phys. Rev. D {\bf 7}, 1888 (1973).

\end{thebibliography}
\end{document}